%% file: main.tex
\documentclass[%
 reprint,
superscriptaddress,
 amsmath,amssymb,
pra,
]{revtex4-2}
\usepackage{graphicx}
\usepackage{dcolumn}
\usepackage{nicematrix,tikz}
\usepackage{hyperref}
\usepackage{soul}

\usepackage{graphicx}
\usepackage{subfig}
\usepackage{booktabs}
\usepackage[export]{adjustbox}
\usepackage{ragged2e}
\usepackage{bm}
\usepackage{xcolor}
\newcommand{\Lim}[1]{\raisebox{0.5ex}{\scalebox{0.8}{$\displaystyle \lim_{#1}\;$}}}

\newcommand{\expnumber}[2]{{#1}\mathrm{e}{#2}}

\begin{document}

\preprint{APS/123-QED}

\title{Compact quantum algorithms for time-dependent differential equations }

\author{Sachin S. Bharadwaj}
\email{sachin.bharadwaj@nyu.edu}
 \affiliation{Department of Mechanical and Aerospace Engineering, New York University, New York 11201 USA}
 
\author{Katepalli R. Sreenivasan}
\email{katepalli.sreenivasan@nyu.edu}
 \affiliation{Department of Mechanical and Aerospace Engineering, New York University, New York 11201 USA}
\affiliation{Courant Institute of Mathematical Sciences, New York University, New York, NY 10012}
 \affiliation{Department of Physics, New York University, New York, NY 10012}
%

\date{\today}

\begin{abstract}
\noindent Many claims of computational advantages have been made for quantum computing over classical, but they have not been demonstrated for practical problems. Here, we present algorithms for solving time-dependent PDEs, with particular reference to fluid equations. We build on an idea based on linear combination of unitaries to simulate non-unitary, non-Hermitian quantum systems, and generate hybrid quantum-classical algorithms that efficiently perform iterative matrix-vector multiplication and matrix inversion operations. These algorithms are end-to-end, with relatively low-depth quantum circuits that demonstrate quantum advantage, with the best-case asymptotic complexities, which we show are near-optimal. We demonstrate the performance of the algorithms by conducting: (a) fully gate level, state-vector simulations using an in-house, high performance, quantum simulator called \textit{QFlowS}; (b) experiments on a real quantum device; and (c) noisy simulations using Qiskit Aer. We also provide device specifications such as error-rates (noise) and state sampling (measurement) to accurately perform convergent flow simulations on noisy devices. The results offer evidence that the proposed algorithm is amenable for use on near-term quantum devices.

\end{abstract}

\maketitle
\section{Introduction}
\label{sec:Intro}
 Simulating nonlinear phenomena is quite arduous. In the case of hydrodynamic turbulence, for instance, the combination of a wide range of scales and the need for fine computational resolution \cite{yeung2020advancing} makes it quite challenging \cite{iyer2021area,buaria2022scaling}. Similar challenges are encountered in other problems such as glassy and molecular dynamics, protein folding and chemical reactions. Simulating such phenomena requires a paradigm shift in computing---and a strong candidate for it is Quantum Computing (QC). QC {may} surpass classical digital computing in some cases \cite{alexeev2021quantum,awschalom2022roadmap}, but it is yet to demonstrate its power in solving practical problems. See, e.g., Refs.~{\cite{bharadwaj2020quantum,bharadwajintroduction,bharadwaj2024simulating}} for a discussion of this point, especially on fluid dynamics. A general point is that most classical systems are nonlinear, whereas quantum algorithms---consisting of quantum gates and quantum circuits---obey laws of quantum mechanics, which are linear and unitary. Reconciling this mismatch makes linearization of the problem inevitable  \cite{lin2022koopman,giannakis2022embedding,liu2021efficient,costa2023further,gonzalez2024quantum,bharadwaj2023quantum,liu2023efficient}, leading to high dimensional linear problems. It is thus necessary to create efficient quantum algorithms to solve high dimensional linear system of equations. Even without linearization, solving nonlinear partial differential equations (PDEs) would still demand, at the least, an efficient way to iteratively operate general non-unitary and non-hermitian matrices that are also endemic to non-hermitian and open quantum systems.  In both cases, one needs a quantum algorithmic framework to perform iterative matrix-vector products and matrix inversion operations efficiently. They are generally of the form
 \begin{equation}
     M\mathbf{u} \textrm{~~~and~~~} M^{-1}\mathbf{u},
 \end{equation} where $M$ and $\mathbf{u}$ represent, respectively, a given matrix operator (e.g., discretized differential operator) and a complex-valued vector---e.g., a velocity field. Indeed, these operations are ubiquitous in computational applications in a wide range of fields.
 
 In this paper, we propose algorithms for solving time-dependent PDEs by reducing them to the above matrix operations and marching them forward in time. In particular, we demonstrate the process by solving a fluid flow problem, including its implementation on an existing quantum computer, and make the case for its suitability on near-term machines. In the ensuing discussion, we first review the related literature and, with that background, highlight the advancement made by the current work. {The proposed procedure opens up a promising new avenue towards realizing quantum acceleration of a wide range of practical problems beyond fluid dynamics.} 
 
\subsection{Related Work}

Quantum computing methods for PDEs, fluid flow problems included, may be classified into three major categories:  
\begin{itemize}
    \item \textbf{Category A} -- Quantum Direct Numerical Simulations (QDNS) \cite{harrow2009quantum,berry2014high,berry2017quantum,subacsi2019quantum,costa2019quantum,liu2021efficient,childs2021high,costa2022optimal,costa2023further,krovi2023improved,liu2023efficient,an2023linear,fang2023time,bharadwaj2023hybrid,an2023quantum,gonzalez2024quantum,dalzell2024shortcut,berry2024quantum,brearley2024quantum,leong2022variational,lubasch2020variational,ingelmann2024two,wright2024noisy,pool2024nonlinear}
\item \textbf{Category B} -- Lattice Boltzmann and Liouville equation methods \cite{todorova2020quantum,budinski2021quantum,bakker2023quantum,li2023potential,itani2024quantum,succi2024three,succi2024ensemble,lin2022koopman,schalkers2024efficient,kocherla2023fully,kocherla2024two,penuel2024feasibility}
\item \textbf{Category C} -- Schr\"odinger equation methods \cite{jin2022quantum,jin2023quantum,lu2024quantum,meng2023quantum,jin2024quantum}
\end{itemize} 

The present work belongs to category A, in which one solves nonlinear continuum equations without modelling or mapping to analogous problems. 
Below, we compare the present work with the state-of-the-art in categories B and C, whose hallmark is that they map the original problem into alternative formulations.

{Category A -- In this category, we consider the Quantum Linear Systems Algorithms (QLSA)---on which the present work is based---and Variational Quantum Algorithms (VQA). VQA approach solves the governing PDE as an optimization problem \cite{leong2022variational,lubasch2020variational,ingelmann2024two,kocher2024numerical,pool2024nonlinear,bravo2023variational}, where a parameterized velocity field is deemed to represent the desired solution when certain parameters correspond to the minimum of a suitably defined cost function. This approach has advantages such as relatively short depth circuits \cite{cerezo2021variational,ingelmann2024two,kocher2024numerical}, the possibility of encoding nonlinearity and minimal measurements (single-valued cost function output) \cite{lubasch2020variational,pool2024nonlinear}. However, they are slow  to converge \cite{ingelmann2024two}, and are prone to the barren-plateau problem, over-parameterization ansatzes \cite{cerezo2021variational,holmes2022connecting,pool2024nonlinear}, as well as the lack of complexity guarantees on quantum advantage \cite{ingelmann2024two,leong2022variational}. 

QLSA offers complexity guarantees on quantum advantage and are devoid of barren-plateaus and heuristics from optimization. The QLSA-based methods entail solving a linear system of equations derived from the (non)linear governing PDEs. Such a construct is obtained by a combination of discretization methods (e.g., finite difference or finite element methods) and linearization techniques such as Carleman or Koopman embedding techniques, yielding a high dimensional linear system of equations, which is then solved using QLSA. Starting from the Harrow-Hassidim-Lloyd (HHL) algorithm \cite{harrow2009quantum}, these sets of algorithms have undergone a considerable evolution \cite{berry2014high,berry2017quantum,subacsi2019quantum,liu2021efficient,childs2021high,krovi2023improved,an2023linear,bharadwaj2023hybrid,fang2023time,an2023quantum,berry2024quantum,costa2022optimal,costa2023further,dalzell2024shortcut,liu2023efficient}, by overcoming several challenges that impeded preceding proposals (see \cite{aaronson2015read, bharadwaj2023hybrid}). This has been aided largely by advances made in improving Hamiltonian simulation algorithms \cite{berry2007efficient,berry2014exponential,berry2015simulating,berry2015hamiltonian,low2017optimal,costa2019quantum,berry2020time,an2021time} as well as the introduction of the concept of Linear Combination of Unitaries (LCU) \cite{childs2012hamiltonian,childs2017quantum}. The central challenges of QLSA include (i) linear, non-optimal dependence on matrix system parameters such as grid size and number of time steps, matrix sparsity and condition number, and the specified accuracy; (ii) exponentially growing query complexity, requiring multiple copies of the initial state and repeated measurements and amplitude amplification of intermediate and final solution states; (iii) specific constraints on the properties of matrix operators such as hermiticity, unitarity, positive-definiteness and on the range of eigenvalues; (iv) stability requirements for time marching problems, and (v) the depth of quantum circuits due to quantum phase estimation methods and inefficient LCU decomposition. The more recent works, particularly those based on LCU, have addressed some of these aspects, while some \cite{bharadwaj2023hybrid} have also (in the context of fluid dynamics) attempted to make the algorithms end-to-end, by addressing problems of state preparation and post-processing quantum information.

{Category B -- The Lattice Boltzmann Method offers an alternative, meso-scale formulation of fluid flows, where a discrete set of probability distributions are tracked on a fixed lattice, with a restricted set of velocity directions and magnitudes. Although the corresponding quantum algorithms exhibit certain innate advantages, they still face certain challenges \cite{todorova2020quantum,budinski2021quantum,lin2022koopman,bakker2023quantum,li2023potential,kocherla2023fully,itani2024quantum,succi2024three,succi2024ensemble,schalkers2024efficient,kocherla2024two,penuel2024feasibility}. Besides the errors from inaccurate quantum circuit modelling of nonlinear collision-streaming operations \cite{itani2024quantum} and physical boundary conditions \cite{todorova2020quantum}, their computational advantage tend to be downgraded because (a) qubit complexity scales linearly with problem size (grid size and time steps) when lattice positions are encoded as binary state vectors, leading to non-unitarity of streaming or collision steps \cite{kocherla2023fully,schalkers2024importance}, (b) repeated measurements are needed at every time step \cite{kocherla2024two,budinski2021quantum} and (c) multiple copies are required of the initial state \cite{itani2024quantum}. Although there have been efforts to ameliorate each of these constraints, a single algorithm that surpasses all these bottlenecks remains elusive. Another strategy would be to solve the linear Liouville equation  \cite{lin2022koopman,succi2024ensemble} that avoids the fundamentally nonlinear governing equation. However, this is still prone to errors from spurious effects that appear with increasing grid resolution. It seems that the best computational complexity for the algorithms in this category emerges when the problem is translated into a linear system of equations, solved subsequently by a Quantum Linear Systems Algorithm (QLSA) \cite{li2023potential,penuel2024feasibility,bakker2023quantum} (category A).}

{Category C -- Algorithms in this category map the original continuum governing PDEs into a Schr\"odinger or a Schr\"odinger-like equation. The algorithm presented in \cite{meng2023quantum} maps the Navier-Stokes equations into an analogous hydrodynamic, nonlinear Schr\"odinger equation by using a Madelung transformation, which is then simulated as a quantum circuit. The upside of this approach stems from its innate quantum mechanical construction, but the overall complexity of the algorithm scales linearly with the problem size and an asymptotic computational advantage remains elusive. A potential alternative is that of Schr\"odingerisation \cite{jin2022quantum,jin2023quantum,lu2024quantum,hu2024quantum,jin2024quantum}, which offers a better computational complexity compared to other works within this category. The approach builds an exact mapping from a classical PDE into a dilated Schr\"odinger equation having an extra dimension with an effective Hamiltonian. However, in the absence of linearization methods, such mappings are restricted to a specific set of PDEs rather than to general nonlinear PDE's such as the Navier-Stokes equations \cite{lu2024quantum}. However, this approach tends to be a natural candidate for analog quantum computers rather than for current gate-based quantum computers, and exhibit a similar computational complexity as some earlier gate-based QLSA proposals. There is still a non-optimal, linear dependence (gate complexity) on the solution accuracy and time 
\cite{hu2024quantum} for simulating problems that are similar to those considered in this paper.}

\subsection{Current Contribution}
{This work presents a set of hybrid, quantum algorithms based on \textcolor{red}{LCU} for solving time-dependent PDEs, with an application to unsteady flows. We make use of a particular LCU decomposition, which has been used earlier in a different context of non-hermitian open quantum systems \cite{schlimgen2021quantum} and quantum chemistry applications, and transform it into a broader framework of a hybrid quantum-classical algorithm for iterative matrix operations. This is accomplished by introducing six Time Marching Compact Quantum Circuit (TMCQC) algorithms to solve the time-dependent PDEs, admitting both explicit and implicit time stepping schemes. These algorithms require only two, or at most four, controlled unitaries for each LCU decomposition leading to relatively low-depth quantum circuits. We show that, of the six TMCQCs proposed, the gate complexity (of basic two- and one-qubit gates) is the best for (TMCQC-5,6), and is $\mathcal{O}\Big(s\epsilon,\log(N_{g}\tau),\textrm{polylog}(\epsilon/\varepsilon_{U},1/\varepsilon_{N}),\log^{-3}((\kappa-1)^{-1})\Big)$. Here, the governing PDEs are discretized using a finite difference method with $N_{g}$ as the grid size, $\tau$ as the number of time steps (thus an overall problem size of $\mathcal{O}(N_{g}\tau)$), $s$ and $\kappa$ being the sparsity and condition numbers of the matrix operators, $\epsilon$ a parameter controlling the LCU decomposition and $\varepsilon_{U}$ and $\varepsilon_{N}$ determine the accuracy of different approximations of the algorithm, as discussed in Section \ref{sec: TMCQC}. We use these algorithms to simulate a linear advection-diffusion problem and use its analytical and classical solutions as reference for estimating the accuracy of the solutions obtained.  The simplicity of this problem makes it an ideal candidate for assessing the performance of the quantum algorithm. We show that the results from the proposed algorithm agree qualitatively and quantitatively with analytical results. }
 
The above complexity scaling is near-optimal in all parameters except $s$ and $\epsilon$. By \textit{optimal} we mean that the algorithmic complexity (qubit, query and two-qubit gate complexities) scale logarithmically
or at most poly-logarithmically in the system parameters. 
By \textit{near-optimal}, we imply that complexity is
optimal in most parameters but a few, in this case $s$ and $\epsilon$. However, $s$ is equal to the order of finite difference scheme and therefore contributes only a small constant prefactor and $\epsilon$, as will shown in Section \ref{sec:Query complexity}, may be maintained close to unity, thus adding only modest overhead to the complexity. There is thus the potential for an overall exponential advantage. The complexity (TMCQC-1,2,4) is either linear in $\tau$, or quadratic at worst, while still scaling optimally in other parameters. The qubit complexity scales as $\mathcal{O}(\log(N_{g}\tau))$ at best and $\mathcal{O}(\log(N_{g}\tau (\log(\varepsilon^{-1}_{N}))/\log((\kappa-1)^{-1})))$ at the worst, which is still optimal in both cases and provides an exponential memory advantage. These complexities are without quantum amplitude amplification, whose inclusion gives an improvement up to quadratic. These are significant advances with respect to existing proposals in categories B, C as well as VQA in category A.

Now with respect to the QLSA approaches, the improvements of the present algorithm are that it (a) avoids expensive Quantum Phase Estimation with a near-optimal dependence on the different parameters of the matrix operators; 
(b) admits non-unitary and non-hermitian matrices, without restrictions on the range of eigenvalues. In comparison with some recent, time-marching proposals \cite{fang2023time,an2023quantum,berry2024quantum}, it uses tools such as the Richardson extrapolation to lower the overall query complexity of the algorithm, thus ameliorating the dependence on the amplification ratio  {\cite{fang2023time}; for the particular problem considered here, in comparison to \cite{an2023quantum,fang2023time,brearley2024quantum}, the present work has a complexity that is logarithmic in $\tau$, $\kappa$ (and $||M||$).} (c) it provides complexity estimates for various parameters of the matrix, and avoids the need to compute integrals by using quadratures; and (e) it provides a simple and straightforward LCU decomposition that can be extended to other problems besides the specific example considered here.

The present end-to-end algorithm includes specific Quantum State Preparation and Quantum Post Processing algorithms to to demonstrate quantum advantage, by introducing both explicit and implicit time stepping methods. We provide a fully gate level implementation of state vector simulations using QFlowS---an in-house, high-performance, quantum simulator \cite{bharadwaj2023hybrid,bharadwaj2024qflows}. We also use a real quantum device (IBM Cairo) and study effects of noise using IBM Qiskit Aer platform and propose circuit designs and strategies for accurate, forward time simulations on near-term quantum devices. By choosing suitable problem sizes, our noisy simulation studies of the proposed quantum circuits indicate that the required gate error rates and execution times for convergent simulations are in the ballpark of the state-of-the-art devices and experimental specifications, thus hinting the possibility of retaining quantum advantage. The study also indicates how noise can be used to our advantage.

In Section \ref{sec: Governing equations} we describe governing PDEs that will be used to assess the performance of the algorithms. Section \ref{sec: LCU} introduces the specific linear combination of unitaries for constructing the time marching quantum circuits; they are discussed in detail in Section \ref{sec: TMCQC}. In Section \ref{sec: End-to-end}, we present end-to-end strategies in view of current and near-term devices, followed in Section \ref{sec: Numerical results} by numerical results on simulators and a real quantum device. Finally, we summarize the discussion and outline our conclusions in Section \ref{sec: Conclusions}.\\

\section{Governing equations}
\label{sec: Governing equations}
The PDEs considered here reflect the momentum and mass conservation (assuming no body forces or source terms) and are of the form 
\begin{align}
        \frac{\partial \textbf{u}}{\partial t} + \mathbf{C}\cdot\nabla \mathbf{u} = \frac{1}{Re} \nabla ^{2} \textbf{u} -  \nabla \textbf{p}, \label{eq:governing1}
\end{align}
\begin{equation}
    \nabla \cdot \textbf{u} = 0,
    \label{eq:incompressibility}
\end{equation}
where $\mathbf{u} = (u,v,w)$ is the velocity, $\mathbf{C}$ is the advection velocity, $\mathbf{p}$ is the pressure, $Re = UL/\nu$ is the Reynolds number---$U$ being a characteristic velocity, $\nu$ the kinematic viscosity and $L$ a characteristic length. 
When $\mathbf{C}=\mathbf{u}$, it represents the full nonlinear Navier-Stokes equations, while if $\mathbf{C}=\text{constant}\neq 0$ one has the linear advection-diffusion equation; $\mathbf{C}=0$ represents the well-known linear Poiseuille/Couette flow equation.  
Here, we particularly consider the one-dimensional, linear, advection-diffusion problem as the running example for all discussions, given by
\begin{equation}
\frac{\partial u}{\partial t} + C\frac{\partial u}{\partial y}  = D \frac{\partial^{2} u}{\partial y^{2}} ,
    \label{eq:advcetion-diffusion flow}
\end{equation}
where the velocity varies only along y (wall-normal direction). 
We set $C=U$ (the constant advection velocity in the x-direction), and we put $D$, representing the diffusion coefficient, to unity. The initial condition $u(L/2,0)$ is chosen as the delta function $\delta(x)$ and the system is subject to a periodic boundary condition $u(0,t) = u(L,t)$. Such a setting also admits an analytical solution \cite{ingelmann2024two}, making it an ideal test-case to evaluate the accuracy of the quantum solutions.  
The algorithmic framework developed here is also agnostic to the linearity or otherwise of the PDEs. For instance, if we were to begin by considering the nonlinear case ($C=\mathbf{u}(x,t)$), a preceding linearization such as a Homotopy Analysis and Carlemann or Koopman method \cite{liu2021efficient,liu2023efficient,costa2023further,giannakis2022embedding,lin2022koopman,bharadwaj2023quantum,gonzalez2024quantum} can be applied to first obtain an approximate, higher dimensional linear system of equations. Such a system can then be solved using the proposed methods.

\section{Linear Combination of Unitaries}
\label{sec: LCU}
The PDEs are first discretized in space and time, using an appropriate finite difference scheme, to obtain a linear system of equations to be solved by the quantum algorithm. A detailed account of this numerical setup is presented in Section \ref{sec: Numerical Setup} of the {\textit{Appendix}}. Computing numerical solutions of the approximate system of equations translates to iterative operations of the form $Mb$ or $M^{-1}b$. Here, $M$ and $b$ represent a general finite difference matrix and the instantaneous velocity field, respectively, both constructed classically with an underlying numerical method. These are just the operations that we now intend a quantum computer to perform efficiently. 
To do this, we design a hybrid quantum-classical workflow as depicted in figure \ref{fig:work flow}.

(a) The first step is to encode the discretized velocity field, $\mathbf{u}(x,t)$, as the amplitudes of an $n_{g}=\log_{2}(N_{g})$ qubit quantum state, $|\Psi\rangle_{u}=\frac{1}{\mathcal{N}}\sum_{j=0}^{N_{g}-1}u_{j}|j\rangle$, where $\mathcal{N}$ is an appropriate normalization constant such that $\vert\vert~|\Psi\rangle_{u}~ \vert\vert=1$. The specific structure of such a vector (for a general $b$) would depend on the numerical scheme. This step is performed by a Quantum State Preparation algorithm, which is made of a quantum circuit that efficiently encodes the input values on to qubits using a set of 2 qubit gates. To be able to achieve quantum advantage it is important that this step is done efficiently, with a complexity less than $\mathcal{O}(N_{g})$. It is, in fact, possible to do so if they meet certain sparsity requirements, or have a specific functional form \cite{bharadwaj2023hybrid,vazquez2022enhancing,mozafari2022efficient,grover2002creating} (denoted as operator $R$ from here on). Here, since we use a delta function as our initial condition, such a state can be easily prepared with a single NOT gate, thus lowering the onus on state preparation. However, as noted in \cite{bharadwaj2023hybrid}, while arbitrary states can be exponentially expensive to prepare, certain generic states (more general than a delta function) with specific features such as convex functional forms, sparsity and random initializations can be prepared efficiently. Furthermore, problems in fluid dynamics such as turbulence are flexible enough to admit a wide range of initial conditions including random states, thus alleviating the state preparation overhead to a certain extent.

(b) The next step is to apply the finite difference operator $M$ on the prepared state as $M|\Psi\rangle$ (or iteratively as $M^{\tau}|\Psi\rangle$ to march $\tau$ time steps). This is a non-trivial task, since $M$ is neither unitary nor hermitian in general. 
To make this possible, we invoke the concept of a linear combination of unitary operators. With this, an arbitrary matrix may be approximated by a weighted sum of unitary operators, which themselves can be broken down into fundamental quantum gates. For this, one could use different algorithms proposed earlier \cite{childs2012hamiltonian,zheng2021universal,berry2014exponential,berry2015simulating}. In this work, however, we redesign a strategy that was previously proposed in the context of simulating non-unitary and non-hermitian quantum systems  \cite{schlimgen2021quantum,berry2015simulating} into a PDE solver algorithm. This redesigned algorithm requires only 2, or at most 4, controlled unitaries for the linear unitary decomposition, reducing the circuit depths drastically, making the overall algorithm more amenable to use on near-term devices. Further, invoking specific techniques that we explain in Section \ref{sec: End-to-end}, these unitaries would then correspond to 15 and 139 two-qubit gates for two and four qubit circuits, with depths of 10 and 95 gates, respectively.

(c) The solutions can then be measured for storage and post-processing classically; however, this would diminish the quantum advantage, since measuring the state is an $\mathcal{O}(N)$ operation. Instead, one could also perform quantum post processing to compute linear or nonlinear observables of the flow field, with the quantum computer outputting a single real value that can measured efficiently. We discuss these in more detail in Section \ref{sec: End-to-end}. The following discussion will describe the proposed decomposition in terms of a linear combination of unitaries.

 \begin{figure}[htpb!]
    \centering
    \includegraphics[trim={0.4cm 0cm 0cm 0cm},clip=true,scale=0.26]{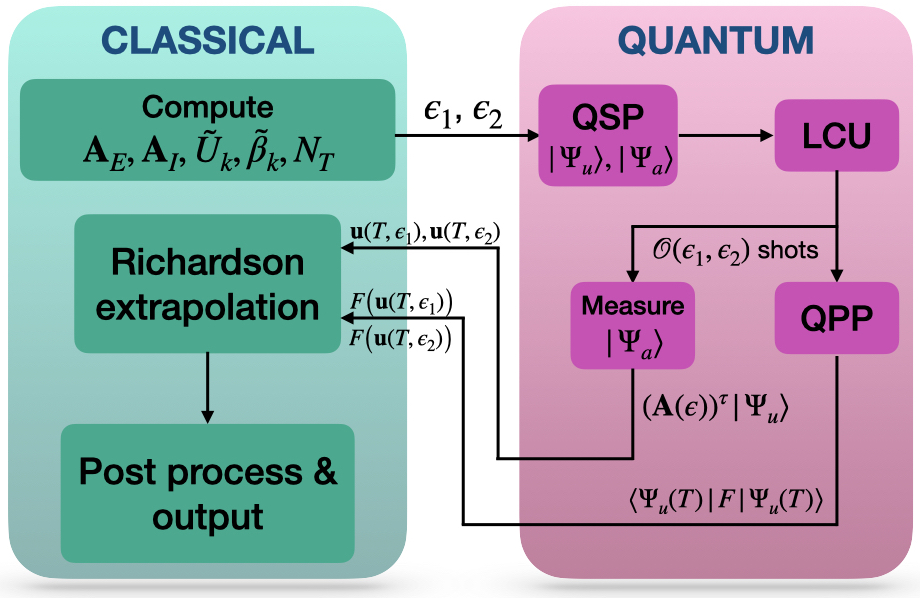}
    \caption{\justifying The figure shows the working flowchart of the hybrid quantum-classical algorithm. The quantum circuit and system parameters are first computed classically and then loaded on the quantum device using quantum state preparation. The forward time simulation is then performed using the algorithm on the linear combination of unitaries. The solutions are then either post-processed quantumly or measured for storage on a classical device. This process is iterated over several times and the measured information is extrapolated classically using Richardson extrapolation, to finally output the solution.}
    \label{fig:work flow}
\end{figure}

\textbf{(1) Four unitaries:} A non-unitary matrix $M$ can be decomposed into symmetric and anti-symmetric matrices, $S$ and $A$, respectively, as
\begin{align}
    S &= \frac{1}{2}(M+M^{\dag}) ~\text{\&}\\
    A &= \frac{1}{2}(M-M^{\dag}),
\end{align}
where $M =  S+A$. Each of these matrices can be further decomposed exactly as
\begin{align}
    S &= \lim_{\epsilon\rightarrow 0}\frac{i}{2\epsilon}(e^{-i\epsilon S }-e^{i\epsilon S})\label{eq:sym} \\ 
    A &= \lim_{\epsilon\rightarrow 0}\frac{1}{2\epsilon}(e^{\epsilon A}-e^{-\epsilon A}),
    \label{eq:antisym}
\end{align} where $\epsilon$ is an expansion parameter. It can be easily shown that $e^{\pm iS}$ and $e^{\pm A}$ are both unitary matrices. We shall now define these operators as $U_{0} = ie^{-i\epsilon S}$, $U_{1} = -ie^{i\epsilon S}$, $U_{2} = e^{\epsilon A}$ and $U_{3} = -e^{-\epsilon A}$, and express $M$ exactly as a weighted sum of purely unitary operators given by
    $M = \sum_{k=0}^{k=3}\beta_{k}U_{k}=\lim_{\epsilon\rightarrow 0}\frac{1}{2\epsilon}(U_{0}+U_{1}+U_{2}+U_{3})$. However, in practice, we approximate $M$ by choosing a \textit{small enough} $\epsilon$, thus resulting in a decomposition with just 4 unitaries given by
    \begin{equation}
        \tilde{M} = \frac{1}{2\epsilon}(U_{0}+U_{1}+U_{2}+U_{3}).
        \label{eq:lcu}
    \end{equation}
    %
Since the goal is to be able to simulate our algorithms on NISQ and near-term devices, it is desired to have a reduction in circuit depth. We now show that we can go a step further to reduce this decomposition to only two unitaries.

\textbf{(2) Two unitaries:}  We can attain this reduction by trading in one extra qubit. For this, consider a simple hermitian dilation of the original matrix $M$ given by
\begin{equation}
    \hat{M} = \begin{pmatrix}
        0 & M\\
        M^{\dag} & 0
    \end{pmatrix}.
\end{equation}
Since $M_{ij} \in \mathbb{R}$, the matrix $\hat{M}$ is symmetric with zero anti-symmetric part. This would leave us with a symmetric, hermitian, and Hamiltonian $\hat{M}$ that can be decomposed into just two unitaries as \begin{equation}
   \hat{M} =\lim_{\epsilon\rightarrow 0}\frac{i}{2\epsilon}(e^{-i\epsilon \hat{M} }-e^{i\epsilon \hat{M}}) = \frac{1}{2\epsilon}(\hat{U}_{0}+\hat{U}_{1}).
   \label{eq: lcu2}
\end{equation} Where, $e^{-i\epsilon \hat{M} } = \hat{U}_{0}$ and $-e^{i\epsilon \hat{M}} = \hat{U}_{1}$. However, to accommodate this reduction, the size of input vector would need to be doubled and given by $\hat{b}=[0,b]$, where one half of the dilated vector is padded with zeros. This doubling would require only one additional qubit. From here on, let $K$ represent the number of unitaries corresponding to either two or four unitaries. We now examine how to implement the decomposition as a quantum circuit.


The basic circuit requires two quantum registers, both set to 0 initially: (1) $|\Psi\rangle_{u}$ -- to store the state vector that is to be operated on, and (2) $|\Psi\rangle_{a}$ -- a register with a total of $n_{a}=\log_{2}(K)$ ancillary qubits (here, $n_{a}\in\{1,2\}$). $|\Psi\rangle_{u}$ is prepared using the operator $R$ which, in our case, is simply a NOT gate to prepare the initial delta function. $|\Psi\rangle_{a}$, on the other hand, is prepared by an operator $V$ into a superposition state proportional to
\begin{equation}
    |\Psi\rangle_{a} = V|0\rangle^{\otimes n_{a}} = \sqrt{1/\beta}\sum_{k}\sqrt{\beta_{k}}|k\rangle,
\end{equation}
where $\beta = \sum_{k}\beta_{k}$. Since all $\vert\beta_{k}\vert$ are equal, the ancillary register can be prepared as a uniform superposition state by simply applying Hadamard gates on each qubit of the register. This preparation step is also efficient, since it requires only an $\mathcal{O}(1)$ depth operation. Now, using register $|\Psi\rangle_{a}$ as the control qubits, we apply the LCU unitaries as a series of uniformly controlled operations $U_{k}$ on $|\Psi\rangle_{u}$ which is represented by the operator, $W=\sum_{k}|k\rangle \langle k|\otimes U_{k}$. Then, $|\Psi\rangle_{a}$ is reset to 0 by applying $V^{\dag}$. Finally the ancillary register is measured in the computational basis, yielding a state proportional to 
\begin{align}
    \nonumber 
    |\Psi\rangle &= R|0\rangle^{\otimes n_{a}}\otimes V|0\rangle^{\otimes n_{g}} \mapsto (V^{\dag}\otimes \mathbb{I})\otimes W(|\Psi\rangle_{a}\otimes |\Psi\rangle_{u}) \\&
    = |0\rangle^{n_{a}}\Big(\sum_{k}\beta_{k}U_{k}\Big)|\Psi\rangle_{u} + |\Psi\rangle_{\perp} \nonumber \\
    &= \frac{1}{\sqrt{\beta}}|0\rangle^{n_{a}}M|\Psi\rangle_{u} + |\Psi\rangle_{\perp}.
    \label{eq: LCU final state}
\end{align}
For brevity, we drop the required normalization constants and $|\Psi\rangle_{\perp}$ corresponds to an orthogonal subspace that stores the unwanted remainder from the above operation of eq.~(\ref{eq:lcu}). The post-selected solution subspace is then re-scaled classically by $\mathcal{O}(\vert\vert|\Psi\rangle_{u}\vert\vert/\epsilon) = \mathcal{O}(1/\epsilon)$ to obtain the actual solution. However, this procedure also shows that, the circuit for linearly combining the unitaries applies the operator $M$ only probabilistically, and the solution subspace of the quantum state is prepared with a small but finite success probability $p_{\textrm{succ}}$. Since we are interested in applying such a decomposition for $\tau$ time steps, $p_{\textrm{succ}}$ would decay as function of $f(2^{-K},\epsilon^{-1},\tau,\vert\vert|\Psi\rangle_{\tau}\vert\vert^{-1})$. The smaller the $p_{\textrm{succ}}$, the larger is the number of repeated circuit simulations (query complexity or shots) required to measure or sample the solution subspace accurately. However, we design the time marching quantum circuits such that the overall query complexity is still kept near optimal, such that the required number of shots does not blow up exponentially. For instance, we use a Richardson extrapolation strategy (outlined in Section \ref{sec: End-to-end}), with which we can do simulations even when $\epsilon\sim\mathcal{O}(1)$.

Finally, to translate these operations outlined above into quantum circuits that work on simulators or real hardware, one would need a further decomposition {of the above operators} into fundamental 1- and 2-qubit gates. This can be done in one of the following ways.

\textbf{(a) Direct Approach} -- This involves decomposing the controlled multi-qubit unitaries into standard two-level unitaries, which in turn require about $\mathcal{O}(N^{2})$ 2-qubit gates \cite{barenco1995elementary,nielsen2010quantum}. Such a decomposition results in deep circuits that might be amenable only on future fault-tolerant devices. Although the chances of achieving a quantum advantage from such a decomposition are bleak to none at present, one could still attempt to use circuit parallelization strategies, proposed in Section \ref{sec: End-to-end}, to ameliorate the large circuit depths. Additionally, since the current decomposition requires only two controlled-unitaries, this reduces the overall circuit depth further, compared to previously proposed unitary methods.

\textbf{(b) QISKIT Transpile} -- IBM's quantum simulators offer an optimized \textit{transpile} functionality (a culmination of various underlying algorithms), that decompose multi-qubit unitaries into single and two qubit gates, picked from a fixed basis of gates available. These transpliations offer multiple levels of optimization with which the circuit depths can be further shortened. This also allows one to design circuits by considering a specific qubit topology on a quantum processor, to extract the best performance. The transpilation feature is quite robust if not the most efficient. We use this feature to implement our linear unitary circuits to conduct real IBM hardware experiments and perform noisy simulations with Qiskit Aer, the results of which are described in Section \ref{sec: Numerical results}. 

\textbf{(c) Hamiltonian Simulation} -- To analyze theoretically the asymptotic gate complexity scaling of the proposed algorithms (in terms of 1- and 2-qubit gates), we assume access to a black-box with simulations based on  Hamiltonian algorithm (described in \cite{berry2015hamiltonian} (see \textit{Lemma 10}) and \cite{childs2017quantum} (see {Sec. 2.1,} \textit{Lemma 8} and proof of \textit{Theorem 3} in {Sec. 3.1})), to efficiently implement the controlled unitaries. In our case, this translates to implementing a Hamiltonian simulation of unitaries of the form $e^{\pm i\epsilon \hat{S}}$ (for $K=2$) and its powers. For an $s-$sparse matrix $\hat{M}$, of size $N\times N$, scaled such that $\vert\vert\cdot\vert\vert\leq 1$, the gate complexity to implement controlled unitaries of the form $e^{\pm i\epsilon \hat{S}}$, up to an admissible error of $\varepsilon_{U}$, would require 
\begin{equation}
    G_{U} = \mathcal{O}\big((s\epsilon+1) (\log N + \log^{2.5}(\epsilon/\varepsilon_{U}))\log(\epsilon/\varepsilon_{U})\big)\label{eq: GU}
\end{equation} one- and two-qubit gates \cite{berry2015hamiltonian}. 
From here on, $G_{U}$ represents the above expression. We remark here that the sparsity is $s=d+2$, where $d$ is the order of finite central difference scheme used. Even if we pick an extremely accurate scheme with the order $d=8$, say, the sparsity would at most be $s=10$, thus contributing only a small pre-factor to the above asymptotic complexity.

\section{Time marching compact quantum circuits}
\label{sec: TMCQC}
As already mentioned, a time marching numerical scheme can be translated into one of the following operations: $\mathbf{u}(\mathbf{x},t+1)=A\mathbf{u}(\mathbf{x},t)$ or $=A^{-1}\mathbf{u}(\mathbf{x},t)$, which are called explicit and implicit schemes, respectively (see Section \ref{sec: Numerical Setup} of the {\textit{Appendix}}). For the present discussion, the general matrix $M$ used before will be replaced by $A$, which now represents a specific numerical setup. We wish to use the framework of linear combination of unitaries, developed in the previous section, to perform efficient time marching simulations, using the above two matrix operations. We now introduce six methods of constructing quantum circuits to perform such a simulation. We shall refer to these as Time Marching Compact Quantum Circuits (TMCQC1-6) and outline their designs. In the discussion immediately below, we make an implicit assumption---which we discuss in more detail towards the end of this section---that the query complexity required to reconstruct the final solution contributes a constant pre-factor to the overall complexity of each method, thus making comparisons between gate complexities and classical time complexities more meaningful.

\subsection{TMCQC1 - Explicit Expansion Circuit}
With explicit time marching (see Section \ref{subsec: Explicit TMCQC} of the {\textit{Appendix}}), integrating the system for $\tau$ time steps requires the application of matrix-vector multiplication operations of the form $(\mathbf{A}_{E})^{\tau}$. In TMCQC1, such an operator is approximated either by two or four unitaries by naively computing the multinomial expansion given by 
(we drop the $1/(2\epsilon)$ for brevity),
\begin{align}
    &(\mathbf{A}_{E})^{\tau} = \Big(U_{0}+U_{1}+U_{2}+U_{3}\Big)^{\tau} \textrm{~~or~~} =\Big(\hat{U}_{0}+\hat{U}_{1}\Big)^{\tau} \nonumber\\
   & \implies \sum_{\substack{v_{1}+v_{2}+\cdots+v_{k}=\tau \\ v_{i}\geq 0 \forall i}} 
    {\tau \choose v_{1},\cdots,v_{K}} U^{v_{1},\cdots,v_{K}} \equiv \sum_{k=0}^{N_{T}-1}
    \tilde{U}_{k},
\label{eq:explicit expansion tau}
\end{align}
where ${\tau \choose v}=\tau!/(v_{1}!,v_{2}!\cdots,v_{K}!)$.\\ 
\textit{Gate complexity -} This multinomial expansion with $K$ monomials has a total of $N_{T}=$$\tau+K-1 \choose K-1$ terms. When $K=2\textrm{~or~} 4$, $N_{T} = $ $\mathcal{O}(\tau)$ or $=\mathcal{O}(\tau^{3})$, respectively. For the rest of the section, we shall use $K=2$ for clarity. Now each term in the above expansion is in turn a product of at most $\mathcal{O}(\tau)$ unitaries. Therefore the overall expansion has at most of $\mathcal{O}(\tau^{2})$ unitaries in total. We shall refer to this as the \textit{LCU depth}. From the previous section we know that, each of these unitaries can be implemented with a gate complexity of $G_{U}$ at best, thus giving a total complexity of $\mathcal{O}(G_{U}\tau^{K})\equiv \mathcal{O}(\log(N_{g}),\textrm{polylog}(\epsilon/\varepsilon_{U}),s\epsilon,\tau^{2})$. Now, comparing this with the classical complexity of explicit time marching via simple matrix-vector multiplications, given by $\mathcal{O}(N_{g}s\tau)$ \cite{shewchuk1994introduction}, we may infer the following. The quantum algorithm is exponentially better in $N_{g}$, but worse in $\tau$ by $\mathcal{O}(\tau)$. This implies that, even though the quantum algorithm is exponentially better in $N_{g}$, in order to compensate for the $\tau$ scaling, one would need to set the CFL stability criterion such that $N_{g}\gg\tau$; this ensures that the advantage gained by the $N_{g}$ scaling outperforms (or at the least matches the loss from) the poorer $\tau$ scaling. In any case, it would suggest that simulating only short $T$ horizons would make this step possible. Further, if we had access to (say) $\mathcal{O}(\tau)$ parallel circuits, we could simulate the unitaries in parallel (as discussed later in Sec.\ \ref{sec: End-to-end}) to lower the gate complexity, thus ameliorating the $\tau$ scaling.

\textit{Qubit complexity -} This approach would require $n_a = \log_{2}(\tau)$ ancillary qubits, $n_g = \log_{2}(N_{g})$ qubits to store the velocity field, giving a total of $n = \log_{2}(N_{g}\tau)$ qubits. 
\subsection{TMCQC2 - Explicit Serial Circuit}
The previous approach can be improved by trading in a few extra qubits, while still keeping the total qubit complexity logarithmic in the problem size. An alternative way to implement the same operation as earlier is to concatenate the LCU decomposition of $A_{E}$ serially, $\tau$ times. However, to keep track of the time steps, we would need an additional register, which we shall refer to as \textit{clock register} \cite{berry2024quantum,fang2023time}, comprising $n_{c}=\log_{2}(\tau)$ qubits. 
The quantum circuit for this is shown in figure \ref{fig:end2end}(a) (for $K=4)$. A step-by-step description of this circuit operation is outlined in Section \ref{subsec: Explicit TMCQC} of the {\textit{Appendix}}. 

\textit{Gate complexity -} It is easy to see that a serial circuit such as this would lead to an LCU depth of $\mathcal{O}(\tau)$, with a total gate complexity of $\mathcal{O}(G_{U}\tau) \equiv \mathcal{O}(\log(N_{g}),\textrm{polylog}(\epsilon/\varepsilon_{U}),s\epsilon,\tau)$. Comparing this with the classical scaling of $\mathcal{O}(N_{g}s\tau)$, one is led to conclude that the quantum algorithm is exponentially better in $N_{g}$ and comparable in $\tau$ and $s$, thus offering a potential advantage for any $\tau \geq 1$. 

\textit{Qubit complexity - } It is clear that the number of qubits required in this case is $n=n_{g}+n_{a}+n_{c}=\log_{2}(N_{g}\tau)+2$. 
\subsection{TMCQC3 - Implicit Expansion Circuit}
We now consider an implicit time marching scheme, which requires one to perform matrix inversion operations. Since we are now faced with an inversion problem, decomposing by linear {combination of} unitaries is more involved. Some approaches proposed earlier include approximating the inverse function $f(x) = 1/x$, in terms of either Fourier or Chebyshev series \cite{childs2017quantum}. The terms in these series are then implemented as unitary gates. Alternatively, we explore here a simpler yet efficient approach by truncating a Neumann series expansion to approximate a matrix inverse operation of the form $\mathbf{A}_{I}=(\mathbf{I} - \alpha \mathbf{A})^{-1}$. The series approximation up to $P$ terms could be written as $(\mathbf{I} - \alpha \mathbf{A})^{-1} = \sum_{p=0}^{P}\mathbf{A}^{p} \approx \tilde{\mathbf{A}}_{I}$, where $\tilde{\mathbf{A}}_{I}$ represents the approximate inverse operator. 
The accuracy of this approximation is given by the truncation error $\varepsilon_{N}$, which clearly depends on the number of terms retained in the series. A detailed account of this method is presented in Section \ref{sec: Trunc Neumann Series} of the {\textit{Appendix}}, where  \textit{Lemma 1} proves this result: Given an $\alpha$ such that the spectral radius $\rho(\mathbf{\alpha A})<1$ (to ensure convergence), and if $P_{min}$ is the number of terms for a truncation error of $\varepsilon_{N}$, the error is bounded from above as
\begin{equation}
    \varepsilon_{N} \leq \mathcal{O}\big(\vert\vert\mathbf{A}\vert\vert^{P_{min}}\big) = \mathcal{O}((\kappa-1)^{P_{min}}).
    \label{eq: trunc error bound}
\end{equation}From this, it can be easily shown that the number of terms required is
 \begin{equation}
    P_{min} = \mathcal{O}\Bigg( \Bigg\lceil\frac{\log(1/\varepsilon_{N})}{\log(1/\vert \vert \mathbf{A}\vert\vert)}\Bigg\rceil\Bigg) =\mathcal{O}\Bigg( \Bigg\lceil\frac{\log(1/\varepsilon_{N})}{\log(1/(\kappa-1))}\Bigg\rceil\Bigg).
    \label{eq:Pmin}
\end{equation} From \textit{Proposition 2} of the Appendix, it is also clear that the error in the velocity solution obtained from such a truncated approximation is similarly bounded. Each term of the truncated series is now further decomposed, using the linear combination unitary method. The final set of unitaries is obtained by computing the multinomial expansion of the new truncated series, while marching forward $\tau$ time steps.

\textit{Gate complexity -} To compute the gate complexity, first consider a $p$\textsuperscript{th} order term in the truncated series. If this term is written in terms of the unitaries decomposition, it will produce $p+K-1 \choose K-1$ terms, which is $\mathcal{O}(p^{K-1})$. Recall that each of these newly produced terms is in fact a product of $p$\textsuperscript{th} order products (at most) of unitaries, thus making the depth $\mathcal{O}(p^{K})$. Next, we sum the series up to $P_{min}$ terms, giving a total of (for $K=2$) $\mathcal{O}(P^{3}_{min})$. 
Finally, since we need to perform time marching, the series obtained above is finally raised to the exponent $\tau$, expanding which gives a total depth of $N_{T}=$${\tau+(P^{\text{min}})^{3}-1\choose (P^{\text{min}})^{3}-1} = \mathcal{O}(\tau^{P^{3}_{min}})$. Therefore the overall gate complexity is $\mathcal{O}(G_{U}\tau^{P^{3}_{min}})$. It is clear that the algorithm is exponential in $\tau$ and also clearly worse (in terms of $\tau$) than the corresponding classical complexity given by $\mathcal{O}(\log(N_{g}s\tau\kappa\log(1/\varepsilon)))$ \cite{harrow2009quantum,shewchuk1994introduction}. Again, although the algorithm is still logarithmic in $N_{g}$, it offers the poorest complexity scaling compared to all other TMCQCs presented here.

\textit{Qubit complexity -} We note that this algorithm requires $n_{g}=\log_{2}(N_{g})$ qubits for storing the grid, $n_{a}=P^{3}_{min}\log_{2}(\tau)$ ancilla qubits for the LCU decomposition, thus giving a total qubit complexity of $\mathcal{O}(P^{3}_{min}\log(N_{g}\tau))$. 

Although we will show that the current method TMCQC3 offers the poorest complexity scaling among the algorithms discussed, the benefit of the Truncated Neumann Series approach will be reaped maximally by the algorithms to be described next. 

\subsection{TMCQC4 - Implicit Serial Circuit} We can improve the previous algorithm by using an additional clock register. The set of unitaries obtained from the truncated Neumann series is now concatenated serially $\tau$ times to perform time marching, with the clock register tracking the time step count. 

\textit{Gate complexity -} The truncated series has a total of $\mathcal{O}(P^{3}_{min})$ terms, so the total depth can be estimated to be $\mathcal{O}(P^{3}_{min}\tau)$. Thus it can be seen readily that the total gate complexity is  $\mathcal{O}\Big(s\epsilon,\log(N_{g}\tau),\textrm{polylog}(\epsilon/\varepsilon_{U},1/\varepsilon_{N}),\log^{-3}((\kappa-1)^{-1})\Big)$.
Comparing this with the classical complexity \cite{harrow2009quantum,shewchuk1994introduction} of $(\log(N_{g}s\tau\kappa\log(1/\varepsilon)))$, one finds that the algorithm (without any parallelization) has a comparable scaling in $\tau$, while it is exponentially better than classical in $N_{g}$ and $\kappa$.

\textit{Qubit complexity -} It can be seen easily that this approach has the total qubit complexity of $\mathcal{O}(\log(N_{g}\tau P_{min}))$.


\subsection{TMCQC5,6 - Explicit \& Implicit One-shot Circuits} Now two final quantum circuit designs are described; as we shall show, they offer the most efficient complexity scaling so far. In contrast to previous methods, these approaches construct a single-large matrix inversion problem (of the size $N_{g}\tau\times N_{g}\tau$), solving which provides at once the solutions for all time steps and hence the term \textit{one-shot}. (This should not be confused with the number of shots required to sample the wavefunctions). This method can be constructed with both explicit ($\mathbf{A}_{EO}\tilde{u} = \mathbf{b}_{EO}$) or implicit ($\mathbf{A}_{IO}\tilde{u} = \mathbf{b}_{IO}$) schemes; details of these construction are outlined in the {\textit{Appendix}}. In order to approximate the inverse, we again use the truncated Neumann series as in TMCQC3,4. Since in this case the solution of the matrix inversion includes the solutions at all time steps, there is no need for any further action. The terms of the truncated series can now be readily written in terms of the unitaries decomposition. Furthermore, since the method avoids an iterative/serial style time marching, the diminution of success probability is also less severe than in previous methods.
\textit{Gate complexity -} This is given by simply computing the terms obtained by just applying the unitaries decomposition to each term of the truncated series. The depth is therefore just $\mathcal{O}(P^{3}_{min})$, which yields a total gate complexity of $\mathcal{O}(G_{U}P^{3}_{min}) \equiv \mathcal{O}\Big(s\epsilon,\log(N_{g}\tau),\textrm{polylog}(\epsilon/\varepsilon_{U},1/\varepsilon_{N}),\log^{-3}((\kappa-1)^{-1})\Big)$. This scaling is nearly optimal in every parameter except $s$, which, as already noted earlier, is at most $s\leq10$. Thus the method is exponentially better than its classical counterpart in terms of all system parameters. 

\textit{Qubit complexity -} The input vector is now a larger dimensional state of size $N_{g}\tau$, which requires $n_{g}=\log_{2}(N_{g}\tau)$ qubits. Further, to apply the unitaries decomposition, we require an ancilla register of size $n_{a}=3\log_{2}(P_{min})$ thus having a total qubit complexity of $\mathcal{O}(\log_{2}(\tau N_{g}P_{min}))$.

The complexities of all circuit designs discussed above are summarized in Table \ref{table:Complexity summary}.
\begin{table*}[htb!]
 \caption{Summary of computational complexity of  quantum algorithms}
 
 \begin{NiceTabular}{c|c|c|c|c}
  \hline \hline
 \Block[tikz={top color=olive!25}]{*-1}{} & \Block[tikz={top color=teal!35}]{*-2}{}  & \Block[tikz={top color=teal!30}]{*-2}{}  &  \Block[tikz={top color=teal!25}]{*-1}{}&
 \Block[tikz={top color=teal!20}]{*-1}{}
 \\
 \textbf{Algorithm} & \textbf{LCU  depth}\footnote{Complexities in terms of number of high-level unitaries. All complexities here consider the case $K=2$.}& \textbf{Gate complexity}\footnote{If the amplitude amplification is required\cite{childs2017quantum}, this is multiplied by another $\mathcal{O}(\kappa)$ factor}& \textbf{Classical complexity}&\textbf{Qubit complexity}\\
 &   &  &  &\\
 \hline \hline
  &   &  &  &\\
 QSP & - & $\mathcal{O}(2kn)$ & -&$\mathcal{O}(\log(N_{g}))$\\
 &   & \textsuperscript{\cite{bharadwaj2023hybrid,mozafari2022efficient}}  &  \\
\hline  \textbf{Explicit}  \\
 \hline
 &   &   &  &\\
TMCQC1 & $\mathcal{O}\big(\tau^{2}\big)$ & $~\mathcal{O}\Big(s\epsilon,\log(N_{g}),\textrm{polylog}(\epsilon/\varepsilon_{U}),\tau^{2}\Big)~$  & $\mathcal{O}(N_{g}s\tau)$&$\mathcal{O}(\log(N_{g}\tau))$\\
  &   &    &  \\
 TMCQC2 & $\mathcal{O}\big(\tau\big)$ & $~\mathcal{O}\Big(s\epsilon,\log(N_{g}),\textrm{polylog}(\epsilon/\varepsilon_{U}),\tau\Big)~$ &$\mathcal{O}(N_{g}s\tau)$&$\mathcal{O}(\log(N_{g}\tau))$\\
 &   &   &  & \\
 TMCQC5 &  $\mathcal{O}\big(P_{min}^{3}\big)$\footnote{$P_{min}=\Bigg\lceil(\log(\varepsilon^{-1}_{N}))/\log((\kappa-1)^{-1})\Bigg\rceil.$} & $~\mathcal{O}\Big(s\epsilon,\log(N_{g}\tau),\textrm{polylog}(\epsilon/\varepsilon_{U},1/\varepsilon_{N}),\log^{-3}((\kappa-1)^{-1})\Big)~$ & $\mathcal{O}(N_{g}s\tau\kappa\log(1/\varepsilon))$ &$\mathcal{O}(\log(N_{g}\tau P_{min}))$\\
 &   &   &  \\
 \hline
 \textbf{Implicit} \\
 \hline
 &   &  &   &\\
 TMCQC3 &  $\mathcal{O}\Big(\tau^{P_{min}^{3}}\Big)$ &$~\mathcal{O}\Big(s\epsilon,\log(N_{g}),\textrm{polylog}(\epsilon/\varepsilon_{U}),\tau^{P_{min}^{3}}\Big)~$  &$\mathcal{O}(N_{g}s\tau\kappa\log(1/\varepsilon))$&$\mathcal{O}\big(P^{3}_{min}\log(\tau N_{g})\big)$\\
 &   &   &  & \\
 TMCQC4 &  $\mathcal{O}\big(\tau P^{3}_{min}\big)$ & $~\mathcal{O}\Big(s\epsilon,\log(N_{g}\tau),\textrm{polylog}(\epsilon/\varepsilon_{U},1/\varepsilon_{N}),\log^{-3}((\kappa-1)^{-1}),\tau\Big)~$ & $\mathcal{O}(N_{g}s\tau\kappa\log(1/\varepsilon))$ &$\mathcal{O}\big(\log(\tau N_{g} P_{min})\big)$\\
 &   &   &  & \\

 TMCQC6 & $\mathcal{O}\big(P_{min}^{3}\big)$  &$~\mathcal{O}\Big(s\epsilon,\log(N_{g}\tau),\textrm{polylog}(\epsilon/\varepsilon_{U},1/\varepsilon_{N}),\log^{-3}((\kappa-1)^{-1})\Big)~$ & $\mathcal{O}(N_{g}s\tau\kappa\log(1/\varepsilon))$&$\mathcal{O}\big(\log(\tau N_{g} P_{min})\big)$ \\
&   &   &  \\

  \hline
  & &  &\\
 QPP& - & $\mathcal{O}(U_{V})+\mathcal{O}((\log N)^{2}/\epsilon_{QPP})$ & - &varies \\
  &    & \textsuperscript{\cite{bharadwaj2023hybrid}}  &   \\
 \end{NiceTabular}
 \label{table:Complexity summary}
 
 \end{table*}
\subsection{Query complexity and success probability}
\label{sec:Query complexity}
The gate complexities outlined above correspond to a single application of the quantum circuit. However, the solution thus prepared has a small, yet finite success probability $p_{\textrm{succ}}$. This implies that one would need to query the circuit repeatedly (in other words, repeat the experiment for $N_{s}$ shots) to reconstruct the solution state by repeated sampling. The overall time complexity of the simulation would therefore be the product of the gate and query complexities. To compute this, we need to first estimate $N_{s}$ for which we first consider eq.~(\ref{eq: LCU final state}) that represents the fundamental action of linear unitaries. Before rescaling the solutions, simply applying the unitaries circuit prepares the state proportional to $2\epsilon A\vert\Psi\rangle$. 
From \textit{Proposition 1} (see {\textit{Appendix}}), without loss of generality, the matrix $A$ can always be scaled by a \textit{small} constant $\delta\lesssim\mathcal{O}(1/\epsilon)$, and the solutions can be simply scaled back after measurements. Therefore the solution subspace produced from eq.~(\ref{eq: LCU final state}) can be said to be proportional to $(2\epsilon\delta/\sqrt{\beta})A\vert\Psi\rangle.$
This implies that a single application of the linear unitaries operation produces the desired state with the success probability  \begin{equation}
   p_{\textrm{succ}} \sim \Bigg(\frac{2\epsilon\delta\vert\vert  A\vert\Psi\rangle\vert\vert}{\sqrt{\beta}}\Bigg)^{2}.
   \label{eq: p_succ simple LCU}
\end{equation}
The above value is for a single linear {combination of} unitaries block. However in essence, each TMCQC provides a different unitary block-encoding of the corresponding $A$ to perform time marching. The $p_{\textrm{succ}}$ of the solution at $t=\tau\Delta t$ thus varies for each TMCQC. The specific dependence arises from the difference in the number of times the unitaries oracle is queried and the corresponding depth, which we shall represent from now on as $G_{\textrm{L}}$. Comparing all the TMCQCs we can easily identify that in TMCQC2 and TMCQC4, the decay in $p_{\textrm{succ}}$ is particularly the most severe since the unitaries are applied in series. Therefore for $\tau$ such applications, the RHS of eq.~(\ref{eq: p_succ simple LCU}) will now be raised to the power $\tau$. For generality, let us represent this exponent with $G_{\textrm{L}}$. A single application of the two-unitary oracle involves preparing a superposition state by Hadamard gates acting on ancilla qubits. This implies that $\beta = 2$. Since the $p_{\textrm{succ}}$ for every time step is independent, the probability thus tends to decay exponentially \cite{berry2024quantum,fang2023time}, compounding for each of the $\tau$ steps. Apart from this, the overall success probability also depends on ratio of the norms of the initial and final time solution states. Thus, we can rewrite $p_{\textrm{succ}}$ as
\begin{align}
   p_{\textrm{succ}} &= \Bigg(\frac{2\epsilon\delta\vert\vert A\vert\vert}{2}\Bigg)^{2G_{\textrm{L}}}\prod_{k=1}^{G_{\textrm{L}}}\frac{\vert\vert~\vert\Psi\rangle_{k}~\vert\vert^{2}}{\vert\vert~\vert\Psi\rangle_{k-1~}\vert\vert^{2}} \nonumber\\ &=  (\epsilon\delta\vert\vert A\vert\vert)^{2G_{\textrm{L}}}\frac{\vert\vert~\vert\Psi\rangle_{\tau}~\vert\vert^{2}}{\vert\vert~\vert\Psi\rangle_{0}~\vert\vert^{2}} .
\end{align}
This implies that it would require at least \begin{equation}
    N_{s} = \Bigg(\frac{1}{\epsilon\delta\vert\vert A\vert\vert}\Bigg)^{2G_{\textrm{L}}}\frac{\vert\vert~\vert\Psi\rangle_{0}~\vert\vert^{2}}{\vert\vert~\vert\Psi\rangle_{\tau}~\vert\vert^{2}} \approx \mathcal{O}\Bigg(\frac{\vert\vert~\vert\Psi\rangle_{0}~\vert\vert^{2}}{\vert\vert~\vert\Psi\rangle_{\tau}~\vert\vert^{2}}\Bigg)
    \label{eq: eta query complexity}
\end{equation} number of shots to recover the solution, up to a constant pre-factor $\eta^{2G_{\textrm{L}}} = (1/\epsilon\delta\vert\vert A\vert\vert)^{2G_{\textrm{L}}}$. Since we can readily choose an $\epsilon$ and $\delta$ such that $\epsilon\delta\vert\vert A\vert\vert \gtrsim 1$ (see \textit{Proposition 1} of {\textit{Appendix}} and its ensuing discussion), such that $\eta \sim \mathcal{O}(1)$ is not an \textit{unfavorably large} constant; this takes the overall complexity closer to optimal. 

We should also note that the complexity discussed above can be further improved quadratically by applying a Grover-like amplitude amplification, as shown in \cite{brassard2002quantum}. However, we continue our discussion without it, to emphasize that the corresponding results are still efficient in some strict sense, even before applying amplitude amplification. This optimistic nature of the discussion can however be quickly diminished when we begin to consider the effects of noise. In practice (as shown later in Section \ref{sec: Numerical results}), to simulate these circuits on currently available IBMQ machines, we need at least $N_{s}=2^{15}$ shots to be able to accurately recover the solution. Nevertheless, the query complexity presented here relates to the asymptotic complexity scaling, which suggests that, when reasonably large problem sizes are solved with the proposed methods, the required shot count will not significantly diminish any available quantum advantage.


In summary, it is ideal to ensure that $\eta \lesssim 1$; otherwise it would lead to a large pre-factor $\mathcal{O}(2^{\log(\tau)/\log(\kappa-1)})$ at best, or $\mathcal{O}(2^{\tau^{2}})$ at worst. A note on condition number dependence is needed here. The discussion so far suggests only logarithmic dependence on the condition number when compared to earlier works that demonstrate an optimal result of linear $\mathcal{O}(\kappa)$ dependence \cite{costa2022optimal}. Although this is indeed the case, it comes at a certain cost described below.
 \begin{itemize}
    \item It is important to note the distinction between quantum linear solvers and time marching solvers (this work). Given this, unlike earlier work, we use a distinct LCU decomposition combined with a specific extrapolation technique that leads to a different complexity scaling. Also, as mentioned in Table \ref{table:Complexity summary}, if an explicit amplitude amplification is required \cite{childs2017quantum}, we pick up a $\mathcal{O}(\kappa)$ factor. However, the question here is whether we could lower the required query complexity without an explicit amplitude amplification, though the use of an amplitude amplification step is not precluded. 
    \item The linear dependence on the condition number as shown in earlier optimal linear solvers is now transformed into a dependence on the quantities $\epsilon$ and $\vert\vert \textbf{A} \vert \vert$ in the following sense. The solution computed by the approximate inverse of the matrix system is dictated by the number of terms $P_\text{min}$, which in turn depends on $\kappa$ and $\vert\vert \textbf{A} \vert \vert$. We observe that the requirement of a bounded $\kappa$ does not stymie the range of system parameters that can be explored. But this transfers the onus onto the norm of a given matrix, $||\mathbf{A}||$. This quantity, however, can be scaled (by the factor $\delta$) and conditioned on this, the overall query complexity is subsequently scaled by a factor $\eta^{2G_{\textrm{L}}} = (1/\epsilon\delta\vert\vert A\vert\vert)^{2G_{\textrm{L}}}$ as mentioned in the main text in Section IV.F. This may be further improved by (say) quantum amplitude amplification to a factor $(1/\epsilon\delta\vert\vert A\vert\vert)^{G_{\textrm{L}}}$. For the case of TMCQC-5,6, where we apply the LCU decomposition once, we have $\eta = (1/\epsilon\delta\vert\vert A\vert\vert)$. This quantity determines the query complexity. This factor tends to be a large constant (corresponding to low probably solution subspace for post selection), thus implying a larger query complexity. Now, one may regard that the complexity of our algorithm scales linearly in $\eta$, which is analogous to the optimal result in terms of condition number $\kappa$, compared to earlier results.
    \item We now mention an advantage here. Unlike the condition number that is controlled via relatively involved preconditioning techniques, the factor $\eta$ can now be lowered and successively made close to unity via iterative applications of Richardson's extrapolation, since the parameter $\epsilon$ can be now controlled; this, in turn, avoids $\eta$ from being an unfavorably \textit{large} constant.
    \item In essence this work is an improvement (even if not exponential) over the previous optimal results in terms of condition number. 
\end{itemize}
To see this, while ensuring that the query complexity is controllable, we need to pay special attention to the following three situations.
\begin{enumerate}
    \item \textit{Stability criteria} -- Flow problems solved by implicit schemes (TMQCQ3,4,6) are not constrained by any stability criteria, which therefore allows more flexibility to choose the parameters. However, explicit schemes as in TMCQC1,2,5 have to satisfy a stability criterion. For the specific case of the advection-diffusion problem being discussed, this is equivalent to $\alpha = D\Delta t/(\Delta x)^{2} \leq 0.5$. This criterion requires the system parameters $U,N_{g},\Delta t$ to be chosen to satisfy $\delta\vert\vert A \vert \vert < 1$ and $\delta\epsilon\vert\vert A \vert \vert \gtrsim 1$. For second-order, central difference, explicit scheme, $\vert\vert A\vert\vert = 1$. In this case, just ensuring $\delta\epsilon \gtrsim 1$ is sufficient. 
    \item \textit{Noise} -- As already noted, it is important to choose a large enough $\epsilon$ to make the solution distinguishable from noise. At the same time, to maintain the accuracy of the solution even for large $\epsilon$ values, we could employ the concept of Richardson-extrapolation to defer our approach to $\lim \epsilon \rightarrow 0$. The accuracy of the extrapolated solutions can be $\mathcal{O}(\epsilon^{4})$ or better. Viewed alternatively, we can lower the shot count $N_{s}$ significantly for the same given accuracy.
    
    \item \textit{Steady state and decaying flows} -- The above two considerations alone are insufficient to maintain a small $N_{s}$. We also need to ensure that the steady state limit  $$\lim_{\tau\rightarrow \infty}\frac{\vert\vert~\vert\Psi\rangle_{0}~\vert\vert}{\vert\vert~\vert\Psi\rangle_{\tau}~\vert\vert}, $$ does not diverge. If the time marching operator is norm-preserving (or the flow is statistically steady), then $\vert\vert~\vert\Psi\rangle_{\tau}~\vert\vert = \vert\vert~\vert\Psi\rangle_{0}~\vert\vert$ trivially. The alternative scenario of the norm-increasing flow is better, where $\vert\vert~\vert\Psi\rangle_{\tau}~\vert\vert \geq \vert\vert~\vert\Psi\rangle_{0}~\vert\vert$, corresponding to flows that experience constantly increasing net influx of energy. Both cases are in practice possible if we consider a constant forcing term in our governing PDEs. However, the more restrictive case is the viscous, dissipative flow whose norm decays with time; in this case the steady state of the flow corresponds to vanishingly small velocity fields due to constant dissipation of energy from the system. However, the advection-diffusion problem considered in this work is not as severe as can be seen from the analytical solution \cite{ingelmann2024two}. The steady state corresponds to a constant, uniform velocity field with a small, yet non-zero norm (in this work, this ratio is = 1/2). For flows whose steady state norms are identically zero, the query complexity grows as $\mathcal{O}(\vert\vert~\vert\Psi\rangle_{\tau}~\vert\vert^{-1})$. In such cases, there could be two alternatives (which should be studied in more detail): (1) The problem of the growing pre-factor $\eta$ can be addressed by choosing a smaller $\epsilon$ such that $\eta^{G_{\textrm{L}}} \sim \mathcal{O}(\vert\vert~\vert\Psi\rangle_{\tau}~\vert\vert^{-1})$. (2) When this choice is not possible, one is limited to solving the problem for only short time horizons, for which the norm decays not too steeply. 
    
    \hspace {0.2cm} The scenario described so far considers the growth in $N_{s}$ as being most severe. If, however, we consider TMCQC5,6 there is no compounding decay in terms of $\epsilon$. The only contributing factor would be from the normalization stemming from the linear decomposition in terms unitaries. However, we show that the unitaries depth is $\mathcal{O}(\log(1/\epsilon)/\log(1/(\kappa-1))^{3}$. For large problem sizes, the corresponding pre-factor in query complexity would not be critical. In any case, from the discussion so far, it is clear that the overall time complexity of the algorithms can be pushed closer to optimal.
    
     
\end{enumerate}


\section{End-to-end algorithm and near-term strategies} 
\label{sec: End-to-end}
A hybrid quantum-classical algorithm requires one to consider: (A) a suitable quantum state preparation, (B) an optimal circuit design for the quantum solver, (C) efficient solution read-out and post processing, and (D) effects of noise and decoherence on real quantum hardware. These factors tend to diminish expected quantum advantage. In this section we outline strategies that could potentially render the overall algorithm to be an end-to-end type, as well as one that can be simulated on noisy near-term quantum devices, while retaining some quantum advantage. A schematic of these methods and typical circuit designs are given in figure \ref{fig:end2end}(b).\\

\begin{figure*}
\subfloat[]{\includegraphics[trim={4cm 11cm 2.5cm 8cm},clip=true,scale=0.6]{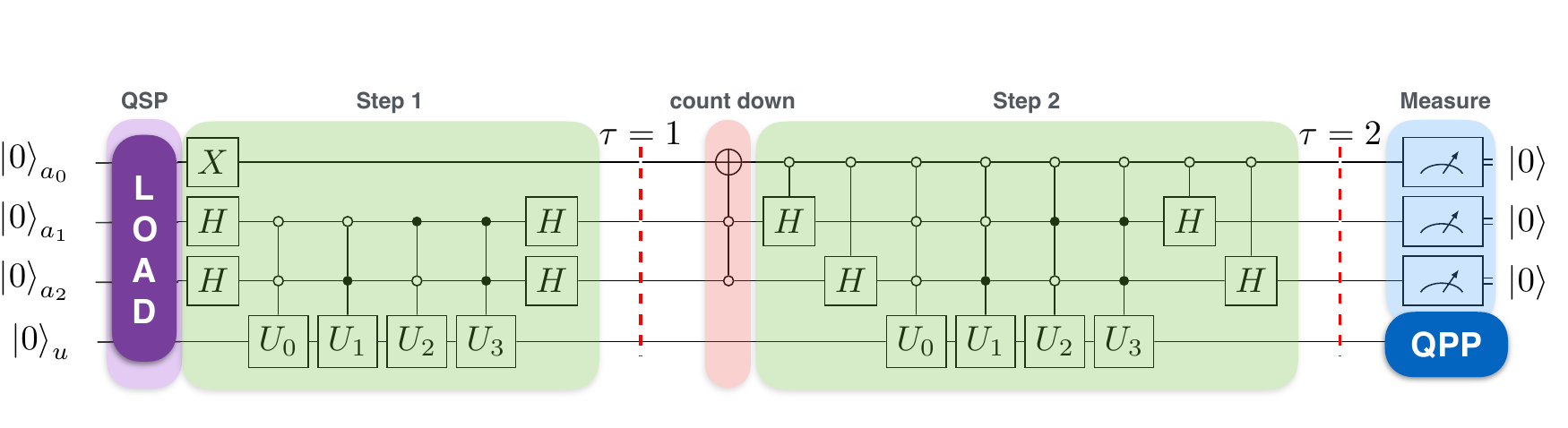}}

\subfloat[]{
\includegraphics[trim={0.0cm 0cm 0cm 0cm},clip=true,scale=0.3]{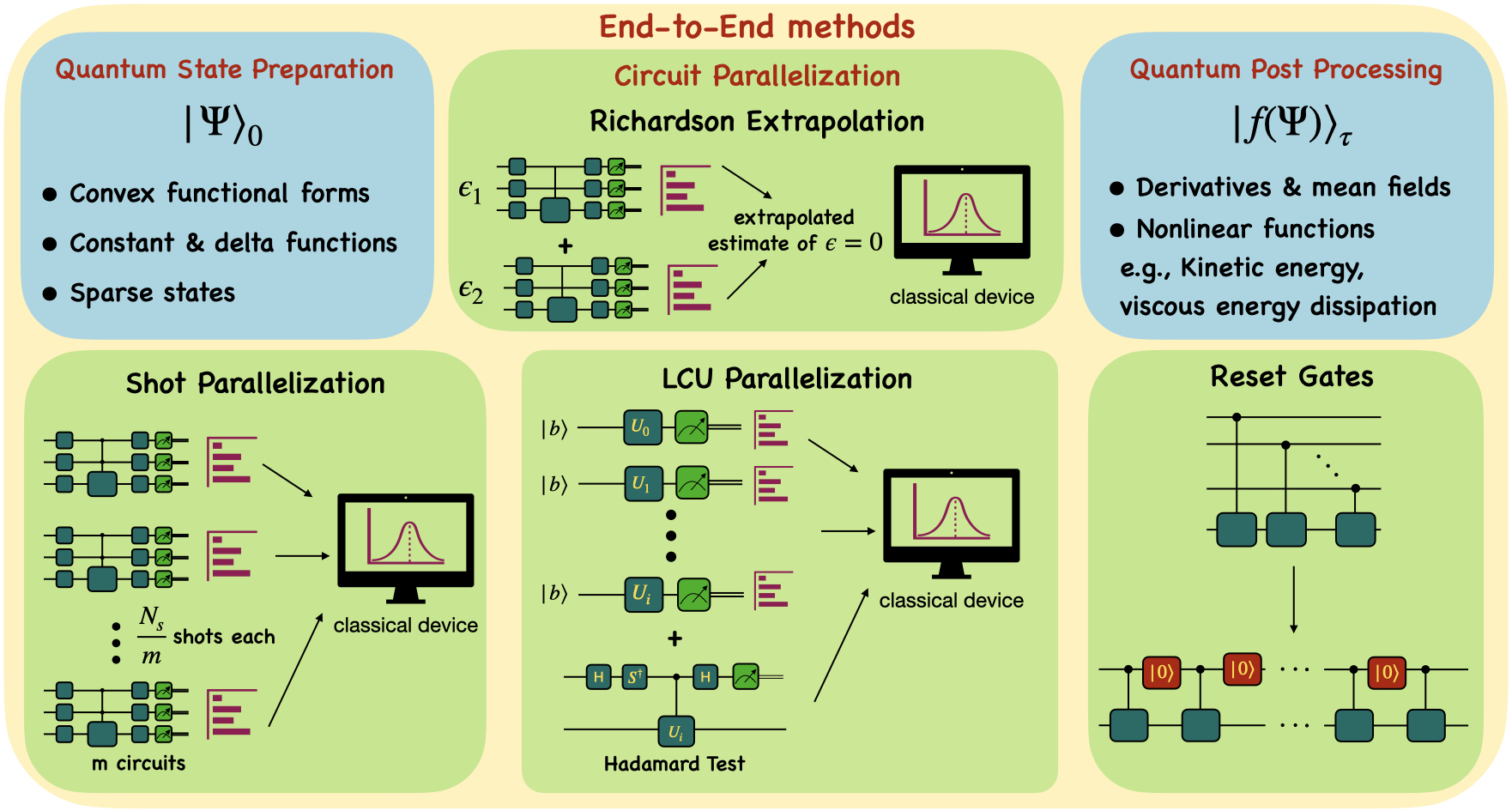}}
    \caption{\justifying (a) 4 qubit iterative LCU circuit to implement a simulation for $\tau=2$ time steps is shown. Here the first register $q(a_{0})$ is the counter qubit, which is initially set to one and successively bit-flipped to count the time steps in a binary fashion. The ancilla qubits $q(a_{1})$ and $q(a_{2})$ are required to implement the linear combination of four unitaries $U_{0}$ to $U_4$. The last register $q(u)$ stores the velocity field at every time step. (b) Schematic of the end-to-end strategies and their quantum circuits that can be used to preserve potential quantum advantage.}
    \label{fig:end2end}
\end{figure*}

\noindent\textbf{(1) Quantum State Preparation } - The first step of a hybrid QCFD algorithm is to encode the initial velocity field into qubit states. In general, for arbitrary states of size $N$, the complexity of state preparation would scale linearly with $N$. This scaling compromises the quantum speed-up that might be achieved from the PDE solver algorithm. However, specific examples might not need such exponential circuit depths. For an input state that has an integrable, convex functional form, the initial state can be prepared with sub-exponential circuit depths as shown in \cite{grover2002creating,vazquez2022enhancing,bharadwaj2023hybrid}. For sparse states, one can prepare the initial state more efficiently with sub exponential circuit depths \cite{mozafari2022efficient,zhang2022quantum,bharadwaj2023hybrid}. Certain cases might, however, require one to trade additional ancillary qubits for shorter circuit depths. Further, since the algorithms proposed here require only a single copy of the initial state for every circuit execution, we can begin with a simple state that is easy to prepare. In fact, most problems in fluid dynamics offer great flexibility in choosing initial conditions, which could be a uniform flow (constant function) that would require a state preparation circuit of just a single layer of Hadamard gates, or a delta function (the case considered here), preparing which requires a single NOT gate. Many times, a random initial conditions are admissible or necessary. More recent efforts demonstrate efficient ways to encode polynomial functions as well \cite{gonzalez2024efficient}.\\ 

\noindent\textbf{(2) Circuit Parallelization and Reset gates } - Circuit depths that can be simulated on current and near-term hardware are rather limited. The limitation arises from the finite, short coherence life-spans of the qubits. Longer circuit executions tend to run into errors based on decoherence. In order to make an attempt at fitting the TMCQCs on such hardware, we propose here a few strategies to parallelize the circuit execution, to both reduce the effective circuit depths and the execution times. 

(A) \textit{Shot parallelization} -  Generally the shot count available on real quantum hardware (here, IBMQ) is limited to about $\sim 3\times10^{4} \textrm{~to~} 5\times10^{5}$ shots. If we consider a simulation requiring a large shot-count of about (say) $N_{s}=2^{20}$ shots or higher, such an execution would not be possible with current hardware. 
A simple workaround to achieve this  would be actually to execute multiple quantum circuits in parallel. For the current example, naively, we can perform two parallel circuit executions with $2^{19}$ shots each. When possible, we could go even further by executing several low shot-count circuits in parallel, with which we can increase the total shot count while improving the sampling accuracy of the quantum state. It is important to note here that this implies a linear space-time trade-off, under the assumption of access to multiple qubits of similar \textit{quality} (calibrated error rates) within the same device. This is not a fundamental assumption but the qubit quality varies in pratice across the device at any given instant of time. To fit our simulations using only the available shot count, we could instead use the Richardson extrapolation to lower the required $N_{s}$. This is discussed below in this section.

(B) \textit{LCU parallelization} - 
Let us consider TMCQC5 and TMCQC6, the best circuits introduced here in terms of optimality. These systems encode a full history state, simulating the entire time evolution in one shot. The required number of ancilla qubits, thus the control qubit operations, depend logarithmically on the depth of LCU decomposition as outlined earlier in Section \ref{sec: TMCQC}. LCU parallelization entails implementing these controlled unitaries in parallel, either as parallel and separate circuits or within the same circuit. For clarity, let us consider a single simulation time step, using either the two or four unitary block encodings. 

\begin{enumerate}
    \item \textit{Naive Parallel unitaries} - We execute each unitary $U_{i}$ (without any controls) in parallel on 2 (or 4) separate circuits, starting with the same initial state. This requires $3n_{g}$ qubits in total. Since each controlled unitary is parallelized, the control qubits can now be discarded. This also implies that the resulting decomposition into two qubit gates will be significantly efficient in gate complexity. The number of parallel circuits, and thus the number gates, is determined by the LCU decomposition. As discussed in Section \ref{sec: TMCQC}, this is logarithmic in system parameters and as a consequence results in gate complexity of the parallel circuits to be efficient. At most a single control qubit might be necessary based on how the partial solutions are combined subsequently. When the resulting parallel (partial) solution states have purely real amplitudes, we can perform a straightforward measurement in the computational basis of each partial solution. However, the sign of negative numbers requires additional measurements, although not a full tomography. Following this, these partial solutions can be combined classically to output the final solution. However, if any of the partial solutions have complex amplitudes, a direct extraction of solutions requires expensive tomography to reconstruct the state. Two possibilities exist. In the first case, we combine those parallel unitaries as a single circuit such that the grouping produces real amplitudes for the partial states. From here, we can proceed with measurement as earlier. This new grouping would, of course, increase the depth of the circuit, which is now at some intermediate level of parallelization.
   Secondly, if we are interested only in computing the expectation value of a certain observable, instead of the full state, it can be done while still retaining the full parallelization. The expectation value of an operator $U_{i}$ can be computed accurately by using a Hadamard Test circuit. A direct Hadamard test outputs $Re(\langle\psi\vert U_{i}\vert\psi\rangle)$. For the imaginary part $Im(\langle\psi\vert U_{i}\vert\psi\rangle)$, a simple modification is done by adding the $S^{\dag}$ gate as shown in figure \ref{fig:end2end}(b).

    If we are given access to $m=G_{\textrm{L}}$ parallel circuits, every parallel circuit would then be significantly shallower, apart from improving the overall complexity itself. This reduction in circuit depths makes it amenable for near-term quantum devices. For example, for the Qiskit Transpile command to solve an $N_{g}=4$ system, if we use a total of 4 qubits as a single circuit, the transpiled depth is $\sim1300$. If we implement it as 4 parallel circuits by eliminating all control operations and qubits, the depth is 10. This is about $\sim$ 130 times shallower in depth. In fact, even if we include a full arbitrary state preparation step, which is at most a circuit of $\mathcal{O}(10)$ depth, the reduction in depth obtained from parallelization can easily compensate for an expensive state preparation. Since these shallower circuits are more accurate with lower effects of noise and decoherence, the total shots required to extract the partial solutions would also drop. We explore this strategy by implementing it on a real IBM quantum device, and present results later in Section \ref{sec: Numerical results}.
    
    \item \textit{Fanout Parallel unitaries} - A more robust parallelization is possible by invoking the idea of \textit{fanout} quantum circuits \cite{nielsen2010quantum,cleve2000fast,moore2001parallel}, though it is hard to realize using near-term devices. Here, the circuit can be parallelized by a single entangled quantum circuit (not separate circuits as earlier) at the cost of extra ancilla qubits \cite{hoyer2005quantum,boyd2023low,zhang2024parallel}. In these types of strategies, the state of a single ancilla register prepared with the target state is then ``basis-type copied" (fanout) to other additional ancilla registers by applying controlled CNOT gates as $\vert \psi\rangle\vert0\rangle\cdots\vert0\rangle \rightarrow \vert \psi\rangle\vert \psi\rangle\cdots\vert \psi\rangle$. Now the unitary operators are applied in parallel on copies of the target state $\vert \psi\rangle$. Especially when the Hamiltonian that is being simulated has certain properties \cite{zhang2024parallel}, or when the set of unitaries can be partitioned into Pauli operators, such fanout schemes can be used, with the aid of Clifford circuits \cite{boyd2023low}, to parallelize the overall algorithm.
\end{enumerate}

\noindent(C) \textit{Reset gates} - Apart from parallelizing the circuits to lower the circuit depth, one can also apply \textit{reset gates} now possible on IBMQ devices to lower the width (ancilla qubit complexity) of the quantum circuit. 
It is common to have multiple ancilla for controlled unitary rotations \cite{nation2021measure}. Instead of having separate control qubits, after application of every controlled operation on the target state, the ancilla register can be reset back to $\vert0\rangle$ state, and can be reused to control the next controlled unitary on the target state as shown in figure \ref{fig:end2end}(b). Such operations can improve the qubit complexity, but care needs to be taken while rescaling the final solution by accounting for any \textit{re-normalized} coefficients of the states that were reset, as well as accounting for the breaking of any important entanglement in the circuit. Another advantage of such resets is that we can reset a qubit quickly before it reaches its coherence time limits. The qubit now spawns a new life for the next controlled operation within its coherence span, thus lowering potential errors due to decoherence.\\

\noindent\textbf{(3) Richardson Extrapolation -- }
The success probability with the algorithms outlined so far can be enhanced via Richardson extrapolation \cite{schlimgen2021quantum}, which offers an elegant way to reduce the required number of shots. Conversely, it could be used to improve the accuracy for a fixed number of total available shots of the sample the solution. This tool allows us to simulate the unitaries decomposition even for $\epsilon \sim \mathcal{O}(1)$, this being crucial to control the query complexity of the overall algorithm, as discussed in the previous section. The concept of this extrapolation is as follows:  Given an operator $U(\Gamma,\epsilon)$, estimating its output at $\Lim{\epsilon\rightarrow 0}$, could be done through extrapolation as shown in \cite{richardson1927viii,schlimgen2021quantum}. From this we can write
\begin{equation}
    U(\Gamma,0) = \frac{U(\Gamma,\epsilon_{1}) - \gamma^{2}U(\Gamma,\epsilon_{2})}{(1-\gamma^{2})},
    \label{eq: richardson U's}
\end{equation}
where $\epsilon_{1}>\epsilon_{2}$ and $\gamma = \epsilon_{1}/\epsilon_{2}$ is the order of extrapolation, where the error of extrapolation scales as $\mathcal{O}(\epsilon^4)$. $\Gamma$ here represents, collectively, any arbitrary set of parameters on which the operator could depend. Higher powers of $\gamma$ lead to higher orders and more accurate extrapolations. As we shall demonstrate later through simulations, even for $\epsilon_{1} and \epsilon_{2}$ close to unity, the solution can be computed accurately through extrapolation. This method could also serve as an aid to possible amplitude amplification procedures \cite{berry2014exponential}. The extrapolation procedure can be viewed alternatively as follows. Given a fixed number of shots $N_{s}$, we can extrapolate the solutions that were obtained with $\epsilon_{1,2}$ and $N_{s}$ shots as 
\begin{equation}
     \vert \mathbf{u}\rangle_{0} = \frac{\vert \mathbf{u}\rangle_{\epsilon_{1}} - \gamma^{2}\vert \mathbf{u}\rangle_{\epsilon_{2}}}{(1-\gamma^{2})},
    \label{eq: richardson}
\end{equation} to produce an extrapolated solution with higher accuracy. The procedure can be repeated for higher orders as well, giving more accurate extrapolations. If we compute the gradient of the above expression with respect to $\gamma$, we note that the gradient is maximum at about 1. Therefore $\gamma$ is a number close to but greater than 1; that is, if $\epsilon_{1,2}$ are close to each other and $\epsilon_{1}>\epsilon_{2}$, then the effect of extrapolation is amplified.  Applying this tool in our TMCQC simulations has the benefits of lowering the required $N_{s}$ for a given accuracy, improving the query complexity, and allowing large $\epsilon$ simulations to help distinguish them from the noise on real hardware devices.\\ 

\noindent\textbf{(4) Quantum Post Processing } - Finally, once the quantum solver stage has prepared the solution state, we can either choose to measure the entire field, which is an $\mathcal{O}(N_{g})$ operation that could compromise the available quantum advantage, or compute important functions or expectation values of flow observables, outputting a single real value (or a few). Measuring this single value (or a few values) protects the quantum advantage by avoiding the tomography of the entire state. We call this Quantum Post Processing. Such meaningful functions of the field could be either mean flow field $\langle\mathbf{u}\rangle = \int_{0}^{L}u(x,t)dx$ or the mean gradient field $\langle\partial\mathbf{u}/\partial\mathbf{x}\rangle$ (and its higher orders). The results for the mean flow are shown in figure \ref{fig:Resolution and RI}(b). The mean gradient can be computed by first using a circuit with linear unitaries to approximate the numerical gradient followed by a string of Hadamard gates applied on all qubits of the target register to compute the sum, which can then be divided by $N_{g}$ classically. The more important functions of the field, however, tend to be nonlinear, for instance, the mean viscous dissipation rate given by $\varepsilon=\nu\langle(\partial u/\partial x)^2\rangle$, where $\nu$ is the viscosity. Such nonlinear functions can be computed using the QPP algorithm described in \cite{bharadwaj2023hybrid}. This method also avoids the need for expensive bit-arithmetic circuits to perform nonlinear transformations, {although it requires a nominal amount of classical pre-computation done hybridly, to prepare the quantum circuits}. The QPP for computing nonlinear observables consists briefly of re-encoding the amplitude and bit encoded values using a fixed number of qubits($n_{\textrm{QPP}}$) that decide the accuracy as
\begin{equation}
    u_{i}\vert i\rangle \rightarrow \vert u_{i}\rangle.
\end{equation} The target values are now represented by an $n_{\textrm{QPP}}$-bit binary approximation. At this point, instead of using bit-arithmetic, since $n_{\textrm{QPP}}$ is known apriori and thus also the corresponding binary basis, we can use this knowledge to generate controlled rotation gates corresponding to each basis state. Therefore, now using these bit-encoded states as control qubits, we can apply controlled rotations on an ancillary register (initially set to 0), where the rotation angles are classically precomputed to be $R_{y}(\theta=\arccos(f(u_{i}))$, where $f(u)$ is the nonlinear function to be applied on the field {or on a coarser subspace of it}. This produces a state proportional to
\begin{equation}
    \vert u_{i}\rangle\vert0\rangle \rightarrow f(u_{i})\vert i\rangle.
\end{equation} 

Such a QPP algorithm produces a single and real measurable, which not only offers insight into the flow field but also lowers the measurement complexity, thus protecting to some degree the available quantum advantage. The limitations of querying for a single value (or a few values) are obvious, though many experiments of the pre-digital era did just that.

\begin{figure*}[htpb!]
    \centering
    \subfloat{
   \hspace{-0.8cm} \includegraphics[trim={0.1cm 15cm 0.2cm 0.0cm},clip=true,scale=0.45]{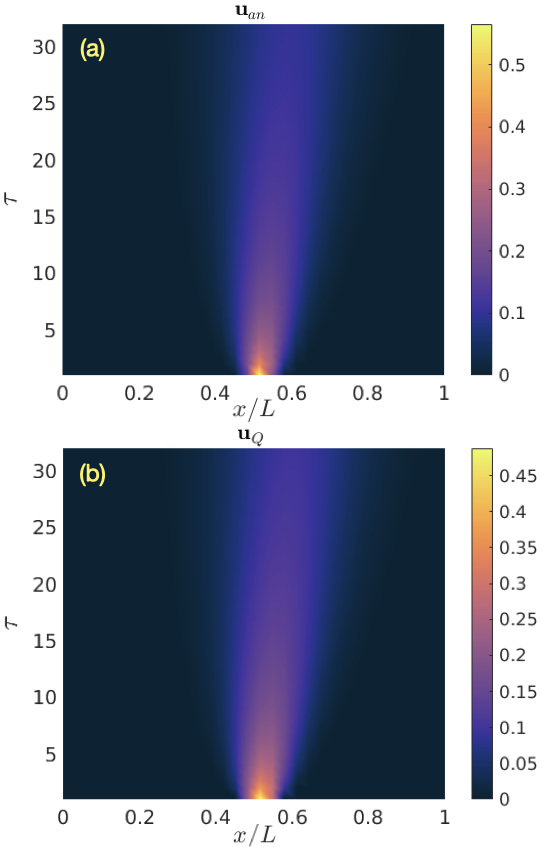}}
\subfloat{\includegraphics[trim={-1cm 0.01cm 0.1cm 0.1cm},clip=true,scale=0.67]{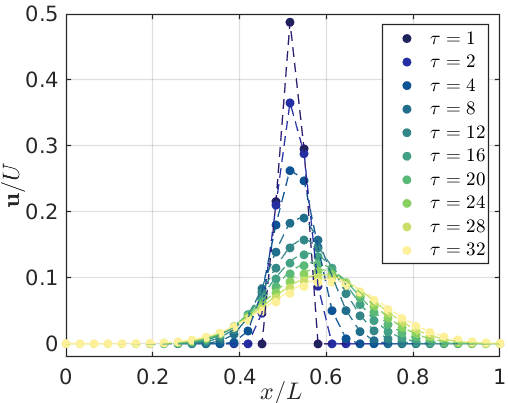}}

\subfloat{
   \hspace{-0.6cm} \includegraphics[trim={0.1cm 0cm 0.2cm 15cm},clip=true,scale=0.45]{uQ_u_an_contour_N32_tau32_dt0-00025_U10_eps0-001.jpeg}}
\subfloat{\includegraphics[trim={-0.5cm 0.01cm 1cm 0.5cm},clip=true,scale=0.68]{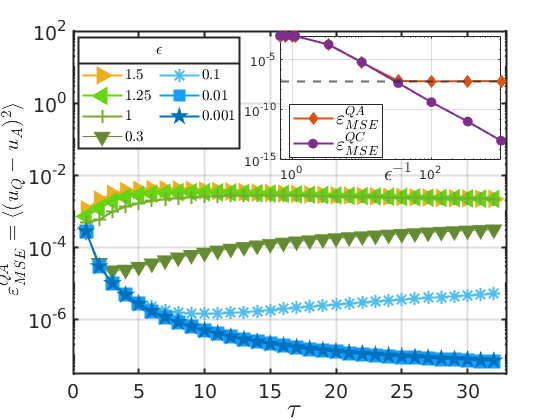}}
    \caption{\justifying (a-b) The figure shows contour shaded plots of the evolution of the velocity field as a function of the $x$ direction (with with $N_{g}=32$) and time steps $\tau$ (up to $\tau=32)$. (a) depicts the analytical solution and (b) the quantum solution computed on QFlowS. The system parameters chosen here are: $dt=\expnumber{2.5}{-4}$, $D=1$, $\alpha=0.256$, $C=10$ and $\epsilon=0.001$. The color bar indicates the magnitude of the velocity field values. (c) Shows the convergent behavior of the mean-squared-error (MSE) in the quantum solution with respect to the analytical solution. The error is plotted as a function of time for different values of $\epsilon$, using up to a total of 12 qubits. \textit{Inset}: Taking the MSE at $\tau=32$, we observe that the error when computed with respect to analytical solution (red) begins to saturate for decreasing values of $\epsilon$, while with respect to the classical solution (violet) it continues to decay with a power-law like behavior. Here, the black dotted-line indicates MSE of the classical solution with respect to analytical solution ($\sim \expnumber{7}{-8}$), which clearly forms the lower bound on the MSE of the quantum solutions for small $\epsilon$ and hence explains the observation. }
    \label{fig:ucontour}
    \end{figure*}

\begin{figure*}[htpb!]
    \hspace{-0.8cm}
     \subfloat[]{
    \includegraphics[trim={0.1cm 0.1cm 0.9cm 0.25cm},clip=true,scale=0.64,valign=c]{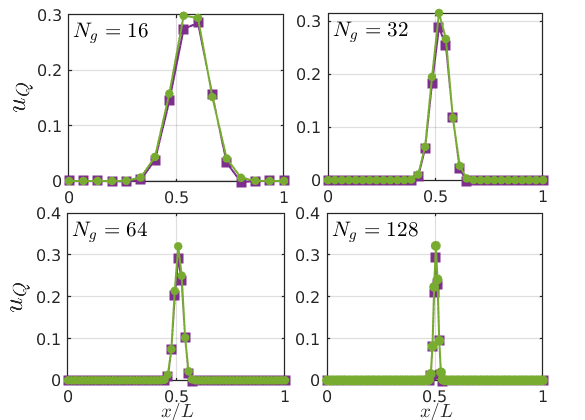}}
    \subfloat[]{
    \includegraphics[trim={0.7cm 0.2cm 1.15cm 0.55cm},clip=true,scale=0.65,valign=c]{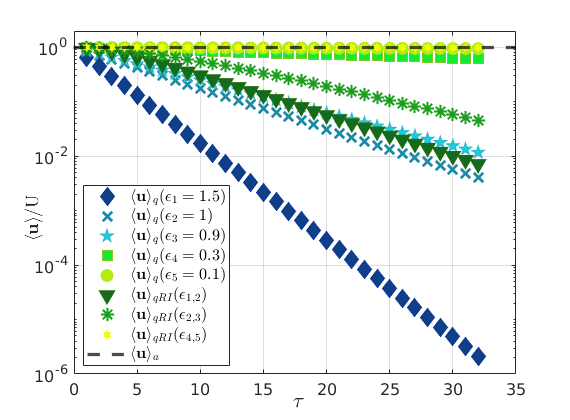}}
    
    \subfloat[]{\adjustbox{valign=c}{
\begin{NiceTabular}{|c c c c c|}
\hline \hline

 $~~N_{g}~~$ & $~~\epsilon_{1}=1~~$  &$~~\epsilon_{2}=0.9~~$  &   $\gamma=1.1\dot{1}$ (RE)  & Accuracy$\uparrow$\\

 \hline \hline

 8  & $\expnumber{3.4}{-03}$ &$\expnumber{2.4}{-03}$  & $\expnumber{2.7}{-04}$ & 88.75\%\\
16 & $\expnumber{1.9}{-03}$ & $\expnumber{1.3}{-03}$  & $\expnumber{6.3}{-05}$ & 95.15\%\\
32 & $\expnumber{9.5}{-04}$ & $\expnumber{6.8}{-04}$ &$\expnumber{3.5}{-05}$ & 94.85\%\\
64 & $\expnumber{4.8}{-04}$ & $\expnumber{3.4}{-04}$ & $\expnumber{1.8}{-05}$ & 94.71\%\\
128 & $\expnumber{2.4}{-04}$ & $\expnumber{1.7}{-04}$ & $\expnumber{9.2}{-06}$ & 96.17\%\\

 \end{NiceTabular}}}

    \caption{\justifying (a) The figure shows the velocity field quantum and analytical solutions plotted for increasing grid resolution $N_{g}\in[16,32,64,128]$ at $\tau=3$. The number of qubits required is $\log_{2}(N_{g})+\lceil\log_{2}(\tau)\rceil + 2$ which is $\in[8,9,10,11]$. Since, we use the iterative TMCQC2, to eliminate any contribution from the serially accumulating error (with increasing $\tau$), and to fairly assess only the effect of the extrapolation, we pick solutions computed at small $\tau$. (Such a constraint is, however, not necessary if one-shot methods TMCQC5,6 are used). We fix $\alpha = 0.256$ and thus the time step size is chosen accordingly for every $N_{g}$. Quantum solutions are computed from extrapolating solutions at $\epsilon_{1}=1$ and $\epsilon_{2} = 0.9$. We note that extrapolation is most effective when $\gamma\rightarrow1$, while still maintaining $\epsilon\sim\mathcal{O}(1)$. (b) Mean velocity field as a function of $\tau$, for $N_{g}=32$, $dt=2.5e-4$, $U=1$ and $D=1$. We note that smaller values of $\epsilon$ produce solutions closer to the analytical value (solid black line) and as large a value as $\epsilon=0.01$ is enough to capture the analytical solution with excellent accuracy. We also note that applying Richardson extrapolation improves the accuracy of the solution, and more so when the unextrapolated values for extrapolation are closer to $\epsilon \sim \mathcal{O}(1)$.
    (c) Shows that the solution obtained from Richardson extrapolation offers higher accuracy than from its parent $\epsilon$ values. The accuracy also clearly increases with increasing grid resolution. The overall accuracy improvement is $\approx 95\%$, for $N_{g}>8$.}
    \label{fig:Resolution and RI}
\end{figure*}

\section{Numerical Results} 
\label{sec: Numerical results}

In this section, we implement the quantum algorithms on simulators as well as a real quantum device. All the TMCQC algorithms proposed in this work have as a common feature, namely the two-unitary or four-unitary decomposition. The circuit designs differ in only the number of total unitaries involved for marching $\tau$ time steps as well as in the number of ancillae qubits required. We explore considerably large parameter spaces that affect the performance of the algorithms, while also maintaining brevity and completeness of discussion. To achieve this, we present results from implementing the TMCQC2 algorithm. Although this algorithm does not enjoy the best asymptotic complexity, it is still an ideal representative of the general circuit structure of all other TMCQCs. Another reason for this choice is that the current capacity of the simulators used here, such as QFlowS and Qiskit, limits the maximum size of circuits that can be simulated. The algorithm chosen conforms to these restrictions while giving extra room to explore a wider range of parameters. We evaluate the correctness and performance of the algorithm and disentangle the interplay between noise, sampling accuracy and the accuracy of the algorithm itself, to estimate the relevant quantities needed for time marching simulations on near-term quantum devices.

\subsection{QFlowS Simulation}
The algorithm must be able to produce physically meaningful and accurate results in the absence of noise or decoherence errors. In to this, we implement a full gate-level simulation of the present flow problem (eq.~(\ref{eq:advcetion-diffusion flow})) on QFlowS \cite{bharadwaj2023hybrid}. QFlowS is an in-house, high-performance quantum simulator, with which we perform ideal state vector simulations, without noise or decoherence. Building noise models into this simulator is an ongoing effort. 

From figure \ref{fig:ucontour}(a,b), we observe that, qualitatively, the quantum solution captures the analytical solution well in both space and time. The initial condition of the flow is a delta function, as shown by the bright yellow spot in figure \ref{fig:ucontour}(a,b). In time, the bright peak slowly diffuses while it also slowly advects in the direction $x>0$. In figure \ref{fig:ucontour} (c), we compare this solution with the classical numerics, since both of them are plagued by the same errors due to the finite difference approximation. The two solutions agree well. This observation is then bolstered quantitatively by computing MSE as shown in figure \ref{fig:ucontour}(d) for different values of $\epsilon$. It can be clearly seen that the MSE converges with time, with the accuracy improving with decreasing values of $\epsilon$, as expected. The more gradual decay in MSE for larger $\epsilon$ can be attributed to two reasons. First, the inaccurate unitary representation of $M$ leads to faster diffusion (numerical diffusion), which leads to faster decay of the solution to zero, as shown by MSE asymptoting with $\tau$. Second, the finite resolution of $x$ and $t$ leads to inaccuracies that accrue quickly, impeding the convergence of MSE. This issue plagues even smaller $\epsilon$. Thus, in practice, for error convergence in serial time marching algorithms (TMCQC1-4), it is crucial to have large enough resolution to keep under check the errors accruing due to large $\epsilon$ approximations, device noise errors and inaccurate quantum state tomography. Apart from better resolution, simulating with larger values of $\epsilon$ is somewhat crucial to lift the solution above the noise level of the hardware, and to lower the query complexity. To do this, we now invoke the Richardson extrapolation to lower shot count required to sample the quantum solution or, conversely, to improve the accuracy of the solution given a fixed number of shots. We first present results using the latter interpretation; the former is discussed later in this section. We extrapolate the solutions computed at larger $\epsilon$ values ($\epsilon\sim\mathcal{O}(1)$), to estimate the solution as $\epsilon\rightarrow 0$ in a deferred manner as described in the previous section. The solutions computed in this way are shown in figure \ref{fig:Resolution and RI}(a). Here, we also capture the effect of increasing grid resolution (up to 128 grid points, with 11 total qubits). The extrapolated quantum solutions show excellent agreement with the exact solution and the accuracy improves with increasing grid resolution as summarized in Table \ref{fig:Resolution and RI}(c). Further, we show in figure \ref{fig:Resolution and RI}(b) the applicability and effectiveness of the extrapolation technique extended to even quantum post processing results, where one measures a function of the solution instead of measuring the entire field itself. 

\begin{figure*}[htpb!]
   
    \subfloat[]{
    \includegraphics[trim={0cm 0cm 0cm 0cm},clip=true,scale=0.33,valign=c]{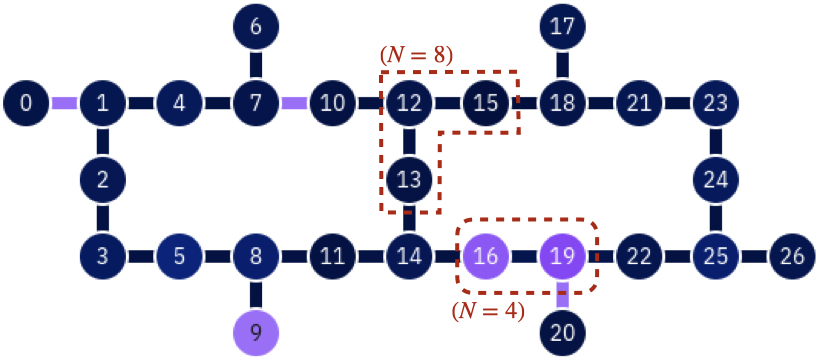}}
    \subfloat[]{\adjustbox{valign=c}{
\begin{NiceTabular}{|lr|}
\hline \hline
 \Block[tikz={top color=gray!25}]{*-1}{} &  
 \Block[tikz={top color=blue!20}]{*-1}{}\\
  \textbf{Feature} & \textbf{IBM Cairo Specification}\\
  & \\
  \hline \hline
  & \\
 Qubits & 27\\
 Processor & Falcon r5.11 (version 1.3.6)\\
 CLOPS & 2400 \\
 Basis gates & CX, ID, RZ, SX, X\\
 Median CNOT error & $\sim 9\times10^{-3}$\\
 Median SX error & $\sim 2\times10^{-4}$\\
 Median readout error & $\sim 1\times10^{-2}$\\
 Median $T_{1}$ & 102.73 $\mu s$\\
 Median $T_{2}$ & 95.42 $\mu s$\\
 \end{NiceTabular}}}
 
    \subfloat[]{\hspace{-0.5cm}
    \includegraphics[trim={0.1cm 0.2cm 0cm 0.1cm},clip=true,scale=0.25,valign=c]{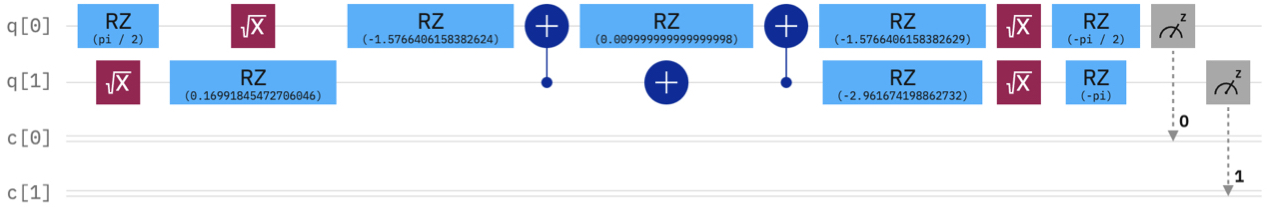}}
    \subfloat[]{\adjustbox{valign=c}{
\begin{NiceTabular}{|c|c|c|}
\hline \hline
 \Block[tikz={top color=gray!25}]{*-1}{} & \Block[tikz={top color=blue!25}]{*-1}{} & 
 \Block[tikz={top color=blue!20}]{*-1}{}\\
 \textbf{Time} (in $\mu s$) & $N_{g}=4$ & $N_{g}=8$ \\
 & &\\
 \hline \hline
 & &\\
 Mean classical time  & 0.9 & 1.13\\
 & &\\
 IBMQ Cairo $T_{\textrm{total}}$ & 27.65\textrm{e}+6 & 28.11\textrm{e}+6\\
 & &\\
 IBMQ Cairo $T_{\textrm{s}}$ & 52.73 & 53.61\\
 & &\\
 IBMQ Cairo $T_{\textrm{s-act}}$ & 2.733 &3.607\\
 & &\\
 \end{NiceTabular}}}
 
     \caption{\justifying (a) The figure depicts IBM Cairo's 27-qubit circuit topology. The qubits highlighted by the red, dotted boxes indicate the specific qubits that were assigned for running the $N_{g}=4,8$ experiments to store the velocity field solution. The colors of the qubits and their interconnects indicate the instantaneous readout assignment error and the CNOT error rates, respectively, with darker shades indicating lower errors;  (b) summarizes the processor specification of IBM Cairo. CLOPS - Circuit Layer Operations per Second. $T_{1}$ and $T_{2}$ refer to maximum energy relaxation time and dephasing/decoherence times of the qubits, respectively. (c) Shows the 2 qubit, Qiksit transpiled circuit of depth 10, for $N_{g}=4$ case on IBM Cairo, using the available gate library. (d) This table summarizes the timing data obtained for $N_{g}=4,8$ experiments. The mean classical time corresponds to the timing statistics to compute the classical solution using the MATLAB software. $T_{\textrm{total}}$ corresponds to the total time taken to iteratively perform $N_{s}=2^{19}$ shots of the same circuit. $T_{\textrm{s}}$ corresponds to the time required per shot computed naively as $T_{\textrm{total}}/N_{s}$. $T_{\textrm{s-act}}$ denotes the actual time taken per shot corresponding to a restless simulation, after subtracting the delay overhead times of $T_{\textrm{delay}} = 50 \mu s$/shot \cite{tornow2022minimum}. }
    \label{fig:ibm cairo topology}
    
\end{figure*}

\begin{figure*}[htpb!]
\subfloat{\includegraphics[trim={1.4cm 0.6cm 3.2cm 1.9cm},clip=true,scale=0.17]{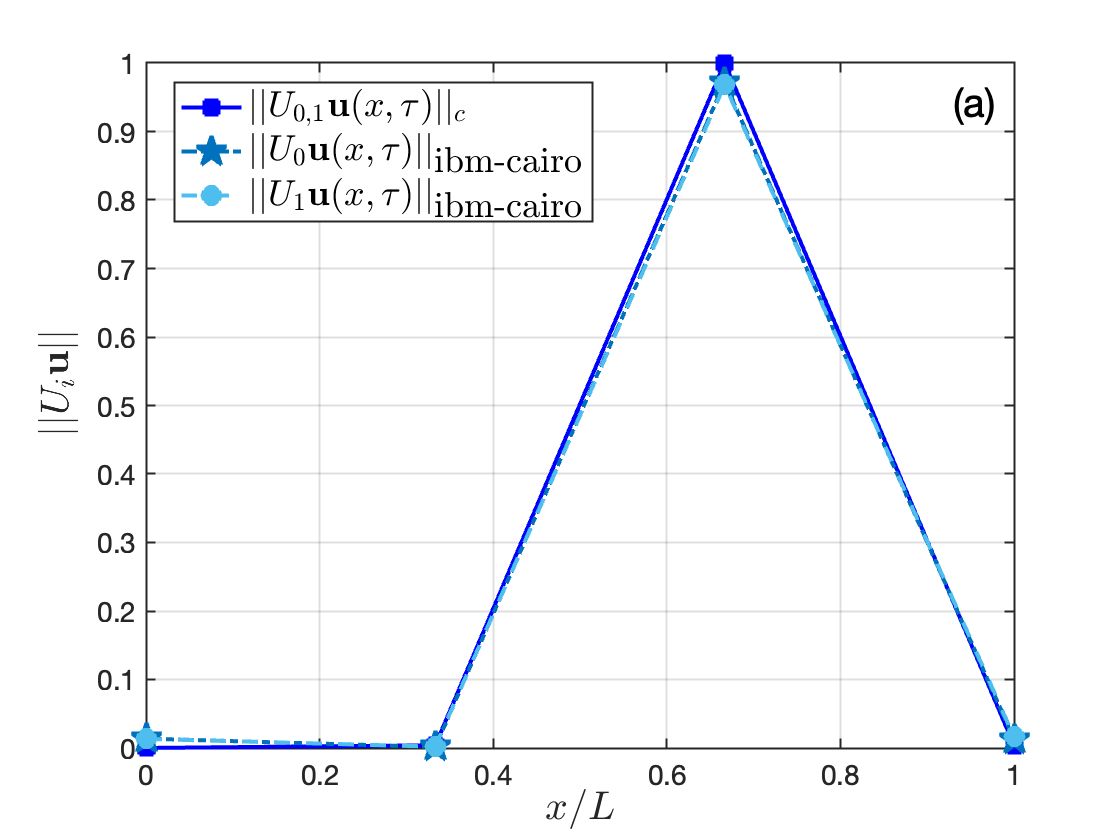}}
\subfloat{\includegraphics[trim={1.4cm 0.6cm 3.2cm 1.9cm},clip=true,scale=0.17]{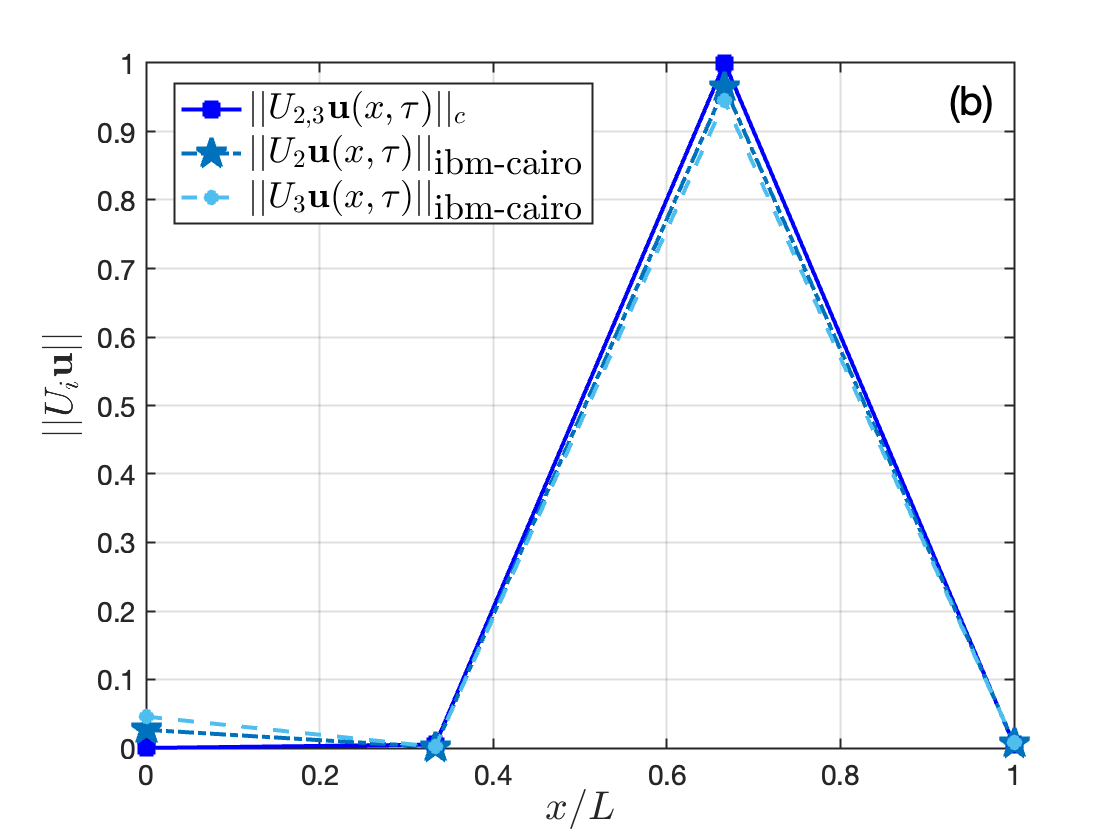}}
\subfloat{\includegraphics[trim={0.5cm 18.5cm 0.2cm 0.0cm},clip=true,scale=0.265]{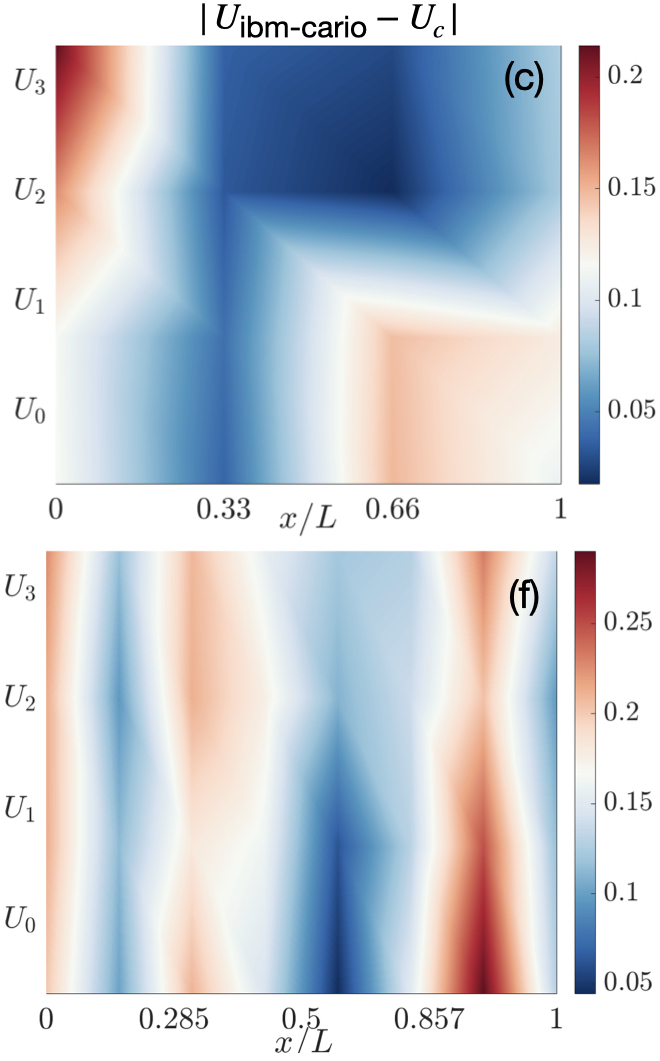}}

\subfloat{\includegraphics[trim={1.4cm 0.6cm 3.2cm 1.9cm},clip=true,scale=0.17]{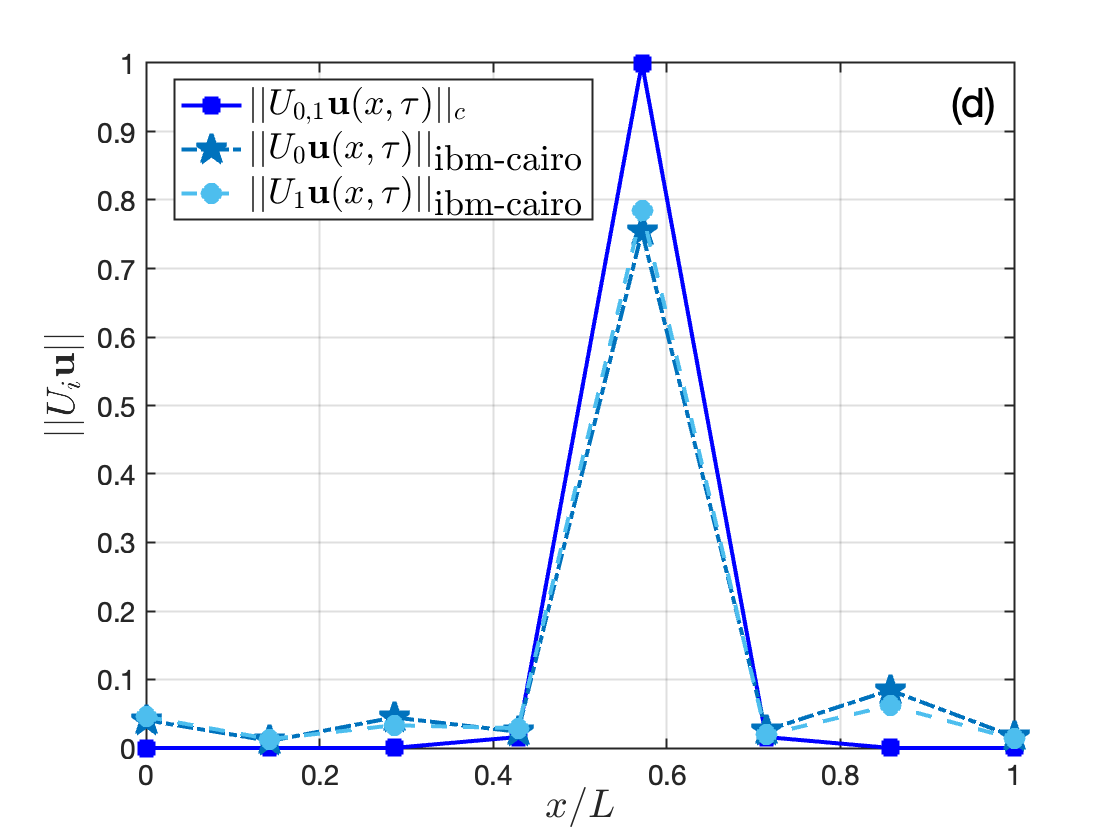}}
\subfloat{\includegraphics[trim={1.4cm 0.6cm 3.2cm 1.9cm},clip=true,scale=0.17]{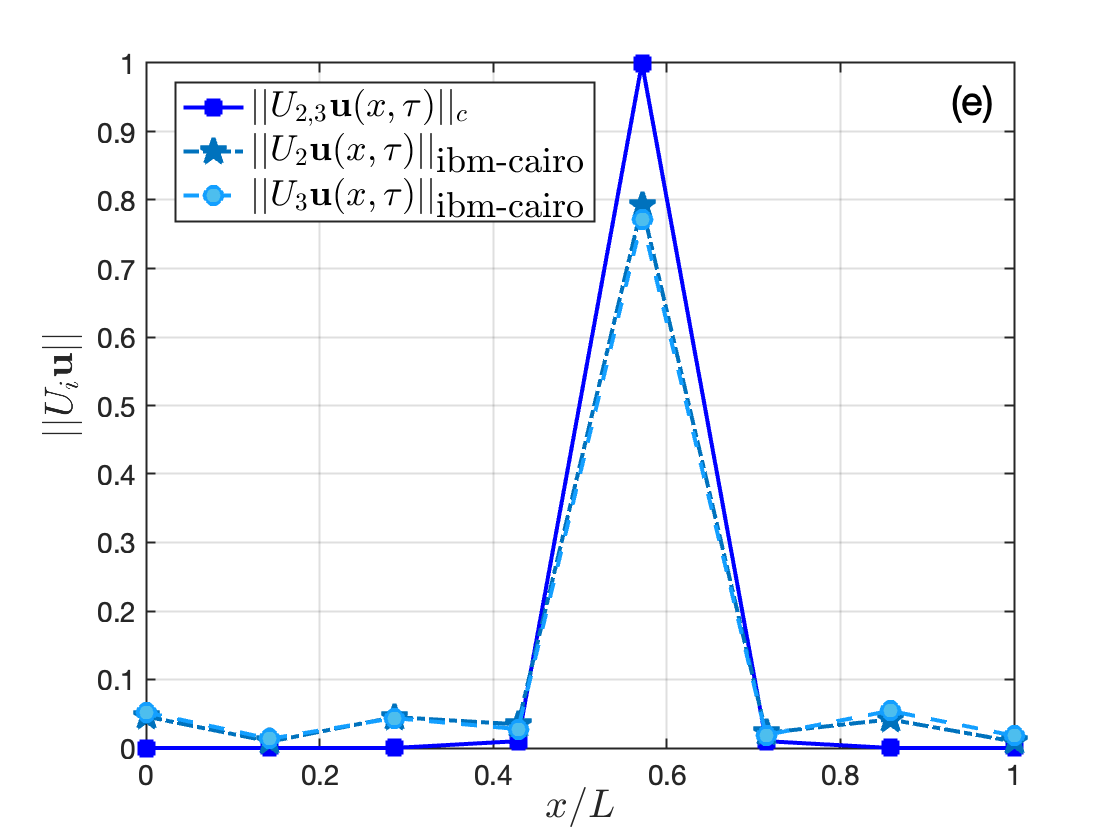}}
\subfloat{\includegraphics[trim={0.0cm 0.2cm 0.2cm 19cm},clip=true,scale=0.262]{U0123Error_IBMCairo_Contours.jpeg}}
     \caption{\justifying (a,b) and (d,e) shows the parallel solutions (star, circle symbols) computed on IBM Cairo quantum device, that represents the action of the unitaries $U_{0,1,2,3}$ on the velocity field, plotted along with the corresponding classical solutions (square symbols) for the $N_{g}=4$ and $N_{g}=8$ cases respectively. The following parameters were chosen for this experiment -- $\epsilon=1$, $dt=\expnumber{2.5}{-5}$, $D=1$ and $C=10$. (c,f) shows the absolute error field of these quantum solutions computed with respect to the classical solutions, plotted for each unitary as function of $x/L$. The colorbar represents the absolute value of the error.}
    \label{fig:Cairo setup}
    
\end{figure*}

\subsection{IBM Quantum Device Experiment} 
Here we discuss the implementation of the unitaries decomposition on an IBM quantum platform {that have quantum processors with up to} 127 qubits. We use the IBMQ Cairo device with 27 qubits, whose circuit topology is shown in figure \ref{fig:Cairo setup}(a) and its specifications (at the time when the experiment was performed) is summarized in figure \ref{fig:Cairo setup}(b). Since the devices are generally calibrated almost every two hours, the exact experimental results might be different at different times. Apart from the gate and measurement errors, one also needs to account for the times $T_{1}$ and $T_{2}$, which indicate the \textit{lifetime} of the qubits, beyond which decoherence sets in, leading to another source of error.. To avoid this, the depth of the circuit needs to be sufficiently shallow for the execution time per shot $T_{\textrm{s}} < \min\{T_{1},T_{2}\}$, whose values are provided in figure \ref{fig:Cairo setup}(b). Considering these error rates and coherence times, the circuit sizes of current devices (within a reasonable error threshold) is rather limited. For this reason, we attempt to implement a parallelized, single time step circuit (for $N_{g}=4$ and \textcolor{red}{$N_{g}=8$}), by decomposing it into basic gates with Qiskit transpile. We show that our unitaries can indeed be simulated with reasonable accuracy on a real quantum device, bolstering the viability of the current algorithm to be scaled up, given adequate resources on near-term devices. This ability also opens a promising avenue to actually perform multi-step, time-marching simulation of PDEs on real quantum devices. We have not used any error correction codes or sophisticated qubit topology optimization. The accuracy of the current results even without these additional features shows that further improvement is possible. These tools become mandatory when simulating larger problem sizes.

To implement our algorithm, we employ the unitaries parallelization and the shot parallelization techniques described earlier to make each parallel circuit as shallow as possible and minimize noise and decoherence effects. To decompose the unitaries in each parallel circuit, we use Qiksit transpilation at two-level optimization, generating a circuit consisting of basis gates available on IBMQ Cairo, listed in figure \ref{fig:Cairo setup}(b). It is worth noting that, apart from gate count reduction, circuits can be optimized further by specifying \textit{coupling maps} (although it is not used here). This allows one to specify a custom qubit topology by picking qubits and interconnects with the least instantaneous errors. Further we observe that, the circuit parallelization used here results in sufficient depth reduction that compensates the overhead due to repeated quantum state preparation. Since each unitary is applied in parallel, it removes the need for control qubits and controlled gates, making the circuits more efficient. In fact, the depth reduction (including state preparation) is by a factor $\sim 130$, which is nearly 99\% in certainty. This situation greatly benefits the one-shot algorithms TMCQC5,6. The depths of the methods proposed here appear to be gradually approaching those of variational quantum algorithms \cite{ingelmann2024two,dalton2024quantifying}. The transpiled circuit of these unitaries for the $N_{g}=4$ case is shown in figure \ref{fig:Cairo setup} (c). The depths for the $N_{g}=4,8$ circuits are 10 and 95, respectively (excluding the measurement operators). For the number of qubits being used and the gate count, these depths are close to the simulation limits on these devices, with acceptable error bounds.

At this stage, it is important to note that, for the two-unitary strategy, we have a decomposition that results from a parent Hermitian matrix. By acting these two unitaries in parallel on our instantaneous velocity field, we obtain the quantum states, whose amplitudes are purely real, with no imaginary part. These parallel solutions can be measured simply in the computational basis (with additional measurements for sign estimation), and added classically to obtain the final solution in a hybrid form. In this case, we need an additional qubit to accommodate the dilated Hermitian matrix. Proceeding to a fully parallel four-unitary, whose operators are half the size of that of the two-unitary case, leads to much shallower parallel circuits. However, the parallel solutions from $U_{0}$ and $U_{1}$ have a finite imaginary part in them (unlike solutions from $U_{2}$ and $U_{3}$), although their partial sum is purely real. If we wish to measure these results separately, it would require a full state tomography only for these two components, while the remaining two components can be directly measured as before. 

Instead, if we are only interested in computing the expectation value of an observable $O$, we could do so by keeping the complex amplitudes and by just performing a Hadamard test, which computes the $Re(\langle\psi\vert O\vert \psi\rangle$. It can be modified easily by adding an additional conjugate phase gate $S^{\dag}$, as shown in figure \ref{fig:end2end}(b), also to obtain the $Im(\langle\psi\vert O\vert \psi\rangle$. Here, for a fully parallel four-unitary case, we present some results from computing $\vert\vert U_{i}\mathbf{u}\vert\vert$, for each unitary (as shown in figure \ref{fig:Cairo setup}), and compare them with classical computations of the same quantity, for $N_{g}=4,8$ cases. To obtain the actual velocity field solution with purely real amplitudes, we also perform an intermediate level parallelization for $N_{g}=4$, by combining unitaries into two parallel circuits. For this, with an extra ancilla qubit, we apply controlled $U_{0,1}$ in one circuit and $U_{2,3}$ in another, in parallel. We then combine these results classically. This circuit, of course, requires one less ancilla compared to a full four-unitary circuit. We also compare this with the parallel two-unitary case. Both yield solutions with comparable accuracy. Of them, the fully parallel four-unitary case suffers far less from noise and decoherence, owing to its short depth circuits, suggesting that accurate computation of expectation values of operators seems more reachable on near-term devices.
\begin{figure}\hspace{-0.5cm}
    \centering
    \includegraphics[trim={0.1cm 0.1cm 0.2cm 0.2cm},clip=true,scale=0.245]{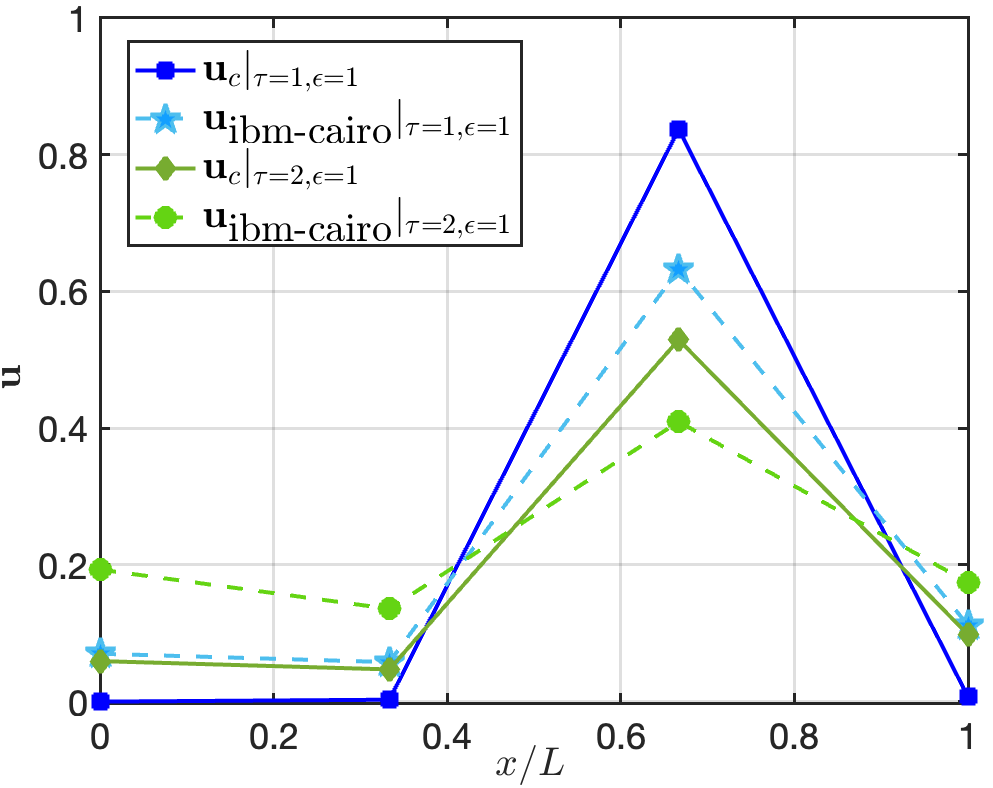}
    \caption{\justifying The fully reconstructed velocity field of the first two time-marching steps (dashed-lines), computed from parallel experiments on IBM Cairo, is plotted along with the classical solution (solid-lines) corresponding to the same input matrix, at $\epsilon=1$. $(\textrm{MSE}\vert_{\tau=1},\textrm{MSE}\vert_{\tau=2})= (0.0152,0.0115)$.} 
    \label{fig:cairo real vel}
\end{figure}
\begin{figure*}[htpb!]
    \subfloat[]{\includegraphics[trim={1.5cm 1.6cm 3.1cm 0.7cm},clip=true,scale=0.185]{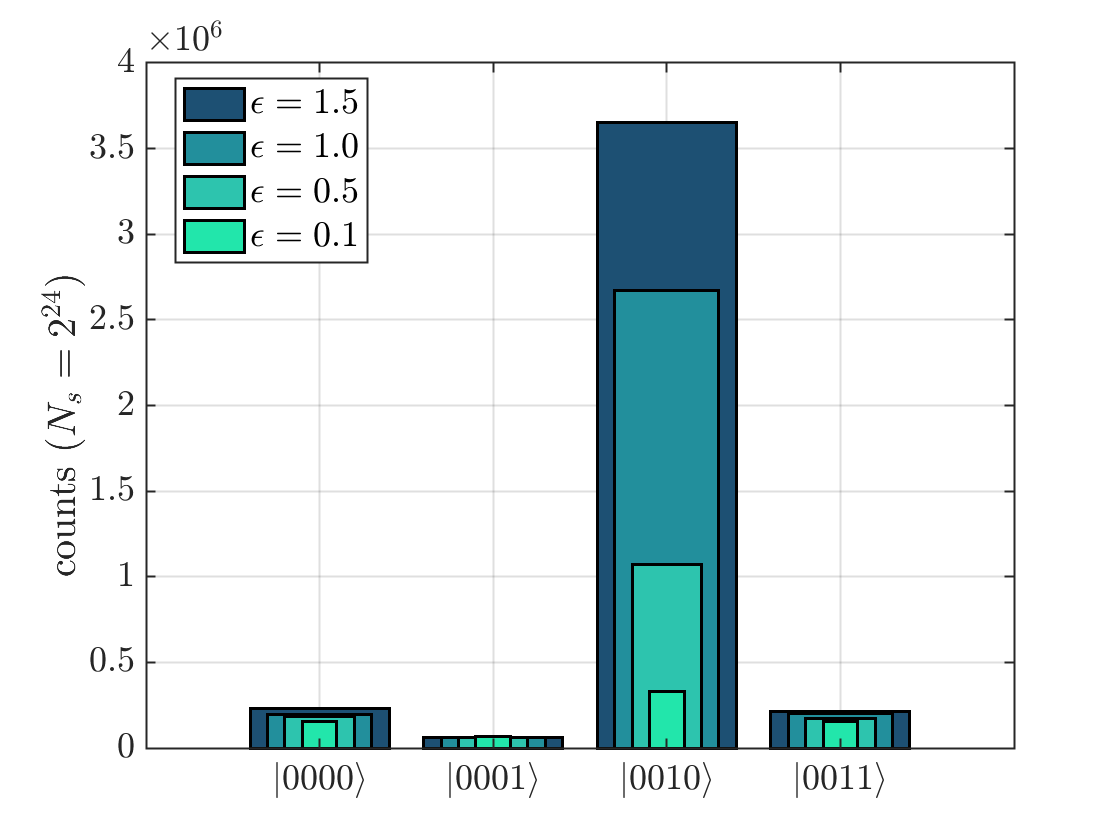}}
     \subfloat[]{
    \includegraphics[trim={0.1cm 0.0cm 0.2cm 0.4 cm},clip=true,scale=0.3]{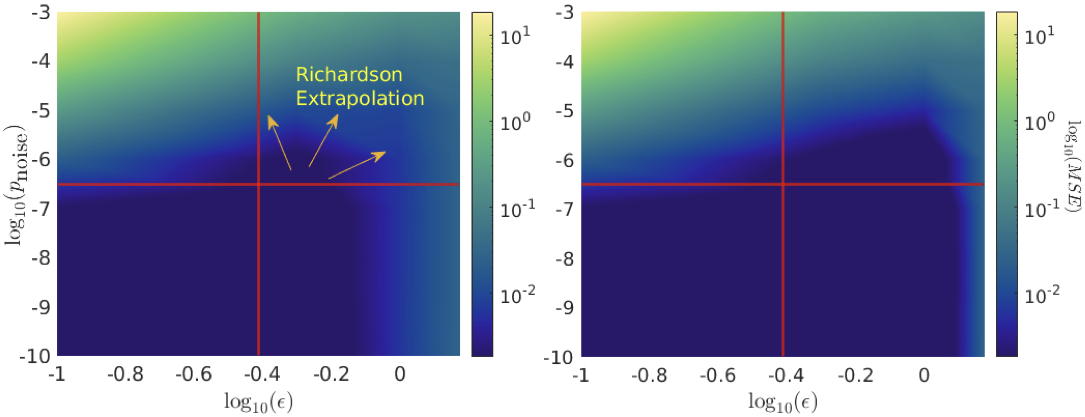}}
    \vspace{-0.45cm}
    \subfloat[]{\includegraphics[trim={0.4cm 0.1cm 1cm 0.6cm},clip=true,scale=0.665]{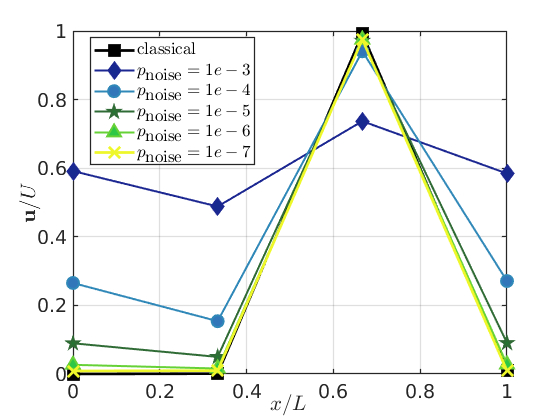}}
    \subfloat[]{
    \includegraphics[trim={0.1cm 0.0cm 0.8cm 3 cm},clip=true,scale=0.655]{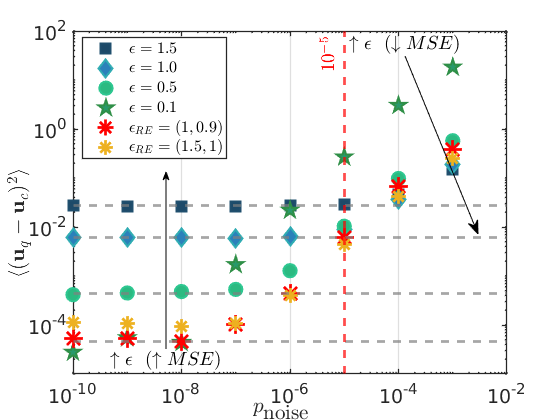}}
    
     \caption{\justifying (a) Shown here are histograms corresponding to the post-selected solution subspace, sampled over a total of $N_{s}=2^{24}$ shots, for $N_{g}=4$ and decreasing values of $\epsilon$. The error rate is set to $\expnumber{1}{-4}$. (b) shows the contour diagram of the mean-squared-error in the noisy quantum solution for a single time step, plotted as function of $p_{\textrm{noise}}$ and $\epsilon$, both in $\log$-scale. The color bar indicates the error magnitude. (c) shows line plots of the velocity field solutions for varying magnitudes of error rates. The thick, black line is the classical solution for reference. (d) depicts the variation of the MSE as a function of error rates, plotted for varying $\epsilon$. The horizontal, gray, dashed-lines indicate the asymptotic zero-noise limit for each $\epsilon$, whereas the vertical, red dashed-line indicates the error-threshold for the onset of error-correction codes. }
    \label{fig:qasm_noisy N4}
\end{figure*}

Further, to ensure that our results are not dominated by the errors due to decoherence of qubits, one requires $T_{\textrm{s}} < \min\{T_{1},T_{2}\}$. We examine this by studying the timing performance of these experiments on IBM Cairo, which is summarized in figure \ref{fig:ibm cairo topology}(d). The measurements show that the naive execution time per shot is $T_{\textrm{s}} \approx \frac{1}{2}\min\{T_{1},T_{2}\}$. It has to be noted that $T_{\textrm{s}}$ is actually computed by naively dividing the overall execution time by $N_{s}$ shots, but this does not reflect the actual time taken per shot ($T_{\textrm{s-act}}$). After every shot, there is a finite delay time before the next shot gets executed, apart from other overhead times for every shot \cite{tornow2022minimum}. By accounting for these overheads, our estimate of $T_{\textrm{s-act}}$ shows that $\min\{T_{1},T_{2}\} \approx 50T_{\textrm{s-act}}$, and therefore the circuit sizes are well within these decoherence limits. We also then compute the mean time required to simulate the same problem classically, on MATLAB {(with Intel(R) Core(TM) i9-9900 CPU @ 3.10GHz processor)}. The results indicate that the timings are rather comparable with the classical algorithm being $\approx 3\times$ faster than $T_{\textrm{s-act}}$. However, it remains to be seen how these numbers would scale for larger problems when compared with high performance classical algorithms. 

Figures \ref{fig:Cairo setup}(a,b,d,e) show that the parallel solutions of $U_{i}\mathbf{u}$, computed on IBM Cairo, capture the classical solutions well. However, the error can be seen to increase with resolution, as a result of the inevitable increase in the circuit depth and the associated errors. The overall error, however, is comparable to the highest accuracy one can expect on that specific quantum device (for these circuit sizes), suggesting that our circuits are able to extract the maximum available accuracy of the machine. The error contours shown in figures \ref{fig:Cairo setup}(c,f) suggest that high amplitude regions of the wavefunction are recovered more accurately than near zero. This is expected since the near-zero values are more prone to noise. Proceeding further, using an intermediate level parallelization with the four-unitary scheme, we show in figure \ref{fig:cairo real vel}, the results of marching two time steps by reconstructing the full solution. It can be seen clearly to capture the corresponding classical results well, having errors comparable to those on the quantum device. The partial sum of parallel circuits used here have depth of about 153 and the CNOT error rate of the device is $\approx \expnumber{9}{-3}$ (experiment performed on March 30, 2024). Considering the measured MSE, the performance as expected shows that our circuits fully extract the available accuracy on these machines. Again, the near-zero values cannot be captured accurately given the overwhelming effects of noise. To overcome this difficulty, instead of a delta function, one could choose an initial condition of the form (say) $u(x,t=0)=0.5\sin(\pi x) + 1$, which also satisfies the boundary conditions. The shifted initial condition avoids near zero solutions. In any case, the results shown here forms a stricter evaluation of the performance.

It is important to reiterate here that the algorithms proposed in this work do not require measurements to be made in between time steps, but the results shown above (from real device experiments) do involve measurements at each time step, considering the currently available circuit sizes. Even though such measurements compromise the quantum advantage, besides adding error because of intermediate measurements, we proceed to do so solely to assess the algorithm's ability to implement the proposed decomposition accurately. However, the accuracy, besides being constrained by the device's limitations, also depends dynamically on the instantaneous magnitude of the solutions, since larger magnitudes can be easily distinguished from the underlying noise. The time-marching performance discussed already and in the next section can thus be considered to represent a very strict lower bound on performance, where the accumulating errors are at their maximum. Nevertheless, the errors in every time step, when compared to the instantaneous classical solutions, are still within admissible values considering the specifications of the machine, thus reflecting the viability of the proposed algorithms. 

Now, if we consider scaling up the problem, we would have to replace these naively transpiled circuits with the transpiled Hamiltonian simulation circuits. At the same time if we invoked the TMCQC5,6 designs, it would result in maximum quantum advantage as theoretically shown here. In this case there will be a total of $\mathcal{O}(P^{3}_{min})$ unitaries that would need to be parallelized. The reduction in the number of CNOTs due to removal of controls on unitaries (and ancilla qubit complexity) by parallelization performs even better at those scales. While we await the realization of that feature as quantum devices emerge with better error-correction subroutines \cite{bravyi2024high,da2024demonstration}, given that asymptotically the proposed TMCQCs require exponentially fewer resources, a decent possibility exists of achieving quantum advantage in the near-term when combined with end-to-end strategies that have been described earlier, along with suitable error correction subroutines \cite{bravyi2024high,da2024demonstration}.

To now disentangle the interplay among noise, $\epsilon$ and finite sampling of quantum states, and to study their effect on the accuracy of solutions, we perform noisy simulations on Qiskit Aer for a range of these parameters. We outline our observations in the following section. For this, we implement circuits that include all controlled operations without any parallelization. These circuits are thus much deeper, yielding rather overly safe estimates of the error rates and $\epsilon$ to achieve convergent simulations. In practice, with careful optimization, parallelization, and basic error-correction, these estimates can be relaxed significantly (by one to two orders of magnitude).

\begin{figure*}[htpb!]
   
    \subfloat[]{\hspace{-0.3cm}\includegraphics[trim={0.2cm 0.1cm 0.2cm 0.2cm},clip=true,scale=0.25]{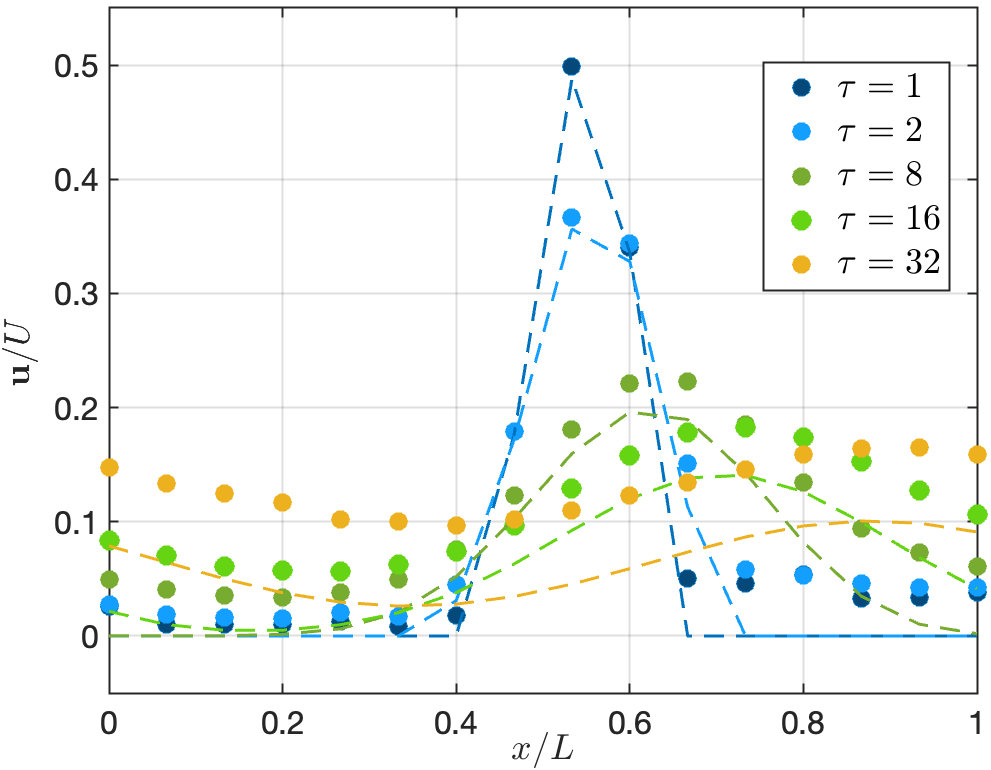}}
    \subfloat[]{\hspace{0.2cm}
    \includegraphics[trim={0.2cm 0.1cm 0.2cm 0.2cm},clip=true,scale=0.252]{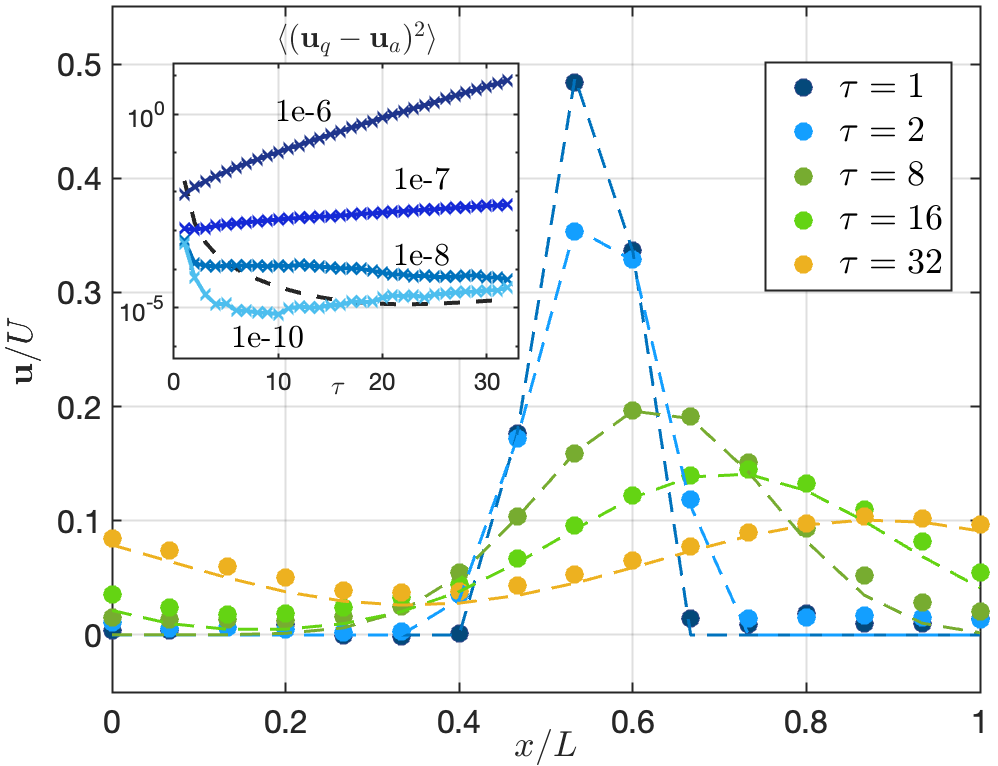}}
    \vspace{-0.2cm}
    \subfloat[]{\hspace{-0.5cm}
    \includegraphics[trim={0.0cm -0.1cm 0.1cm 0.1cm},clip=true,scale=0.25]{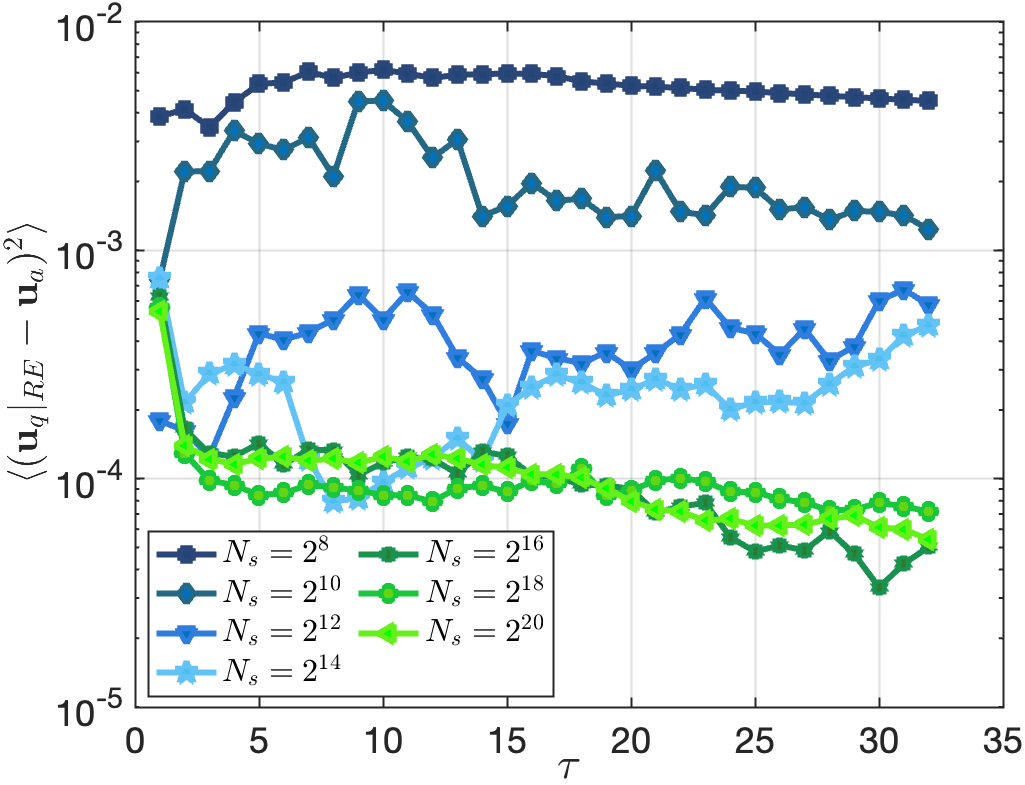}}
    \subfloat[]{
    \includegraphics[trim={0.0cm 0.2cm 0.3cm 0.2cm},clip=true,scale=0.249]{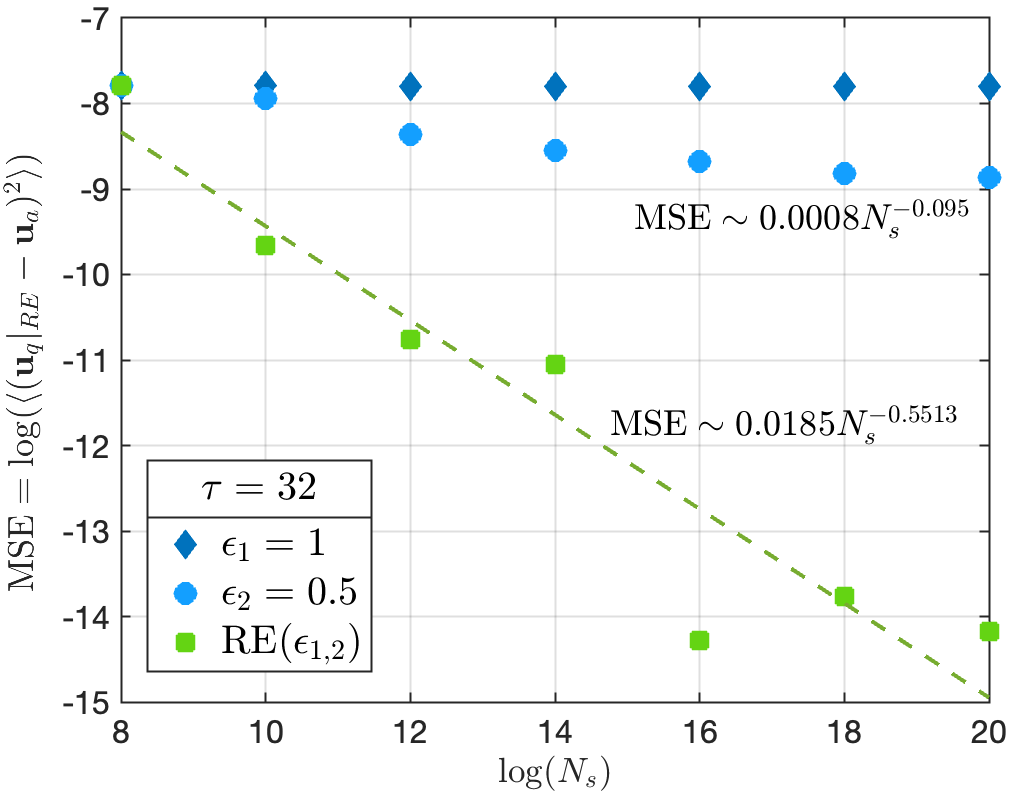}}
   
     \caption{\justifying Figures (a,b) show the time evolution of the velocity field (filled-symbols) with $N_{g}=16$ and $\Delta t=0.001$, at error rates $p_{\textrm{noise}} = 10^{-7}$ and $10^{-8}$ respectively, up to $\tau=32$. The dashed lines represent the corresponding classical solution plotted for reference. \textit{Inset--} of (b) shows the mean-squared-error convergence as a function of time, of solutions obtained after extrapolation $RE(\epsilon_{1},\epsilon_{2})=(1.0,0.5)$, plotted for varying $p_{\textrm{noise}}$ values. The black dashed-line indicates the best possible scenario, from a no-noise state vector simulation. (c) also shows the error convergence of the extrapolated flow field solutions as a function of time, plotted for varying number of shots $N_{s}$. (d) shows a $\log-\log$ plot representing the decay of the mean-squared-error (with respect to the analytical solution) at $\tau=32$, as a function of increasing sampling sizes $N_{s}$.}
    \label{fig:qasm_noisy N16}
\end{figure*}
\subsection{Qiskit Aer - Noisy Simulations}
The influence of noise on the accuracy of results needs to be assessed on the viability of our algorithms on near-term devices and on the algorithmic design for maximum accuracy. We use Qiskit Aer to build a custom noise model into our algorithm, allowing us to simulate noisy circuits. In this work we use the bit-flip error noise model to set error rates (failure rates) of measurement operations ($p_{\textrm{meas}}$) and reset operations ($p_{\textrm{res}}$) and gate action ($p_{\textrm{gate}}$). Here, we set all the error rates to be equal ($=p_{\textrm{noise}}$) and vary them together. In practice, the reset and measurement error rates differ from gate error rate by one or two orders of magnitude. However, noting that the noise models in Aer do not show appreciable differences when they are set equal to each other. The error rate can be thought of as the probability with which a certain operation/gate would fail. The error rate on current devices (at least those that were available to us) are at best $p_{\textrm{noise}}=10^{-3}$. Further, it has been shown earlier in \cite{gottesman2022opportunities,aliferis2005quantum}, that if these error rates can be reduced to threshold values of $\in[10^{-4}$ and $10^{-5}]$, there exist error-correction algorithms with which one may perform simulations of arbitrarily long circuits. This seems reasonable because quantum error correction methods deal with gate errors that dominate other sources such as measurement errors that are dealt with using other methods.  In fact, more recent progress \cite{bravyi2024high} has shown that, using carefully designed error correction codes, these error rates can be further reduced to about $p_{\textrm{noise}}\in[10^{-7}, 10^{-12}]$. These algorithms make use of \textit{logical} qubits that are, in turn, composed of several qubits, resilient as an aggregate to errors due to noise and decoherence. Therefore, in this work we study the effect of noise by varying error rates between $p_{\textrm{noise}}\in[10^{-3}, 10^{-10}]$. We consider the $N_{g}=4,16$ cases with the same initial and boundary conditions as before.

We recall that the TMCQCs rely on two and four unitaries decomposition to approximate the time marching matrix. This approximation is exact for $\lim\epsilon\rightarrow 0$. However, choosing small values of $\epsilon$ not only diminishes the probability of the post-selected solution subspace, but also the magnitude of the solution amplitudes itself, which will then be scaled down by $\epsilon$. Distinguishing these small magnitudes becomes particularly hard with noise and finite sampling errors due to limited shot count ($N_{s}$) on current devices. The effects of noise can be varied even if one were to provide a large enough $N_{s}$. Let us consider the following two extreme scenarios. 

\textbf{High noise} - With a $p_{\textrm{noise}}=10^{-4}$ (close to the threshold \cite{gottesman2022opportunities,aliferis2005quantum}) and for $N_{g}=4$, we can see from the histogram in figure \ref{fig:qasm_noisy N4}(a), the shot counts of the solution subspace amplitudes become progressively lower with decreasing $\epsilon$. We can also observe that the counts of the small amplitude values seem to saturate and not go to 0. This is a typical signature of a finite noise simulation, where even the near-zero values are lifted to larger finite noise amplitudes. Such errors due to noise can accumulate rather quickly in the time marching solutions, so that their long-time accuracy is not feasible. From this point of view, noise appears to have an adverse impact on the accuracy of simulations. 

\textbf{Zero noise} - For an ideal device with zero noise, it is obvious that the shot count of solutions, especially the ones with small amplitudes, would tend continually towards zero with decreasing $\epsilon$. In fact, for a small enough $\epsilon$, the shot count would be exactly zero for several time steps (leading to divergence) and would therefore require an extremely large $N_{s}$ to even obtain a single sample. This works against our desire to obtain quantum advantage. In this case, the divergence of the error with time steps is essentially due to velocity solutions that remain zero at some grid points for several time steps. From this perspective, it appears that noise can, in fact, be used to advantage, since it amplifies near-zero signals (but these near-zero values would be lost in the noise). Therefore, even though the solutions might have higher errors for the initial few time steps, they could stabilize subsequently when the magnitudes become large enough.

It is now important to remind ourselves that, on current and near-term devices, it is impossible to escape the effects of noise. However, the previous discussion motivates the need to find a middle ground with \textit{finite} noise, just \textit{small} enough $\epsilon$ and $N_{s}$, such that the effective solution possesses maximum accuracy. A simple step to address this issue would be to begin with a shifted initial condition as described earlier ({in Section \ref{sec: Numerical results}.B}), which would make the solutions avoid near-zero values to begin with. Here, however, we still use delta functions to provide a strict analysis on the subject. Further, we can invoke the idea of Richardson extrapolation introduced earlier ({in Section \ref{sec: End-to-end}}) to perform simulations with larger $\epsilon$ values to approximate solutions at smaller $\epsilon$. This tool would also lower the errors in the solution or, alternatively, lower $N_{s}$ for a desired accuracy. {In fact this tool can even be extended to perform error mitigation as shown in \cite{temme2017error}.} Now, to understand the interplay between these quantities we perform our analysis in two steps. First, we fix $N_{s}=2^{24}$, which is large enough to avoid significant undersampling errors. We then study the connection between $p_{\textrm{noise}}$ and $\epsilon$, together with Richardson extrapolation. Second, we fix the error rates and study the connection between $N_{s}$ and $\epsilon$.

For the first step, the error phase diagram is shown in figure \ref{fig:qasm_noisy N4}(b), demarcated into four quadrants. The Northwest (NW) quadrant is least favorable, since we have small $\epsilon$ values coupled with large $p_{\textrm{noise}}$, leading to the solutions being completely overwhelmed by noise, thus showing the least accuracy or maximum error. In the Northeast (NE) quadrant, $\epsilon$ values are also larger and, therefore, even though $p_{\textrm{noise}}$ is large, the solutions are somewhat more expressive in the presence of noise, leading to a slightly better accuracy. The solutions here are still erroneous, since they correspond to an inaccurate unitaries decomposition with large $\epsilon$ values. Next, the Southeast (SE) quadrant shows an improvement over the previous case, since $p_{\textrm{noise}}$ is now small. Finally, the Southwest (SW) quadrant offers the highest accuracy (dark-blue zone) with small $\epsilon$ and $p_{\textrm{noise}}$, both of which are favorable features. The objective would now be to grow the expanse of this dark-blue zone to the rest of the phase-diagram. Eastward with larger $\epsilon$ values, we can invoke the Richardson extrapolation. By simply performing extrapolations with $\epsilon$ values lying in the range $\in[0.5,1.5]$, we see that the accuracy can be improved in the SE quadrant, as shown in the right panel in figure \ref{fig:qasm_noisy N4}(b). Expanding northward would require error mitigation schemes. In fact, to estimate observables of the solutions at the zero noise limit, we can invoke another layer of Richardson extrapolation as shown in \cite{temme2017error}, known as the Zero Noise Extrapolation (ZNE). Exploration in that direction forms an important part of our future work. 

We can see the effect of extrapolation itself a bit more clearly, by plotting the extrapolated solutions (with $(\epsilon_{1},\epsilon_{2})=(1.5,1)$) of the velocity field for varying error rates, as shown in figure \ref{fig:qasm_noisy N4}(c). 
It can be clearly seen that accuracy progressively improves with lower error rates and begins to converge at about $p_{\textrm{noise}}\approx10^{-7}$. Also, the solution at $p_{\textrm{noise}}=10^{-3}$ can be comparable to the solutions obtained from experiments on a real device (IBM Cairo), as detailed in the previous section. To quantify the error and magnify the trends, we plot MSE as a function of $p_{\textrm{noise}}$. In the high noise region, it can be seen that smaller $\epsilon$ values clearly lead to larger errors and this trend reverses in the low noise region. As $p_{\textrm{noise}}$ becomes very small, the errors for each $\epsilon$ asymptote to the corresponding no-noise limits. We also compute errors in solutions computed via Richardson extrapolation (shown in red and yellow stars). As we lower the $p_{\textrm{noise}}$, the effect of extrapolation becomes more pronounced. At the threshold of approximately $p_{\textrm{noise}}=10^{-5}$, the improvement in accuracy after extrapolations is already $\approx 98\%$ (or $\approx$ 70 times), indicating the effectiveness of the extrapolation on near-term devices.

So far we have studied a single application of the unitaries decomposition. To successfully implement a time-marching simulation, it is also important to examine the time evolution of errors. To ensure convergence, it is important to minimize the overall error at every time step. For this, first we increase the grid resolution to $N_{g}=16$ to lower the errors of grid resolution. Then we use Richardson extrapolation with $(\epsilon_{1},\epsilon_{2})=(1,0.5)$ to further improve accuracy. We then perform forward time marching simulations up to $\tau=32$ time steps, for varying levels of $p_{\textrm{noise}}$. The time evolution of the flow field, shown in figures \ref{fig:qasm_noisy N16}(a,b), demonstrate that the extrapolated solutions capture the flow physics accurately. To do better, we plot the error convergence of extrapolated solutions as in {the inset of figure \ref{fig:qasm_noisy N16}(b)}. From this it can be seen that the error at each time step for $p_{\textrm{noise}}=10^{-6}$ is still large, leading to non-convergent behavior. However, for $p_{\textrm{noise}}<10^{-6}$, the error seems to be under control and the solution begins to converge. It is useful to remind ourselves of the circuits being simulated here are fully controlled circuits without any parallelization or optimization. The circuit depths are $\approx 36963$ with a total of about $\approx 21270$ CNOT gates and these numbers can be reduced substantially by methods described {in Sec.\ \ref{sec: End-to-end}}). The error convergence can then occur at $p_{\textrm{noise}}$ well above $10^{-6}$, closer to the theoretical threshold \cite{gottesman2022opportunities,aliferis2005quantum}. Furthermore, increasing the grid resolution can also significantly improve the error at every time step, and therefore reduce the onus on the $p_{\textrm{noise}}$ values required for convergence. In any case, the results presented here can be viewed as strict estimates of the error rates needed for convergence. These estimates suggest that the algorithm presented here has a potential for actual implementation on near-term quantum devices possessing logical qubits and error-correction algorithms. Recent results from \cite{bravyi2024high} are already in the same ball-park. 

Finally, we fix the error rates at $p_{\textrm{noise}} = 10^{-8}$ and vary $N_{s}$ to study the effect of finite sampling of qubit states. We again plot the error convergence of extrapolated solutions as a function of time, but now with varying $N_{s}$ as shown in figure \ref{fig:qasm_noisy N16} ({c}). Although the overall error evolution might seem to converge, they converge to lower values with increasing $N_{s}$, as expected. To quantify this better and assess the advantage of the Richardson extrapolation, we plot the decay of the MSE with $N_{s}$ with and without extrapolation at $\epsilon_{1}=1$ and $\epsilon_{2}=0.5$, all for $\tau=32$. They correspond to errors accumulated over all previous time steps, thus also include the effect of time marching. We see that the MSE of the extrapolated solutions has a power-law decay with $N_{s}$, of the form $\sim 0.0185N^{-0.5513}_{s} \approx 1/\sqrt{N_{s}}$, which, in fact, is the Law of Large Numbers. This behavior shows that the extrapolated solutions, even after accumulating errors from several time steps, still follows the $\approx 1/\sqrt{N_{s}}$ scaling, unlike the un-extrapolated ones. This suggests that the extrapolated solutions have overcome the errors from inaccurate decomposition (due to large $\epsilon$ values) and only errors from under-sampling persist. For the solutions with no extrapolation, the MSE decays as $\sim 0.0008N^{-0.095}_{s}$ as shown in figure \ref{fig:qasm_noisy N16} ({d}). These trends show the advantage of extrapolation; for a fixed accuracy, it reduces the required number of shots by orders of magnitude.

In summary, we provide estimates of $p_{\textrm{noise}}$, $\epsilon$ and $N_{s}$, in conjunction with extrapolation techniques, to perform accurate and convergent time marching simulations with the proposed quantum algorithms. Such estimates are scarce in the literature for quantum algorithms solving PDEs in a time marching fashion. The current estimates also suggest that the algorithms introduced here have a promising potential to be implemented on near-term devices \cite{bravyi2024high} with improved accuracy. 

\subsection{Applications}

(1) \textit{Nonlinearity} -- We have so far demonstrated the working and performance of the proposed algorithms by taking a specific example of a linear, advection-diffusion problem. However, we emphasize that these algorithms are agnostic to the nonlinearity of the governing PDEs, as long as they can be reduced to a set of iterative matrix inversion or matrix-vector multiplication operations; they can be solved similarly by matching any of the TMCQC forms. Such efforts would typically involve employing a linearization technique such as Carleman \cite{liu2021efficient,gonzalez2024quantum}, Koopman \cite{giannakis2022embedding,lin2022koopman} or Homotopy \cite{bharadwaj2023quantum} methods, which embed a finite dimensional nonlinear problem as an \textit{infinite} dimensional linear problem. This generates a linear system of equations of large dimensions that can be solved using the TMCQCs presented in this work. The degree of nonlinearity that can be solved would strongly depend on the linearization used, and the complexity of the overall algorithm remains to be explored. In any case, a more generic linear system of equations arising from suitable transformations applied on an originally linear or nonlinear PDE without a known analytical solution would require the following consideration. For an arbitrary matrix equation, the truncated Neumann series would correspond to a general upper bound given by eq. (\ref{eq:factor2}). Irrespective of the value of $\Gamma$, this would still imply only a logarithmic contribution to how $P_\textrm{min}$ scales and would be given as $ P_{min} = \mathcal{O}\Bigg(\Bigg\lceil\Bigg( \frac{\log(1/\Gamma\varepsilon_{N})}{\log(1/(\kappa-1))} \Bigg)\Bigg\rceil\Bigg).$ This therefore indicates that the current methodology admits the possibility of using a linear embedding scheme, and solve general nonlinear PDEs. To this end, in an upcoming manuscript, we propose a Homotopy Analysis Algorithm to solve nonlinear PDEs using the present algorithm \cite{bharadwaj2023quantum}, thus broadening the applicability of the proposed method to practically interesting problems. Extending the present method and evaluating its performance on such applications forms the central goal of ongoing investigations.


(3) \textit{Beyond fluid dynamics} -- As mentioned already, these algorithms fundamentally offer an elegant and efficient way to apply iterative matrix-vector multiplications or inversions, as quantum circuits. Therefore, the scope of the tool proposed here is well beyond solving fluid flow problems. In fact, such operations are ubiquitous in machine learning, quantum sensing, interior point methods, image processing and so on, thus making the proposed algorithm versatile. Exploration of these aspects is a worthy enterprise, given the potential and viability of the algorithms on near-term devices and the complexity guarantees.

\section{Conclusions and Outlook}
\label{sec: Conclusions}
In summary, we have introduced a set of Time Marching Compact Quantum Circuit algorithms that require shallow circuit depths and a minimal number of unitaries to simulate unsteady PDEs governing fluid flow problems. The algorithm can be used to solve linear and nonlinear problems, if they can be cast into any one of the TMCQC formats. These algorithms have, at most, a gate complexity of  $\mathcal{O}\Big(s\epsilon,\log(N_{g}\tau),\textrm{polylog}(\epsilon/\varepsilon_{U},1/\varepsilon_{N}),\log^{-3}((\kappa-1)^{-1})\Big)$. Further, by appropriately initializing the numerical setup for the flow problem, the query complexity can also be minimized, contributing a constant prefactor in the overall time complexity. Therefore, we can say that the overall asymptotic time complexity has a near-optimal scaling (logarithmic to polylogarithmic) in $N_{g},\tau,\epsilon_{U},\epsilon_{N}$ and $\kappa$, though exponential in sparsity $s$. Since the matrices used in this work are sparse and tri-diagonal, the sparsity would contribute only a constant prefactor. At the worst, the complexity of the algorithms is still near-optimal in all system parameters, but exponential in $\tau$ and $s$. 

Our end-to-end strategies and quantum subroutines include quantum state preparation, circuit parallelization, Richardson extrapolation, and Quantum Post Processing---using which the overall algorithm can conserve the available quantum advantage. The proposed TMCQCs are then simulated on QFlowS, which is an in-house high performance quantum simulator. The results from these ideal (no-noise) simulations show that the algorithm can accurately capture the flow physics of the chosen problem, qualitatively and quantitatively.

We have then experimented unitaries decomposition by transpilation on a real quantum device (IBM Cairo) after ensuring that the circuits conform to the quantum volume requirements of the device. Again, the results from these experiments (subject to both noise and decoherence) show that the algorithm captures the flow physics, by reaching the maximum possible accuracy attainable on the device for given circuit sizes. The parallelization techniques were effective in lowering circuit depths and gate counts. These results suggest that our algorithm can be implemented on real devices. Further, when the simulation of even larger circuit sizes becomes possible, the currently transpiled circuits can be replaced by the improved Hamiltonian simulation circuits to attain the asymptotic complexities derived here.

With reference to the currently available devices and near-term devices on the horizon, it is also important to determine what specifications would be necessary to carry out a full scale simulation. To do this, and also to better understand the interplay between noise, state sampling, $\epsilon$ and their collective effect on time marching simulations, we have performed a comprehensive set of noisy simulations using Qiskit Aer. Our analysis suggests that, with Richardson extrapolation, we can carry out accurate simulations even with $\epsilon\sim\mathcal{O}(1)$, but error-convergent results are possible only for $p_{\textrm{noise}} < 10^{-6}$. Our analysis thus describes the estimates of the required cut-offs on each of these quantities in designing algorithms offering error-convergent time-marching simulations. We also highlight the role of extrapolation in reducing the errors even with large $\epsilon$ simulations, or in lowering the shot counts by orders of magnitude to achieve a given accuracy. The extrapolation itself provides up to $\sim 98\%$ improvement in performance. From our experimental results as well as the recent progress made with logical qubits and quantum error-correction codes \cite{bravyi2024high}, it appears that our estimates present a decent possibility for implementing the proposed algorithms to be implemented at full scale on near-term quantum devices; it also appears that these algorithms have a potential for demonstrating quantum advantage on near-term devices despite their limitations. Clearly, when more powerful devices become available, further effort is essential in carefully designing these algorithms, in tandem with error-correction subroutines, to achieve quantum advantage.  

In essence, the importance and applicability of the tools presented in this work extend beyond solving fluid dynamical PDEs. At a fundamental level, the algorithms offer a way for performing iterative, matrix-vector multiplication and inversion operations on exponentially large data dimensions, while also considering limitations on current and near-term devices. Apart from its utility in the previously proposed QCFD approaches, such a tool is necessary in nearly most numerical methods and therefore caters to a range of applications including machine learning, image processing, optimization and quantum sensing, to name a few. As noted earlier, the proposed algorithm can be used to solve nonlinear PDEs as well, once they are appropriately cast into a linear setting via linearization techniques. 

{Further investigations that are relegated to future include the exploration of higher order finite difference schemes and spectral methods, assessing the computational complexity requirements when the matrix $M(t)$ is also a time-dependent and the PDE is inhomogeneous, the exploration of applying the VQA approaches to compute cost functions. Exploring these possibilities and different applications in greater detail, along with experiments on other quantum computers, such as those with photonic or ion-trap qubits, is an important part of future work.}

{Ultimately, QC will profitably integrate with classical computing, taking advantage of machine learning tools, but such scenarios demand that we understand the strengths and limitations of QC on its own.}

\vspace{0.15cm}

\noindent\textbf{Acknowledgements} We thank {Balu Nadiga, Stephan Eidenbenz,} Dhawal Buaria, András Gilyén, Srinivasan Arunachalam, Philipp Pfeffer, Julia Ingelmann, Jörg Schumacher, Akash Rodhiya and Wael Itani for helpful discussions. {We wish to acknowledge New York University's Greene supercomputing facility on which part of these simulations were performed and} the IBM Quantum resource access provided by Los Alamos National Laboratory through the Oak Ridge OLCF allocation, for which we are grateful to Stephan Eidenbenz and Balu Nadiga. The views and results of this paper are those of the authors, and do not reflect the
official policy or position of IBM or the IBM Quantum team. The work was supported by New York University. \\

{\noindent \textbf{Data Availability} All the data is included in the manuscript.}\\

\textbf{Code Availability} The code will be made available following any reasonable request made to the authors.
\appendix
\input{supplementary}
\bibliography{refs}

\end{document}

%% file: supplementary.tex
\section{NUMERICAL SETUP}
\label{sec: Numerical Setup}
We outline here the numerical discretization of the governing equations using finite differences and the different methods to set up time marching equations that can be simulated using the quantum algorithms proposed in this work. We consider the 1D linear Advection-Diffusion flow ($C=U$) under periodic boundary conditions, described in the main body of the paper. We further discuss implications of using Dirichlet boundary conditions.

\subsection{Spatial discretization} 
We review the discretization of the spatial dimension of the flow domain using central finite difference schemes. We consider the domain of length $L$, to be discretized into $N_{g}$ equidistant grid points with a spacing of $\Delta x=h=L/N_{g}$ ($x_{i} = x_{0}+i\Delta x$), creating a discretization error $\sim \mathcal{O}(\Delta x^{2})$. The velocity field at each point and at time $t$ is given by $\mathbf{u}=[u(0,t), u(\Delta x,t),\cdots,u((N_{g}-1)\Delta x,t)]$. To approximate the spatial derivatives in eq.(\ref{eq:advcetion-diffusion flow}), we employ the well known 2\textsuperscript{nd} order central difference scheme
\begin{equation}
     D\frac{\partial^{2} u}{\partial x^{2}} -U\frac{\partial u}{\partial x} \approx    D\frac{u_{i+1}- 2u_{i} + u_{i-1} }{(\Delta x)^{2}} 
     - U \frac{u_{i+1} - u_{i-1} }{2\Delta x}.
     \label{eq:CDS}
\end{equation}

\subsection{Temporal discretization} To integrate in time, the temporal domain $t \in [0,T]$ is discretized into $\tau = T/\Delta t$ time steps ($t_{j}=t_{0}+j\Delta t$) using two different schemes, both admitting an error  $\sim \mathcal{O}(\Delta t)$.

\noindent  (1) \textbf{Forward Euler or Explicit method}: Applying this discretization to the time derivative together with eq.(\ref{eq:CDS}) gives 
\begin{align}
&\frac{u_{i}^{j+1}-u_{i}^{j}}{\Delta t} = D\frac{u^{j}_{i+1}- 2u^{j}_{i} + u^{j}_{i-1} }{(\Delta x)^{2}} 
     - U \frac{u^{j}_{i+1} - u^{j}_{i-1} }{2\Delta x}, \nonumber \\
     &\implies u_{i}^{j+1} = \left(\alpha - \chi \right) u_{i+1}^{j} + \left( 1- 2 \alpha \right) u_{i}^{j} +\left( \alpha + \chi \right) u_{i-1}^{j},
    \end{align}
where $\alpha = \frac{D \tau}{(\Delta x)^2} $ and $\chi = \frac{U\tau}{2 \Delta x} $ are the stability and convective parameters, respectively. For the problem under discussion, $\alpha \leq \alpha_{\textrm{cfl}} = 1/2$ is required to ensure stability of the solution (von Neumann stability criteria). This can now be cast into a matrix equation of the form
\begin{equation}
     \mathbf{u}^{j+1} = (\mathbf{I} - \mathbf{A}_{{ex}})\mathbf{u}^{j} = \mathbf{A}_{E}\mathbf{u}^{j}.
\end{equation}
(a) Under \textit{periodic boundary conditions} the matrix takes the form
\begin{equation}
    \mathbf{A}_{E} = \begin{bmatrix}
1-2\alpha & \alpha-\chi & 0 &\cdots & 0 & \alpha+\chi\\
\alpha+\chi & 1-2\alpha & \alpha-\chi & &  & 0 
\\
0 & \alpha+\chi & 1-2\alpha & \alpha-\chi & &  0
\\
\vdots  & & \ddots    & \ddots  & \ddots & \vdots\\
0 &  &  & \alpha+\chi & 1-2\alpha & \alpha-\chi\\
\alpha-\chi & 0 & \cdots & 0 & \alpha+\chi & 1-2\alpha\\

\end{bmatrix}.
\end{equation}
(b) Under \textit{Dirichlet boundary conditions}, we only solve for velocity at the interior points of the domain and not at the boundaries (since this is already given). Then the matrix would be given by
\begin{equation}
    \mathbf{A}_{E} = \begin{bmatrix}
1-2\alpha & \alpha-\chi & 0 &\cdots & 0 & 0\\
\alpha+\chi & 1-2\alpha & \alpha-\chi & &  & 0 
\\
0 & \alpha+\chi & 1-2\alpha & \alpha-\chi & &  0
\\
\vdots  & & \ddots    & \ddots  & \ddots & \vdots\\
0 &  &  & \alpha+\chi & 1-2\alpha & \alpha-\chi\\
0 & 0 & \cdots & 0 & \alpha+\chi & 1-2\alpha\\

\end{bmatrix}.
\end{equation}
\noindent (2) \textbf{Backward Euler or Implicit method}: In this method, the discretized governing equation would take the form 
\begin{align}
&\frac{u_{i}^{j+1}-u_{i}^{j}}{\Delta t} = D\frac{u^{j+1}_{i+1}- 2u^{j+1}_{i} + u^{j+1}_{i-1} }{(\Delta x)^{2}} 
     - U \frac{u^{j+1}_{i+1} - u^{j+1}_{i-1} }{2\Delta x}, \nonumber \\
     &\implies u_{i}^{j} = \left(- \alpha + \chi \right) u_{i+1}^{j+1} + \left( 1 + 2 \alpha \right) u_{i}^{j+1} +\left( -\alpha - \chi \right) u_{i-1}^{j+1}.
    \end{align}
The above equation can be written as matrix equation of the form
\begin{equation}
     \mathbf{u}^{j} = (\mathbf{I} - \mathbf{A}_{{im}})\mathbf{u}^{j+1} = \mathbf{A}_{I}\mathbf{u}^{j+1}.
\end{equation} Therefore $\mathbf{u}^{j+1} = \mathbf{A}^{-1}_{I}\mathbf{u}^{j}$. This scheme is known to be \textit{unconditionally stable} for any size of $\Delta t$ {and is not subject to any stability criteria as in the explicit case.} However, to obtain the solution at every time step requires inverting the matrix $\mathbf{A}_{I}$. In this work we employ the truncated Neumann series to approximate this inverse (described later {in Section \ref{sec: Trunc Neumann Series} of the Appendix}), which requires the condition $\vert\vert \mathbf{A}_{{im}}\vert\vert < 1$, for the series to converge. Therefore, $\alpha$ and $\chi$ need to be chosen accordingly.

(a) Under \textit{periodic boundary conditions}, the matrix takes the form
\begin{equation}
    \mathbf{A}_{I} = \begin{bmatrix}
1+2\alpha & -\alpha + \chi & 0 &\cdots & 0 & -\alpha-\chi\\
-\alpha-\chi & 1+2\alpha & -\alpha+\chi & &  & 0 
\\
0 & -\alpha-\chi & 1+2\alpha & -\alpha+\chi & &  0
\\
\vdots  & & \ddots    & \ddots  & \ddots & \vdots\\
0 &  &  & -\alpha-\chi & 1+2\alpha & -\alpha+\chi\\
-\alpha+\chi & 0 & \cdots & 0 & -\alpha-\chi & 1+2\alpha\\

\end{bmatrix}.
\end{equation}
(b) Under \textit{Dirichlet boundary conditions} the matrix is given by
\begin{equation}
    \mathbf{A}_{I} = \begin{bmatrix}
1+2\alpha & -\alpha + \chi & 0 &\cdots & 0 & 0\\
-\alpha-\chi & 1+2\alpha & -\alpha+\chi & &  & 0 
\\
0 & -\alpha-\chi & 1+2\alpha & -\alpha+\chi & &  0
\\
\vdots  & & \ddots    & \ddots  & \ddots & \vdots\\
0 &  &  & -\alpha-\chi & 1+2\alpha & -\alpha+\chi\\
0 & 0 & \cdots & 0 & -\alpha-\chi & 1+2\alpha\\

\end{bmatrix}.
\end{equation}

With this discretization, the following sections describe specific aspects of different quantum algorithms proposed for time marching.

\textit{NOTE --} The case of $\chi=0$ corresponds to the Poiseuille (pipe) flow \cite{bharadwaj2023hybrid}, where the matrices $\mathbf{A}_{E}, \mathbf{A}_{I}$ are symmetric for Periodic as well as Dirichlet boundary conditions.

\subsection{Explicit method - TMCQCs}
\label{subsec: Explicit TMCQC}
(a) \textit{Explicit expansion method}: The explicit forward time integration up to $\tau$ time steps translates to the application of the operator $(A_{E})^{\tau}$. A given matrix is first decomposed into a linear combination of unitaries, and the sum is raised to the power $\tau$ and expanded, yielding a new sum of unitaries with appropriate coefficients.

(b) \textit{Explicit iterative method}: To advance by $\tau$ time steps using the iterative method requires one to iteratively invoke the unitaries circuit to apply the original decomposition 
$\tau$ times. To extend the single step operation outlined {in Section \ref{sec: LCU} of the main text}, to perform $\tau$ iterations we introduce an additional ancillary register $q_{a0}$ with $n_{t}=\log(\tau)$ \textit{countdown} qubits. Let us consider a general non-unitary, non-hermitian matrix $M$, and consider for clarity the case for $\tau=2$ shown in figure \ref{fig:end2end}(a). 
Including the additional ancillary qubit ($n_{t}=1$), with the first application of the LCU operators as before (marked by red-dashed line ($\tau=1$ on figure \ref{fig:end2end}(a), we are left with the state proportional to 
$\vert1\rangle\frac{1}{\sqrt{\beta}}|0\rangle^{n_{a}}M|\Psi\rangle_{u} + \vert0\rangle|\Psi\rangle_{\perp}$, where
\begin{align}
    &|\Psi\rangle_{\perp} = \frac{1}{\sqrt{\beta}}\Big( \overbrace{(U_{0}+U_{1}+U_{2}+U_{3})}^{\text{M}}|00\rangle +\nonumber\\\nonumber &\underbrace{(U_{0}-U_{1}+U_{2}-U_{3})}_{\text{A}}|01\rangle +\underbrace{(U_{0}+U_{1}-U_{2}-U_{3})}_{\text{B}}|10\rangle \\  &+\underbrace{(U_{0}-U_{1}-U_{2}+U_{3})}_{\text{C}}|11\rangle \Big).
\end{align} The countdown qubits are all set to unity initially as the state $|\tau-1\rangle$ which, after every time step, are reduced by one in binary, using bit flips to finally reach $|0\rangle$, thus having counted $\tau$ time steps. Now we use the ancillary qubit $q_{a0}$ to tag the subspace of the wave function, thus associating it with a time step counter and apply a controlled NOT gate with $q_{a1}$ and $q_{a2}$ (requiring both to be 0), which yields the state
    $\vert1\rangle|0\rangle^{n_{a}}M|\Psi\rangle_{u} + \vert0\rangle|\Psi\rangle_{\perp}$. 
We then reapply the unitaries circuit similar to the first iteration, but this time, all the operations will be controlled additionally by $q_{a0}$ (requiring to be set to 1, marking the solution subspace) as shown in figure \ref{fig:end2end}(a) by the red-dashed line ($\tau=2$). This gives \begin{align}
    \vert \Psi \rangle = &\vert1\rangle\vert00\rangle\frac{1}{\sqrt{\beta'}}M^{2}\vert b\rangle +  \vert1\rangle\overbrace{\frac{1}{\sqrt{\beta'}}\Big(|01\rangle A+|10\rangle B +|11\rangle C \Big)M\vert b\rangle}^{\vert\Psi\rangle_{\perp_{1}}} \nonumber \\ &+  \vert0\rangle\underbrace{\frac{1}{\sqrt{\beta}}\Big(|01\rangle A+|10\rangle B +|11\rangle C \Big)\vert b\rangle}_{\vert\Psi\rangle_{\perp_{2}}}.
\end{align}
Finally we apply a NOT gate on $q_{a0}$ giving the state
    $\vert \Psi \rangle = \vert0\rangle\vert00\rangle\frac{1}{\sqrt{\beta'}}M^{2}\vert b\rangle + \vert1\rangle\vert\Psi\rangle_{\perp_{1}}  + \vert1\rangle\vert\Psi\rangle_{\perp_{2}}$,
and then measure the first three qubits in the computational basis. When one lands an output of (for the first three qubits), $\vert000\rangle$ upon measurement, we are left with a state proportional to $M^{2}\vert b\rangle$. 

(c) \textit{Explicit one-shot method}: Alternatively, a matrix inversion problem of the form
\begin{equation}
    \mathbf{A}_{EO}\tilde{u} = \mathbf{b}_{EO}
\end{equation}
can be setup to solve for the velocity field at all time steps together. $\mathbf{A}_{EO}$ has a double-banded structure as written in eq.\ref{eq:FE_OBM}, $(A_{EO})_{ij} = \mathbf{I}~ \forall ~ i=j $ (see below). However, $\forall~ i\leq \tau$, $ (A_{EO})_{ij} = -(\mathbf{I} - \mathbf{A}_{e}) ~\text{for}~ j=i-1 $ and $\forall~ i\geq \tau$,  $ (A_{EO})_{ij} = -\mathbf{I} ~\text{for}~ j=i-1 $. Further, $(b_{EO})_{i} = \{u_{in} \forall ~ i=0 $; $= \mathbf{f}\Delta t 
~\forall~ 0<i\leq \tau$; and $= 0~ \forall ~ i>\tau\}$. Here, $\textbf{f}=0$ identically. Following this, the inverse of matrix $\mathbf{A}_{EO}$ is approximated either as a truncated Neumann series (see later sections, for details on error bounds) or as a truncated Fourier series approach as outlined in \cite{childs2017quantum}, which also provides error bounds on the truncation. Here, we consider the former approach. First, we rewrite the inverse as $(\textbf{I}-(\textbf{I}-\mathbf{A}_{EO}))^{-1}$. 
Then, under the constraint of $\vert\vert\textbf{I}-\mathbf{A}_{EO}\vert\vert<1$, the matrix inverse is given by
\begin{equation}
    \mathbf{A}^{-1}_{EO} \approx  \sum_{p=0}^{P}(\textbf{I}-\mathbf{A}_{EO})^{p} = \sum_{p=0}^{P}(\tilde{\mathbf{A}}_{EO})^{p}.
\end{equation} 
\begin{figure*}
\begin{equation}\begin{bmatrix}
\mathbf{I} & 0 & \cdots &  &  &  &0\\
-(\mathbf{I} - \mathbf{A}_{e}) & \mathbf{I} &  & &  & 
\\
0 & \ddots & \ddots    &\\
\vdots &  & -(\mathbf{I} - \mathbf{A}_{e}) & \mathbf{I} & &\\
 & & &-\mathbf{I} & \mathbf{I} & &\\
 &  &  &   & \ddots & \ddots\\
0 &  & \cdots & & 0 & -\mathbf{I} & \mathbf{I} \\

\end{bmatrix} \begin{bmatrix}
u^{0}\\u^{1}\\ \vdots\\ u^{\tau}\\ u^{\tau+1}\\ \vdots\\ u^{\tau+\textcolor{blue}{c}}\\ 
\end{bmatrix} = \begin{bmatrix}
u^{in}\\b_{0}\\ \vdots\\ b_{\tau-1}\\ 0\\ \vdots\\ 0\\ 
\end{bmatrix} 
\label{eq:FE_OBM}
\end{equation}
\begin{equation}
\begin{bmatrix}
\mathbf{I} & 0 & \cdots &  &  &  &0\\
-\mathbf{I} & (\mathbf{I} - \mathbf{A}_{i}) &  & &  & 
\\
0 & \ddots & \ddots    &\\
\vdots &  & -\mathbf{I}  & (\mathbf{I} - \mathbf{A}_{i}) & &\\
 & & &-\mathbf{I} & \mathbf{I} & &\\
 &  &  &   & \ddots & \ddots\\
 0&  & \cdots & &0  & -\mathbf{I} & \mathbf{I} \\

\end{bmatrix} \begin{bmatrix}
u^{0}\\u^{1}\\ \vdots\\ u^{\tau}\\ u^{\tau+1}\\ \vdots\\ u^{\tau+\textcolor{blue}{c}}\\ 
\end{bmatrix} = \begin{bmatrix}
u^{in}\\b_{0}\\ \vdots\\ b_{\tau-1}\\ 0\\ \vdots\\ 0\\ 
\end{bmatrix} 
\label{eq:BE_OBM}
\end{equation}
\end{figure*}
\subsection{Implicit method - TMCQCs}
\label{subsec: Implicit TMCQC}

(a) \textit{Implicit iterative method}: This method again, requires one to simply multiply the initial state by the matrix $\mathbf{A}_{NI}$, $\tau$ times iteratively or serially giving
\begin{equation}
    u^{\tau} = (\mathbf{A}_{NI})^{\tau}u^{0},
\end{equation} where the above matrix $\mathbf{A}_{NI}$ is a truncated Neumann series approximation to the inverse of $\mathbf{A}_{I}$, given by
\begin{equation}
    \mathbf{A}_{NI} = \mathbf{A}^{-1}_{I} \approx  \sum_{p=0}^{P}\mathbf{A}^{p} 
\end{equation} computed up to $P$ terms. This approximation is convergent only when $\alpha$ is chosen such that $\vert\vert\alpha\mathbf{A} \vert\vert<1$.

(b) \textit{Implicit expansion method}: This method proceeds exactly as the explicit expansion method, except that the matrix $\mathbf{A_{NI}}$ is used to compute the expansion.

(c) \textit{Implicit one-shot method}: One can also alternatively set up a matrix inversion problem of the form
\begin{equation}
    \mathbf{A}_{IO}\tilde{u} = \mathbf{b}_{IO},
\end{equation}
to solve for the velocity field at all time steps together. The matrix will be of the form shown in eq.\ref{eq:BE_OBM}, where $(A_{IO})_{ij} = (\mathbf{I} - \mathbf{A}_{i}) ~ \forall ~ i=j $ and $1<i<\tau$. However, $\forall~ i\leq \tau$, $ (A_{IO})_{ij} = -\mathbf{I} ~\text{for}~ j=i-1 $ and $\forall~ i\geq \tau$,  $ (A_{IO})_{ij} = -\mathbf{I} ~\text{for}~ j=i-1 $. Further, $(b_{IO})_{i} = \{u_{in} \forall ~ i=0 $; $= -\mathbf{f}\Delta t 
~\forall~ 0<i\leq \tau $; and $= 0~ \forall ~ i>\tau\}$. Here again, $\textbf{f}=0$ identically. Further the inverse of matrix $\mathbf{A}_{IO}$ is again given by the truncated Neumann approximation as,
\begin{equation}
    \mathbf{A}^{-1}_{IO} \approx  \sum_{p=0}^{P}(\textbf{I}-\mathbf{A}_{IO})^{p} = \tilde{\mathbf{A}}_{IO}
\end{equation} 

\section{Truncated Neumann Series}
\label{sec: Trunc Neumann Series}

The matrix inversion problem of the kind $(\mathbf{I}-\mathbf{M})^{-1}$, under the condition $\rho(\mathbf{M})<1$ (where $\rho$ is the spectral radius), can be rewritten using the Neumann power series as follows:
\begin{equation}
    (\mathbf{I}-\mathbf{M})^{-1} = \sum_{p=0}^{\infty}\mathbf{M}^{p}.
    \label{eq: Neumann series}
\end{equation}
To approximate the inverse using such a summation, we compute a truncated sum of eq.(\ref{eq: Neumann series}) up to $p=P_{min}-1$ terms. Then it can be shown that for Laplacian type operators outlined earlier, the error bound of truncation $\varepsilon_{N}$ is given by \textit{Theorem 1}.

\textbf{Lemma 1} \textit{Consider a central finite difference matrix $\mathbf{M}\in\mathbb{R}^{N\times N}$, corresponding to a flow field discretized into $N$ grid points, integrated up to time $T$ (with $\tau$ time steps), with an $\alpha = D\tau/(\Delta x)^{2}$ and $\chi = U\tau/(2\Delta x)$, chosen such that $\vert\vert\mathbf{M}\vert\vert<1$. If the inverse $(\mathbf{I}-\mathbf{M})^{-1}$, is approximated by the truncated Neumann series as $\sum_{p=0}^{P_{min}-1}\mathbf{M}^{p}$, the truncation error of $\varepsilon_{N} = \vert\vert (\mathbf{I}-\mathbf{M})^{-1} - \sum_{p=0}^{P_{min}-1}\mathbf{M}^{p}\vert\vert$ is bounded from above by
\begin{equation}
    \varepsilon_{N} \leq \mathcal{O}( \vert\vert \mathbf{M}\vert\vert^{P_{min}}) = \mathcal{O}((\kappa-1)^{P_{min}}),
    \label{eq: eps_N bound}
\end{equation}
where $\kappa =  \vert\vert \mathbf{I}-\mathbf{M}\vert\vert \cdot \vert\vert (\mathbf{I}-\mathbf{M})^{-1}\vert\vert \leq 2$ is the condition number and $P_{min}$ is the number of terms needed to admit a specified $\varepsilon_{N}$ is bounded below accordingly by
\begin{equation}
    P_{min} = \mathcal{O}\Bigg(\Bigg\lceil\Bigg( \frac{\log(1/\varepsilon_{N})}{\log(1/(\kappa-1))} \Bigg)\Bigg\rceil\Bigg).
    \label{eq: pmin}
\end{equation}}

\textbf{Proof -} Starting with the Neumann series representation as in eq.
\ref{eq: Neumann series}, consider a truncated series $\mathbf{R}_{P_{min}}$ computed up to $P_{min}$ terms
\begin{equation}
    \mathbf{R}_{P_{min}} = \mathbf{I}+\mathbf{M}+\mathbf{M}^{2}+\cdots +\mathbf{M}^{P_{min}-1}.
\end{equation}
The truncation error would therefore be given by
\begin{align}
    \varepsilon_{N} &= \vert\vert (\mathbf{I}-\mathbf{M})^{-1} - \mathbf{R}_{P_{min}}\vert\vert = \vert\vert \sum_{p=0}^{\infty}\mathbf{M}^{p} - \sum_{p=0}^{P_{min}-1}\mathbf{M}^{p}\vert\vert \nonumber \\ \nonumber
    &=  \vert\vert \mathbf{M}^{P_{min}}+\mathbf{M}^{P_{min}+1} + \cdots \vert\vert \\ 
    &=  \vert\vert \mathbf{M}^{P_{min}}\Big(\sum_{p=0}^{\infty}\mathbf{M}^{p}\Big) \vert\vert
    =  \vert\vert \mathbf{M}^{P_{min}}(\mathbf{I}-\mathbf{M})^{-1}) \vert\vert \\
    &\leq  \vert\vert \mathbf{M}^{P_{min}}\vert\vert \cdot \vert\vert (\mathbf{I}-\mathbf{M})^{-1}) \vert\vert
    \label{eq:product terms}
\end{align}
Consider the first factor in this eq.(\ref{eq:product terms}), given the submultiplicativity of the matrix norm we have
\begin{equation*}
= \vert\vert \mathbf{M}^{P_{min}}\vert\vert
    \leq \prod_{P_{min}}\vert\vert \mathbf{M}\vert\vert = \vert\vert \mathbf{M}\vert\vert^{P_{min}}
    \label{eq:factor1}.
\end{equation*}

For the second factor in eq.(\ref{eq:product terms}), again noting that $\vert\vert \mathbf{M}\vert\vert \leq \vert\vert \mathbf{M}\vert\vert_{\infty}$, and additionally by invoking the Varah bound \cite{varah1975lower}, we obtain 
\begin{align}
    &\vert\vert (\mathbf{I}-\mathbf{M})^{-1}) \vert\vert_{\infty} \leq \frac{1}{\Gamma} \textrm{~where,} \nonumber \\ &\Gamma = \min_{i}\Big\{\vert (\mathbf{I}-\mathbf{M})_{i,i}\vert - \sum_{i\neq j}\vert (\mathbf{I}-\mathbf{M})_{i,j}\vert\Big\} \nonumber \\\implies &\vert\vert (\mathbf{I}-\mathbf{M})^{-1}) \vert\vert_{\infty}
    \leq \frac{1}{\Gamma(\alpha,\chi)}
    \label{eq:factor2}.
\end{align}
Now given a 2\textsuperscript{nd} order scheme, $\Gamma(\alpha,\chi)$ depends on the choice of boundary conditions and magnitudes of $\alpha$ and $\chi$. For both boundary conditions, $\Gamma = 1$ when $\chi \leq \alpha$, while for $\chi > \alpha$, $\Gamma=1-2(\alpha-\chi)$. The case $\chi=0$ corresponds to the Poiseuille flow problem, which has a symmetric finite difference matrix for both boundary conditions. 

In this work we consider the case $\chi\leq \alpha$ for the following reason. First, we note that $Pe = UL/D$ represents the Peclét number, which is the ratio of the advective to diffusive time scales. In other words, $Pe\rightarrow 0$ corresponds to a flow that is mainly diffusive, while for $Pe\rightarrow \infty$ the flow is dominated mainly by the advection process. Rewriting the inequality between $\chi$ and $\alpha$, we note that $\chi>\alpha$ and $\chi \leq \alpha$ correspond to $Pe>2N$ and $Pe\leq 2N$, respectively. Since the former entails only advective processes, we choose the latter to allow both diffusive and advective processes together. Now, since for this case $\Gamma=1$, from eq.(\ref{eq:factor1}) and eq.(\ref{eq:factor2}), we get 
\begin{equation}
\varepsilon_{N} \leq \vert\vert \mathbf{M}\vert\vert^{P_{min}}
    \label{eq:eps_N bound stage 1}.
\end{equation}
To compute the bound for $\vert\vert \mathbf{M}_{\infty}\vert\vert$, we proceed as follows. The resource requirement for a matrix inversion depends on the condition number $\kappa = \vert\vert \mathbf{I}-\mathbf{M}\vert\vert \cdot \vert\vert (\mathbf{I}-\mathbf{M})^{-1}\vert\vert$. So let us consider the inequality
\begin{align}
    \vert\vert (\mathbf{I}-\mathbf{M})) \vert\vert &\leq \vert\vert (\mathbf{I}-\mathbf{M})) \vert\vert_{\infty} \nonumber\\
    \frac{\kappa}{\vert\vert (\mathbf{I}-\mathbf{M})^{-1}\vert\vert}  &\leq \vert\vert \mathbf{I} \vert\vert_{\infty} + \vert\vert \mathbf{M} \vert\vert_{\infty} = 2.
    \label{eq:kappa leq 2}
\end{align}
The above equations follow again from eq.(\ref{eq:factor2}) when $\Gamma=1$.
Now we note that 
\begin{equation}
    0 \leq \vert\vert \mathbf{M} \vert\vert \leq 1 .
    \label{eq: norm M bounds}
\end{equation}
This is a necessary condition for inversion and therefore the eigenvalues are bound between $-1 \leq \lambda \leq 1$. Since we know $\kappa\geq 1$ always, we get from eq.(\ref{eq:kappa leq 2}):
\begin{equation}
    1 \leq \kappa \leq 2.
    \label{eq: kappa bounds}
\end{equation}
By subtracting equation (\ref{eq: norm M bounds}) from (\ref{eq: kappa bounds}) we note 
\begin{equation}
    \vert\vert \mathbf{M} \vert\vert \leq \kappa -1,
\end{equation}
and thus prove that
\begin{equation}
    \varepsilon_{N} \leq \mathcal{O}((\kappa-1)^{P_{min}}).
\end{equation}
By rearranging the above equation, we show that
\begin{align}
    P_{min} &=\mathcal{O}\Bigg( \Bigg\lceil\frac{\log(1/\varepsilon_{N})}{\log(1/ \vert \vert \mathbf{M}\vert\vert)}\Bigg\rceil\Bigg) \nonumber\\ &=\mathcal{O}\Bigg(\Bigg\lceil\Bigg( \frac{\log(1/\varepsilon_{N})}{\log(1/(\kappa-1))} \Bigg)\Bigg\rceil\Bigg).
\end{align}
The bound clearly shows that, for both increasing accuracy (small $\varepsilon_{N}$) and larger the $\kappa$ requires a larger $P_{min}$ to estimate the matrix inverse.

\begin{figure}
    \centering
    \includegraphics[trim={0.3cm 0.3cm 0.3cm 0.1cm},clip=true,scale=0.25]{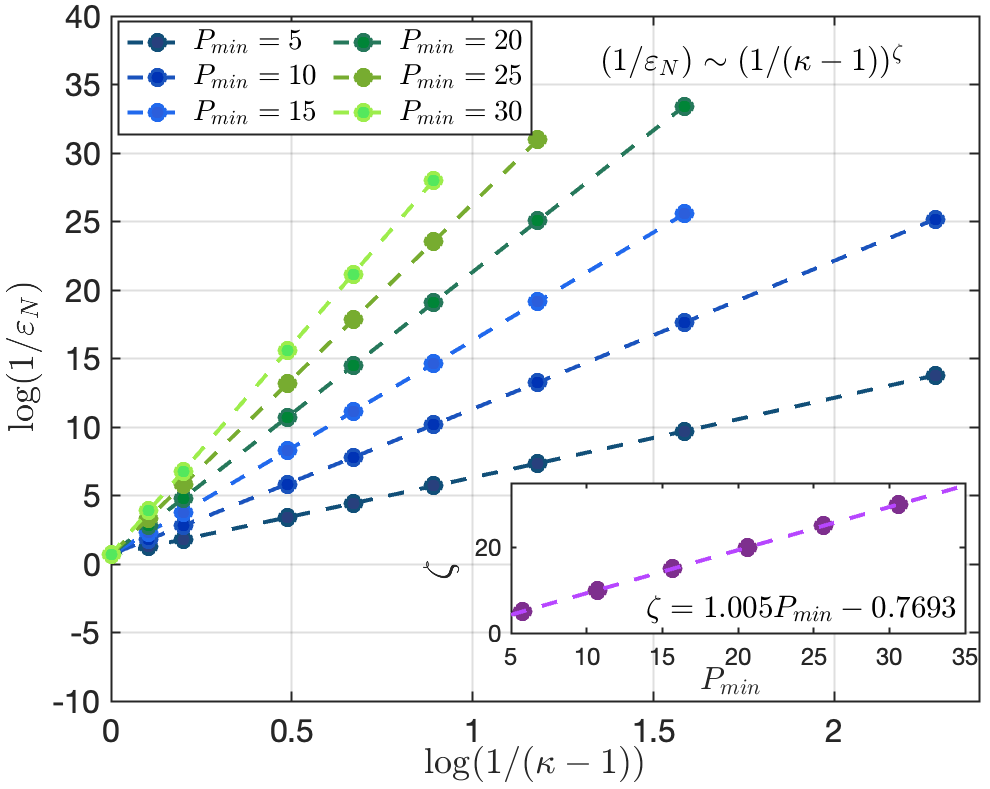}
    \caption{\justifying The power-law like variation of truncation error with condition number is shown, for increasing values of $P_{min}$. These values correspond to an implicit time marching problem for $N_{g}=32$, $D=1$ and $C=10$. The inset shows the scaling between the chosen $P_{min}$ and the exponents of the power-law fits ($\zeta$), indicating a direct correspondence.}
    \label{fig: kappa error and Pmin}
\end{figure}

As a confirmation of this result, we numerically compute the dependence of $\varepsilon_{N}$ with $\kappa$ for varying cut-off $P_{min}$ values. The power-law fits to these relations are shown in figure \ref{fig: kappa error and Pmin}. The fits confirm the theoretical result, eq.(\ref{eq: eps_N bound}).

\textbf{Proposition 1} \textit{A suitable combination of $\alpha$ and $\chi$ exists such that for both explicit and implicit time schemes, the corresponding finite difference matrix operators $\mathbf{M}$ (or $\mathbf{(I-M)})$ can be re-scaled by a positive real number $\delta \lesssim \mathcal{O}(1/\epsilon)$ such that $Q = \epsilon\delta\vert\vert \mathbf{M}\vert\vert \gtrsim 1$, while still ensuring $\vert\vert \delta \mathbf{M}\vert\vert = \delta\vert\vert \mathbf{M}\vert\vert<1$} and $\alpha<\alpha_{\textrm{cfl}}$ for implicit and explicit schemes, respectively.

In \textit{Proposition 1}, the effect of such a scaling in a quantum algorithm is that the resulting solutions simply scale similarly. These can be easily rescaled back classically, to obtain the original scale solution. 
A few clarifications at this point motivate this scaling. To accurately approximate the matrix through linear combination of unitaries, one requires small $\epsilon$. However, such small $\epsilon$ would require a large $N_{s}$ to extract the solution. To ameliorate this, we use the Richardson extrapolation (see Sections \ref{sec: End-to-end} \& \ref{sec: Numerical results}) to produce accurate solutions even with larger $\epsilon$ values, as discussed further in \textit{Lemma 2}. The method essentially offers a deferred approach to the $\lim\epsilon\rightarrow 0$ case. It is also necessary to distinguish the solution from the noise on the real quantum devices. The error from a simple second order extrapolation would generally be of the order $\mathcal{O}(\epsilon^{2})$. In Section \ref{sec: Numerical results}, for instance, we present results from extrapolation with $\epsilon$ as large as 0.5 up to 1.5, which demonstrate high accuracy. For an $\epsilon$ that gives reasonably accurate solutions, we also need to make sure that the corresponding $N_{s}$ is a small number (see eq.(\ref{eq: eta query complexity}). To do this we pick a $\delta$ to scale $A$ such that $\delta\epsilon||A||\gtrsim 1$. The $\delta$ should also ensure that $||A||<1$ for the explicit case, while the stability criteria is satisfied for the implicit case.

\textbf{Lemma 2} \textit{The time marching operators as approximated by a second order, Richardson extrapolated, the block encoding of linear combination of unitaries for (a) TMCQC1,2 (b) TMCQC3,4 and (c) TMCQC5,6 admit an error $\varepsilon = \vert\mathbf{X} - \mathbf{Y}\vert_{\textrm{max}}$,of at most (a) $\mathcal{O}(\tau\epsilon^{4})$, (b) $\mathcal{O}(\tau P^{3}_{min}\epsilon^{4})$ and (c) $\mathcal{O}(P^{2}_{min}\epsilon^{4})$.}\\

\noindent\textbf{Proof} -- The above follows directly from error from unitaries block-encoding, error from Hamiltonian simulation and the query complexity associated with implementing these steps in each TMCQC method. Without loss of generality, we can set the error due to unitary decomposition $\varepsilon_{LCU}$ = $\varepsilon_{U} = \varepsilon$ (Ham-Sim error). We first note that the two-step and four-step unitary decomposition as in eq.(\ref{eq:lcu},\ref{eq: lcu2}) contributes an error of the order $\varepsilon = \mathcal{O}(\epsilon^2)$. Further applying the second order Richardson extrapolation approximates the solution with an error $\mathcal{O}(\epsilon^{4})$. The overall error for each case is computed as follows:
    
    (a) \textit{TMCQC1,2} -- This method involves applying the unitaries encoding $\tau$ times iteratively for $\tau$ time steps. Therefore the cumulative error can be easily seen to be $\mathcal{O}(\tau\epsilon^{4})$.
    
    (b) \textit{TMCQC3,4} -- In this case the LCU is applied $\tau P^{3}_{min}$ times, yielding an overall error of $\mathcal{O}(\tau P^{3}_{min}\epsilon^{4})$.
    
    (c) \textit{TMCQC5,6} -- For these one shot methods, the final matrix needs to be inverted and approximated as a truncated Neumann series with $P_{min}$ terms. Every p-th term in the series that is given by $\mathbf{M}^{p}$, admits an error of $p\epsilon^{4}$. This error is then summed up to $P_{min}$ terms, giving an overall error of $\mathcal{O}(P^{2}_{min}\epsilon^{4})$.

From the above and eq.(\ref{eq:Pmin}), we can connect the overall block-encoding error with the truncated Neumann series error as follows for each case:

\vspace{0.15cm}

(a) $\varepsilon = \mathcal{O}(\tau\epsilon^{4})$ -- (\textit{\textrm{TMCQC}1,2})  

(b)$\varepsilon = \mathcal{O}\Bigg(\tau\epsilon^{4}\Bigg\lceil\Bigg( \frac{\log(1/\varepsilon_{N})}{\log(1/(\kappa-1))} \Bigg)\Bigg\rceil^{3}\Bigg)$ --(\textit{\textrm{TMCQC}3,4} )

(c)$\varepsilon = \mathcal{O}\Bigg(\epsilon^{4}\Bigg\lceil\Bigg( \frac{\log(1/\varepsilon_{N})}{\log(1/(\kappa-1))} \Bigg)\Bigg\rceil^{2}\Bigg)$ --
(\textit{\textrm{TMCQC}5,6}) 

We therefore note that, one can always choose an appropriately \textit{small} $\epsilon$ such that $\varepsilon \leq 0.5$, which is used as the premise for \textit{Proposition 2} (see below) to show that, when these approximated operators are acted on the quantum states, the corresponding error in the solution is also bounded similarly. Further given the $\epsilon^{4}$ factor, $\epsilon$ need not be extremely small. In other words, even if one chooses a relatively large $\epsilon$ to ensure that the effect of noise does not dominate the solution, $\varepsilon < 0.5$ can still be satisfied.

\textbf{Proposition 2} (see \textit{Proposition 9} of \cite{childs2017quantum})
\textit{Given a Hermitian operator $\mathbf{J}$ such that $\vert\vert\mathbf{J}^{-1}\vert\vert \leq 1$ and $\bar{\mathbf{J}}$ is an approximation of $\mathbf{J}$ such that $\vert\vert\mathbf{J}-\bar{\mathbf{J}}\vert\vert \leq \varepsilon < 0.5$, the corresponding solution states $\vert\mathbf{u}\rangle = \mathbf{J}\vert\mathbf{u}\rangle/\vert\vert\mathbf{J}\vert\mathbf{u}\rangle\vert\vert$ and $\vert\bar{\mathbf{u}}\rangle = \bar{\mathbf{J}}\vert\bar{\mathbf{u}}\rangle/\vert\vert\bar{\mathbf{J}}\vert\bar{\mathbf{u}}\rangle\vert\vert$ satisfy $\vert\vert~\vert\vert\mathbf{u}\rangle-\vert\bar{\mathbf{u}}\rangle~\vert\vert \leq 4\varepsilon$}.